\documentclass[12pt]{iopart}

\usepackage{amssymb}
\usepackage{txfonts}
\usepackage{dcolumn}
\usepackage{bm}
\usepackage{color}
\usepackage{ulem}
\usepackage{graphicx}
\usepackage{braket}
\usepackage[perpage]{footmisc}

\begin{document}

\title[Topological spin crystals by itinerant frustration]{Topological spin crystals by itinerant frustration}

\author{Satoru Hayami and Yukitoshi Motome}

\address{Department of Applied Physics, University of Tokyo, Bunkyo, Tokyo 113-8656, Japan}
\ead{hayami@ap.t.u-tokyo.ac.jp}
\vspace{10pt}
%\begin{indented}
%\item[]August 2017
%\end{indented}

\begin{abstract}
Spin textures with nontrivial topology, such as vortices and skyrmions, have attracted attention as a source of unconventional magnetic, transport, and optical phenomena. Recently, a new generation of topological spin textures has been extensively studied in itinerant magnets; 
in contrast to the conventional ones induced, e.g., by the Dzyaloshinskii-Moriya interaction in noncentrosymmetric systems, they are characterized by extremely short magnetic periods and stable even in centrosymmetric systems. Here we review such new types of topological spin textures with particular emphasis on their stabilization mechanism. 
Focusing on the interplay between charge and spin degrees of freedom in itinerant electron systems, we show that itinerant frustration, which is the competition among electron-mediated interactions, plays a central role in stabilizing a variety of topological spin crystals including a skyrmion crystal with unconventional high skyrmion number, meron crystals, and hedgehog crystals. 
We also show that the essential ingredients in the itinerant frustration are represented by bilinear and biquadratic spin interactions in momentum space. 
This perspective not only provides a unified understanding of the unconventional topological spin crystals but also stimulates further exploration of exotic topological phenomena in itinerant magnets. 
\end{abstract}

%
% Uncomment for keywords
%\vspace{2pc}
%\noindent{\it Keywords}: XXXXXX, YYYYYYYY, ZZZZZZZZZ
%
% Uncomment for Submitted to journal title message
%\submitto{\JPA}
%
% Uncomment if a separate title page is required
%\maketitle
% 
% For two-column output uncomment the next line and choose [10pt] rather than [12pt] in the \documentclass declaration
%\ioptwocol
%

%\tableofcontents

%%%%%%%%%%%%%%%%%%%%%%%%%%%%%%%%%
\section{Introduction}
\label{sec:intro}

Noncollinear and noncoplanar spin textures have drawn considerable interest in condensed matter physics, since they often give rise to topologically nontrivial quantum states and associated unconventional phenomena. 
Such intriguing aspects are brought by chirality degrees of freedom consisting of multiple-spin products: the spin vector chirality defined by a vector product of spins, $\mathbf{S}_i \times \mathbf{S}_j$, and the spin scalar chirality by a triple scalar product, $\mathbf{S}_i \cdot (\mathbf{S}_j \times \mathbf{S}_k)$~\cite{Nagaosa_RevModPhys.82.1539,Xiao_RevModPhys.82.1959,batista2016frustration}. 
The noncollinear and noncoplanar spin configurations with nonzero vector and scalar chiralities can result in unconventional electronic structures and transport properties through the spin Berry phase~\cite{berry1984quantal,Loss_PhysRevB.45.13544,Ye_PhysRevLett.83.3737,tatara2002chirality,Katsura_PhysRevLett.95.057205,taguchi2001spin}. 
The quantum topological Hall effect is one of such unconventional phenomena, directly reflecting the nontrivial topology in the electronic band structure modified by noncoplanar spin textures; notably, the Hall coefficient can be quantized at an integer value when the system becomes a topologically nontrivial Chern insulator~\cite{Ohgushi_PhysRevB.62.R6065,Shindou_PhysRevLett.87.116801,Martin_PhysRevLett.101.156402}.

\begin{figure}[h!]
\begin{center}
\includegraphics[width=1.0 \hsize]{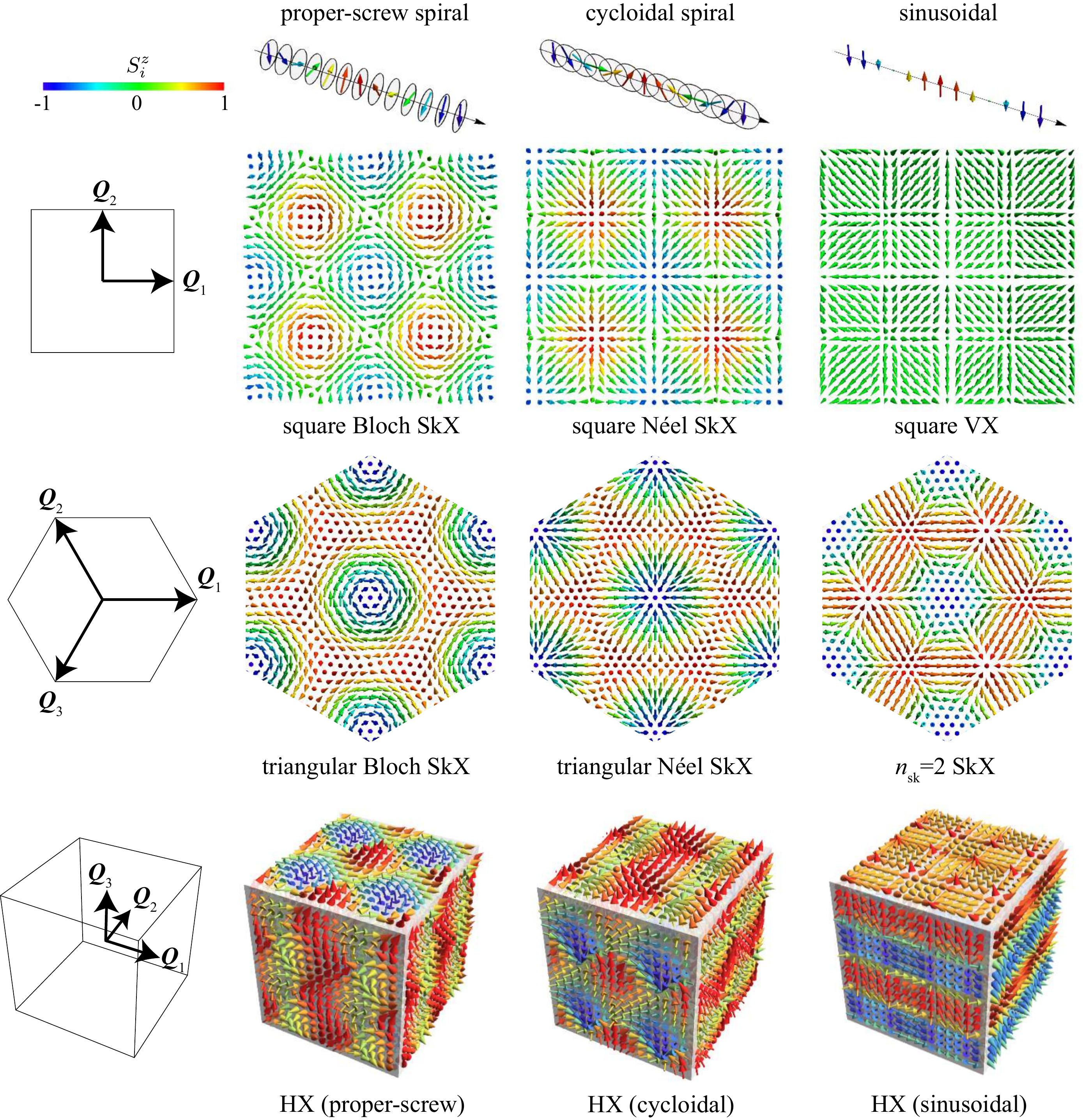} 
\caption{
\label{fig:Intro_crystal}
Schematic pictures of various topological spin crystals given by the superposition of the proper-screw spirals, cycloidal spirals, and sinusoidal waves on square, triangular, and cubic lattices. 
SkX, VX, and HX stand for the skyrmion crystal, vortex crystal, and hedgehog crystal, respectively. 
The leftmost panels represent the wave vectors for each superposition. 
In the right three columns, the color of the arrows represents the $z$ component of spins. 
}
\end{center}
\end{figure}

Theoretically, various noncollinear and noncoplanar spin textures can be engineered by taking superpositions of spin density waves, which are called multiple-$Q$ magnetic states~\cite{Bak_PhysRevLett.40.800,Shapiro_PhysRevLett.43.1748,bak1980theory,Forgan_PhysRevLett.62.470,batista2016frustration}. 
We present several examples in figure~\ref{fig:Intro_crystal}. 
In the two-dimensional case, a superposition of proper-screw spirals can constitute a periodic array of the Bloch-type skyrmions, which is called the Bloch skyrmion crystal (SkX)~\cite{Muhlbauer_2009skyrmion,yu2010real,seki2012observation}, while a superposition of cycloidal spirals can lead to a N\'eel SkX~\cite{kezsmarki_neel-type_2015,Kurumaji_PhysRevLett.119.237201}, as shown in figure~\ref{fig:Intro_crystal}. 
Meanwhile, multiple-$Q$ sinusoidal waves can give rise to a vortex crystal (VX) with a periodic array of coplanar spin vortices and another type of the SkX with the skyrmion number of two ($n_{\rm sk}=2$)~\cite{Ozawa_PhysRevLett.118.147205,Hayami_PhysRevB.95.224424}. 
In each case, square- and triangular-lattice-type spin superstructures are obtained by superpositions of two orthogonal and three $120^\circ$ waves, respectively. 
On the other hand, in the three-dimensional case, multiple-$Q$ magnetic states may give rise to periodic arrays of another topological objects, the magnetic hedgehogs~\cite{Binz_PhysRevB.74.214408,Park_PhysRevB.83.184406,Yang2016,tanigaki2015real,kanazawa2017noncentrosymmetric,fujishiro2019topological,Okumura_PhysRevB.101.144416}. 
In the bottom of figure~\ref{fig:Intro_crystal}, we display three examples of such hedgehog crystals (HXs) composed of three proper-screw, cycloidal, and sinusoidal waves. 
Reflecting the noncollinear and noncoplanar spin configurations, all these spin textures exhibit unusual multiferroic phenomena and quantum transports through the spin Berry phase mechanism~\cite{Neubauer_PhysRevLett.102.186602,Kanazawa_PhysRevLett.106.156603,nagaosa2013topological,okamura2013microwave,Mochizuki_PhysRevB.87.134403,Hamamoto_PhysRevB.92.115417,Gobel_PhysRevB.99.060406,Zou_PhysRevLett.125.267201}. 
In this review, we call such multiple-$Q$ magnetic states with nontrivial topology as ``topological spin crystals"~\cite{fujishiro2020engineering}
\footnote{
We use the term ``topological" 
in a broader sense in the following to include the magnetic spin textures in which the spin chirality is nonzero locally but the integrated value over the whole system is canceled out.
}.

Several stabilization mechanisms have been proposed for such topological spin crystals. 
A famous one is based on the Dzyaloshinskii-Moriya (DM) interaction in noncentrosymmetric magnets where spatial inversion symmetry is broken in the lattice structure~\cite{dzyaloshinsky1958thermodynamic,moriya1960anisotropic}. 
As the DM interaction has a form of $\mathbf{D}_{ij} \cdot \mathbf{S}_i \times \mathbf{S}_j$, where $\mathbf{D}_{ij}$ is the DM vector set by the lattice structure, it tends to twist the spin configurations. 
Indeed, competition between the DM and ferromagnetic interactions gives rise to instabilities toward the SkXs in an external magnetic field~\cite{rossler2006spontaneous,Yi_PhysRevB.80.054416,Butenko_PhysRevB.82.052403,Wilson_PhysRevB.89.094411,Mochizuki_PhysRevLett.108.017601,Banerjee_PhysRevX.4.031045,Gungordu_PhysRevB.93.064428,Rowland_PhysRevB.93.020404,Leonov_PhysRevB.96.014423}. 
Besides, the topological spin crystals are also stabilized by various other mechanisms, such as the long-ranged magnetic dipole interactions~\cite{lin1973bubble,malozemoff1979magnetic,Garel_PhysRevB.26.325,takao1983study,Ezawa_PhysRevLett.105.197202}, frustrated exchange interactions~\cite{Okubo_PhysRevLett.108.017206, leonov2015multiply,Lin_PhysRevB.93.064430,Hayami_PhysRevB.93.184413}, and multiple-spin interactions~\cite{Momoi_PhysRevLett.79.2081,Kurz_PhysRevLett.86.1106,heinze2011spontaneous,Yoshida_PhysRevLett.108.087205}. 
These mechanisms have been discussed to unveil the microscopic origin of a variety of topological spin crystals discovered in experiments~\cite{Tokura_doi:10.1021/acs.chemrev.0c00297}, such as the SkXs in B20 compounds~\cite{ishikawa1976helical,beille1983long,Muhlbauer_2009skyrmion,yu2010real,yu2011near}, other intermetallic compounds~\cite{tokunaga2015new,karube2016robust,Li_PhysRevB.93.060409,karube2018disordered,Karube_PhysRevB.98.155120}, oxides~\cite{seki2012observation,Adams2012,Seki_PhysRevB.85.220406,Kurumaji_PhysRevLett.119.237201}, sulfides~\cite{kezsmarki_neel-type_2015}, and monolayers~\cite{heinze2011spontaneous,romming2013writing}, antiskymions in Heusler compounds~\cite{nayak2017discovery,peng2020controlled}, and meron crystals in a magnetic alloy~\cite{yu2018transformation}. 

In this article, we give an overview on yet another mechanism, {\it itinerant frustration}, which is partly related to the frustrated exchange interactions and the multiple-spin interactions. 
This is a mechanism inherent to itinerant magnets where the spin and charge degrees of freedom of electrons are coupled by electron correlations. 
In such spin-charge coupled systems, the itinerant electrons induce effective magnetic interactions which tend to twist the spin configurations. 
The most well-known interaction is the Ruderman-Kittel-Kasuya-Yosida (RKKY) interaction derived by the perturbation in terms of the spin-charge coupling~\cite{Ruderman,Kasuya,Yosida1957}. 
The RKKY interaction is long-ranged and oscillating with a period set by the characteristic Fermi wave number in itinerant electrons; 
thus, it favors a single-$Q$ spiral state characterized by the Fermi wave number. 
Nonetheless, since there are in general several symmetry-related Fermi wave vectors according to the symmetry of the Fermi surface, the instability toward the single-$Q$ spiral state may occur at the multiple wave vectors simultaneously. 
This is a kind of frustration characteristic of itinerant electron systems, in the sense that the electron-mediated interaction leads to the degeneracy between different single-$Q$ spirals states. 
We call this the itinerant frustration. 
In this situation, higher-order contributions beyond the RKKY interaction, which are given in a form of multiple-spin interactions, lift the degeneracy and may stabilize multiple-$Q$ spin states. 
Among many contributions, it has been unveiled that an effective positive biquadratic (four-spin) interaction plays an important role in stabilizing topological spin crystals~\cite{Akagi_JPSJ.79.083711,Akagi_PhysRevLett.108.096401,Hayami_PhysRevB.90.060402,Ozawa_doi:10.7566/JPSJ.85.103703,Hayami_PhysRevB.94.024424,Hayami_PhysRevB.95.224424,lounis2020multiple,hayami2020multiple}.
As the underlying mechanism is generic in itinerant magnets and is irrespective of lattice structures, it has garnered attention for understanding the microscopic origins of various topological spin crystals, especially the recently discovered ones that are hard to understand by the conventional scenarios because of the extremely short magnetic periods and the centrosymmetric lattice structures. 
Indeed, the itinerant frustration has been intensively discussed, e.g., for the VXs in MnSc$_2$S$_4$~\cite{Gao2016Spiral,gao2020fractional}, CeAuSb$_2$~\cite{Marcus_PhysRevLett.120.097201,Seo_PhysRevX.10.011035}, and Y$_3$Co$_8$Sn$_4$~\cite{takagi2018multiple}, the SkXs in Co-Zn-Mn alloys~\cite{karube2018disordered}, EuPtSi~\cite{kakihana2017giant,kaneko2018unique,tabata2019magnetic}, Gd$_2$PdSi$_3$~\cite{kurumaji2019skyrmion,Hirschberger_PhysRevLett.125.076602,Hirschberger_PhysRevB.101.220401,Nomoto_PhysRevLett.125.117204,moody2020charge}, Gd$_3$Ru$_4$Al$_{12}$~\cite{hirschberger2019skyrmion,Hirschberger_10.1088/1367-2630/abdef9}, and GdRu$_2$Si$_2$~\cite{khanh2020nanometric,Yasui2020imaging}, and the HXs in MnSi$_{1-x}$Ge$_{x}$~\cite{tanigaki2015real,kanazawa2017noncentrosymmetric,fujishiro2019topological,Kanazawa_PhysRevLett.125.137202} and SrFeO$_3$~\cite{Ishiwata_PhysRevB.84.054427,Ishiwata_PhysRevB.101.134406,Rogge_PhysRevMaterials.3.084404,Onose_PhysRevMaterials.4.114420}. 

The purpose of this article is to review the theoretical findings of topological spin crystals in itinerant magnets. 
A key concept is the itinerant frustration arising from the itinerant nature of electrons. 
Starting from the definition of the itinerant frustration in comparison with the conventional frustration in the short-range exchange interactions in insulating magnets, we review a variety of the multiple-$Q$ instabilities caused by the itinerant frustration. 
Our emphasis is laid on the importance of the effective long-range biquadratic interaction which lifts the degeneracy at the level of the bilinear interaction and stabilizes noncollinear and noncoplanar spin textures. 
We discuss that the effective bilinear-biquadratic spin model in momentum space provides a canonical model for understanding the multiple-$Q$ instabilities in itinerant magnets. 
Furthermore, we show that the model and its extensions successfully explain the origins of multiple-$Q$ topological spin crystals recently discovered on various lattice structures, both centrosymmetric and noncentrosymmetric, and with unusually short magnetic periods. 

The organization of this paper is as follows. 
In section~\ref{sec:Localized versus itinerant frustration}, after reviewing the frustration in insulating magnets, we introduce the concept of itinerant frustration inherent to itinerant magnets. 
We discuss their similarities and differences by exemplifying the magnetic interactions in real and momentum spaces. 
We also present the expressions of the effective multiple-spin interactions derived from the perturbation expansion in terms of the spin-charge coupling in itinerant magnets, which 
are relevant to the multiple-$Q$ topological spin crystals. 
In section~\ref{sec:Multiple-$Q$ topological spin crystals in itinerant magnets}, we review the multiple-$Q$ topological spin crystals found in the Kondo lattice model by focusing on three types of the Fermi surface instabilities: the perfect nesting, the nesting by the multiple connections of the Fermi surfaces in the extended Brillouin zone, and a more generic situation where the bare susceptibility has multiple maxima connected by the lattice symmetry. 
In section~\ref{sec:Effective spin model in itinerant magnets}, we present the effective spin model with the bilinear and biquadratic interactions in momentum space, which reproduces well the multiple-$Q$ topological spin crystals discovered in the Kondo lattice model. 
We show that the effective spin model provides a powerful framework not only to understand the microscopic origin of the multiple-$Q$ topological spin crystals found in experiments but also to encourage a further exploration of exotic topological states, since it enables us to investigate a wide parameter region systematically by smaller computational costs than those for the original itinerant electron problems.
We also present a plethora of the multiple-$Q$ topological spin crystals by incorporating various additional interactions, such as the anisotropic interaction, single-ion anisotropy, and the Dzyaloshinskii-Moriya interaction, in the effective spin models in section~\ref{sec:Extensions of the effective spin model}. 
Section~\ref{sec:summary} is devoted to the summary and future perspective.

\section{Localized versus itinerant frustration}
\label{sec:Localized versus itinerant frustration}

Frustration is a conflict of competing interactions, which has often been discussed in insulating magnets on geometrically-frustrated lattice structures~\cite{liebmann1986statistical,ramirez1994strongly,diep2004frustrated,lacroix2011introduction}.  
There, the frustration arises from the competing short-range exchange interactions. Meanwhile, the concept of frustration can also be introduced in itinerant magnets, where the conflict occurs between long-range interactions mediated by itinerant electrons. 
This is termed as the itinerant frustration in this article. 
In the following, we discuss similarities and differences between the localized and itinerant frustration, by describing the essence of the frustration in insulating magnets in section~\ref{sec:Frustration in insulating magnets} and in itinerant magnets in section~\ref{sec:Frustration in itinerant magnets}. 
In section~\ref{sec:Multi-spin interactions beyond the RKKY interaction}, we outline the derivation of the effective multiple-spin interactions which play an important role in lifting the degeneracy in the itinerant frustration.

\subsection{Frustration in insulating magnets}
\label{sec:Frustration in insulating magnets}

\begin{figure}[h!]
\begin{center}
\includegraphics[width=0.75 \hsize]{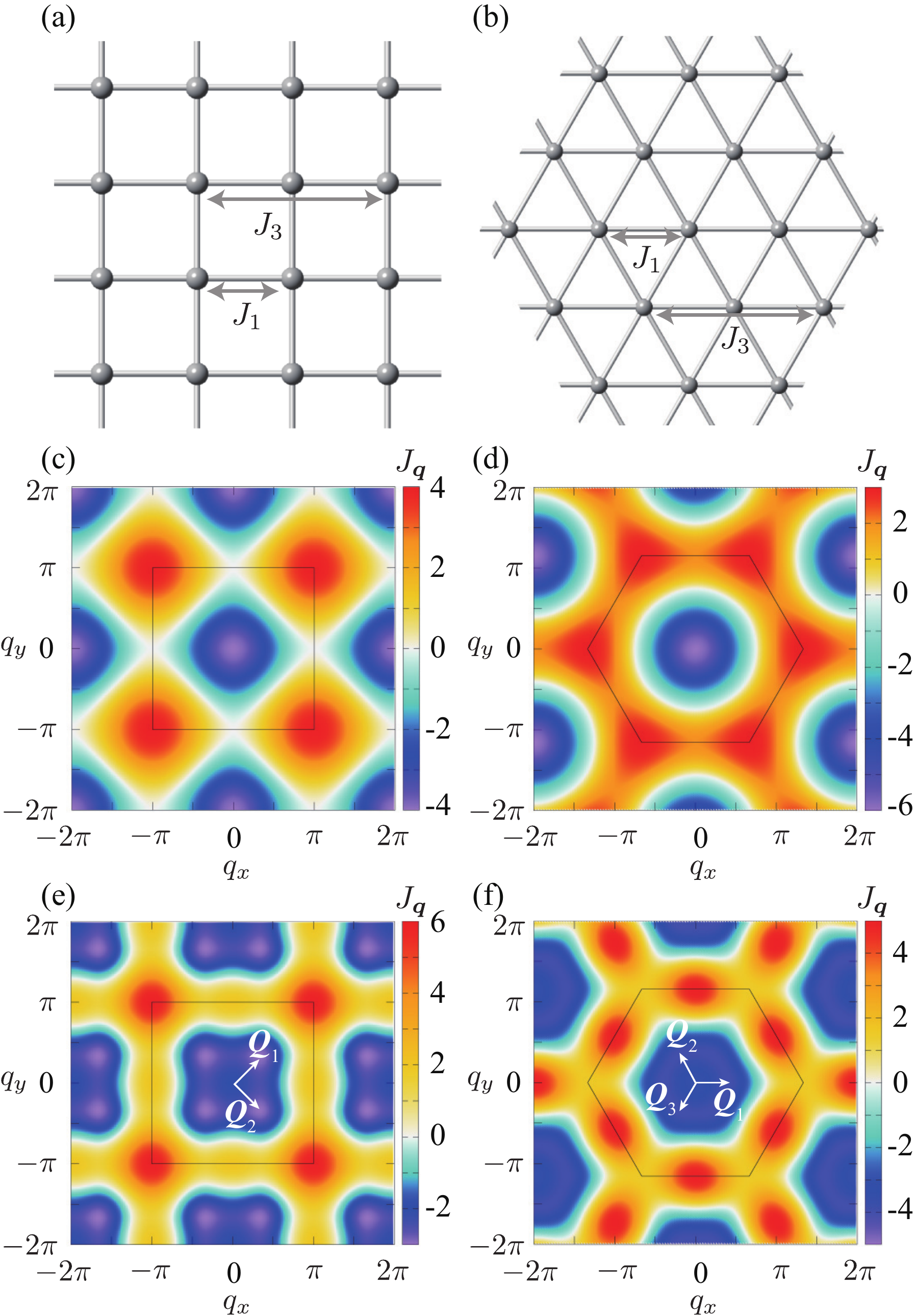} 
\caption{
\label{fig:Jq}
(a) Square and (b) triangular lattice structures. 
$J_1$ and $J_3$ represent the nearest- and third-neighbor exchange interactions, respectively.  
(c)-(f) The contour plots of $J_{\bm{q}}$ for the square lattice model with (c) $J_3=0$ and (e) $J_3=0.5$, and the triangular lattice model with (d) $J_3=0$ and (f) $J_3=0.5$. 
In all the cases, we take $J_1=-1$; see (\ref{eq:Jq_SL}) and (\ref{eq:Jq_TL}).  
The squares in (c) and (e) and the hexagons in (d) and (f) represent the first Brillouin zone. 
$\bm{Q}_{\nu}$ in (e) and (f) are the wave vectors where $J_{\bm{q}}$ is minimized. 
}
\end{center}
\end{figure}

Before introducing the itinerant frustration, let us start by briefly reviewing the frustration in insulating magnets. 
A simple model for the insulating magnets is given by the Heisenberg Hamiltonian as
\begin{equation}
\label{eq:Ham_spin}
\mathcal{H}^{\rm Heis}=   \sum_{ij} J_{ij} \bm{S}_i \cdot \bm{S}_j, 
\end{equation}
where $\bm{S}_i$ denotes the localized spin at site $i$ and $J_{ij}$ is the exchange coupling constant between $i$ and $j$th spins. 
For simplicity, we treat the spins as the classical vectors normalized as $|\bm{S}_i|=1$. 
The Fourier transform of the Hamiltonian in (\ref{eq:Ham_spin}) is expressed as 
\begin{equation}
\label{eq:Ham_spinq}
\mathcal{H}^{\rm Heis}=   \sum_{\bm{q}} J_{\bm{q}} \bm{S}_{\bm{q}} \cdot \bm{S}_{-\bm{q}}, 
\end{equation}
where 
\begin{equation}
\label{eq:Jq}
 J_{\bm{q}} = \sum_{ij} J_{ij}e^{-i \bm{q}\cdot (\bm{r}_i -\bm{r}_j)}, 
\end{equation}
and $\bm{S}_{\bm{q}}$ is the Fourier transform of $\bm{S}_i$; $\bm{r}_i$ is the position vector for site $i$. 

In the model in (\ref{eq:Ham_spinq}), owing to the constraint $\sum_{\bm{q}}|\bm{S}_{\bm{q}}|^2=N$, the minimization of $J_{\bm{q}}$ gives the ground state. 
We consider two examples on the square and triangular lattices, as shown in figures~\ref{fig:Jq}(a) and \ref{fig:Jq}(b), respectively. 
In both cases, we assume the nearest- and third-neighbor exchange interactions, $J_1$ and $J_3$, respectively. 
For the square lattice case, $J_{\bm{q}}$ is explicitly written as 
\begin{equation}
\label{eq:Jq_SL}
J_{\bm{q}}= 2J_1 (\cos q_x + \cos q_y)+2J_3 (\cos 2q_x + \cos 2q_y), 
\end{equation}
while for the triangular lattice case, it is written as 
\begin{eqnarray}
\label{eq:Jq_TL}
J_{\bm{q}}= & &2J_1 \left[\cos q_x + \cos \left(\frac{q_x}{2}+\frac{\sqrt{3}q_y}{2}\right) + \cos \left(\frac{q_x}{2}-\frac{\sqrt{3}q_y}{2} \right)\right] \nonumber \\
&+&2J_3 \left[\cos 2q_x + \cos \left(q_x+\sqrt{3}q_y\right) + \cos \left(q_x-\sqrt{3}q_y \right)\right], 
\end{eqnarray}
where the lattice constant is taken to be unity for each lattice. 

Figure~\ref{fig:Jq}(c) shows the contour plot of $J_{\bm{q}}$ by taking $J_1=-1$ and $J_3=0$ on the square lattice. 
In this case, $J_{\bm{q}}$ has the minimum at $\bm{q}=\bm{0}$, indicating that the ground state of the system becomes ferromagnetic. 
When the sign of $J_1$ is reversed to be positive, the sign of $J_{\bm{q}}$ is also reversed, and hence, $J_{\bm{q}}$ has the minimum at the Brillouin zone edge, $\bm{q}=(\pi,\pi)$. 
Thus, in this case, the ground state is given by a collinear antiferromagnetic state with staggered spin order. In these two situations, the interaction energy is optimized for all the bonds simultaneously, and as a result, the ground state is unique (without degeneracy), i.e., there is no frustration in the system. 

In the triangular lattice case, while the situation is the same for the ferromagnetic case with $J_1=-1$ as shown in figure~\ref{fig:Jq}(d), it is impossible to optimize the energy on all the bonds simultaneously for the antiferromagnetic case with $J_1=1$. 
In this case, the conflict of the interaction is relieved by three-sublattice ordering with 120$^{\circ}$ noncollinear spin configuration. 
This corresponds to the ordering vector at the Brillouin zone edges, $\bm{q}=(4\pi/3,0)$, where the reversed $J_{\bm{q}}$ is minimized in figure~\ref{fig:Jq}(d). 
In this situation, therefore, the ground state is also unique and there is no frustration in the system.

Figures~\ref{fig:Jq}(e) and \ref{fig:Jq}(f) show the results when we introduce $J_3=0.5$ for the square and triangular lattice cases, respectively. 
In these cases, $J_{\bm{q}}$ show the minimum at the multiple wave vectors: 
$\bm{Q}_1 = (\pi/3, \pi/3)$ and $\bm{Q}_2 = (\pi/3, -\pi/3)$ for the square lattice case, and $\bm{Q}_1 = (2\pi/5, 0)$, $\bm{Q}_2 = (-\pi/5, \sqrt{3}\pi/5)$, and $\bm{Q}_3 = (-\pi/5, -\sqrt{3}\pi/5)$ for the triangular lattice case. 
The momenta are related with each other by the rotational symmetry of the square and triangular lattices. 
In these cases, therefore, the spiral state with one of the wave vectors, $\bm{Q}_{\nu}$, has the same energy with that with other $\bm{Q}_{\nu'}$ ($\nu,\nu'=1,2$ for square and $1,2,3$ for triangular), leading to the degeneracy in the ground state. 
This is an example of the frustration in insulating magnets
\footnote{We note that the frustration is often used for the cases with a macroscopic number of degenerate ground states in the classical systems. We, however, use it in a broader sense here by including the cases with the degeneracy associated with the lattice rotational symmetry, for illustrating the analogy to the itinerant frustration in section~\ref{sec:Frustration in itinerant magnets}.}. 
 
In this circumstance, there is a chance to stabilize a multiple-$Q$ state by taking into account additional other interactions to the model in (\ref{eq:Ham_spin}), 
such as the magnetic anisotropy~\cite{leonov2015multiply,Lin_PhysRevB.93.064430,Hayami_PhysRevB.93.184413,Lin_PhysRevLett.120.077202,Binz_PhysRevLett.96.207202,Binz_PhysRevB.74.214408,Park_PhysRevB.83.184406,zhang2017skyrmion}, bond-dependent interactions in the form of compass and Kitaev type~\cite{Michael_PhysRevB.91.155135,Lee_PhysRevB.91.064407,Lukas_PhysRevLett.117.277202,Rousochatzakis2016,yao2016topological,Chern_PhysRevB.95.144427,Maksimov_PhysRevX.9.021017,amoroso2020spontaneous}, and higher-order multiple-spin interactions derived by higher-order exchange processes beyond the Heisenberg one~\cite{takahashi1977half,yoshimori1978fourth,Momoi_PhysRevLett.79.2081,Bulaevskii_PhysRevB.78.024402,Hoffmann_PhysRevB.101.024418,li2021spin}.
Also, thermal fluctuations~\cite{Okubo_PhysRevB.84.144432,Okubo_PhysRevLett.108.017206,Rosales_PhysRevB.87.104402}, quantum fluctuations~\cite{Kamiya_PhysRevX.4.011023,Wang_PhysRevLett.115.107201,Marmorini2014,Ueda_PhysRevA.93.021606}, and disorder by impurities~\cite{Maryasin_PhysRevLett.111.247201,maryasin2015collective,Lin_PhysRevLett.116.187202,Hayami_PhysRevB.94.174420,hayami2019magnetic} play a role in stabilizing such multiple-$Q$ states. 
We note that some of these attempts have been devoted to understanding of the multiple-$Q$ instability in itinerant magnets by taking the localized spin models as effective models; see also section~\ref{sec:Remarks}.

\subsection{Frustration in itinerant magnets}
\label{sec:Frustration in itinerant magnets}

In contrast to the insulating case, itinerant magnets naturally have the instability toward multiple-$Q$ topological spin crystals even without additional interactions, fluctuations, and so on (several examples will be shown in section~\ref{sec:Multiple-$Q$ topological spin crystals in itinerant magnets}). 
The instability is understood from effective multiple-spin interactions rooted in the kinetic motion of itinerant electrons. 
In this section, we introduce the concept of itinerant frustration and show how such an instability arises from it. 

To illustrate the situation, it is convenient to introduce a model in which itinerant electrons are coupled with localized spins via the exchange coupling. 
The model is called the $s$-$d$ model or the Kondo lattice model, whose Hamiltonian is given by 
\begin{equation}
\label{eq:Ham_KLM}
\mathcal{H}^{\rm KLM} = -\sum_{i, j,  \sigma} t_{ij} c^{\dagger}_{i\sigma}c_{j \sigma}
+J_{\rm K} \sum_{i, \sigma, \sigma'} c^{\dagger}_{i\sigma} \bm{\sigma}_{\sigma \sigma'} c_{i \sigma'}
\cdot \bm{S}_i, 
\end{equation}
where $c^{\dagger}_{i\sigma}$ ($c_{i \sigma}$) is a creation (annihilation) operator of an itinerant electron at site $i$ and spin $\sigma$. 
The first term represents the kinetic motion of itinerant electrons with the transfer integral $t_{ij}$ between sites $i$ and $j$; the nearest-neighbor hopping $t_1=1$ is set as an energy unit. 
The second term represents the exchange coupling between itinerant electron spins and localized spins; $\bm{\sigma}=(\sigma^x,\sigma^y,\sigma^z)$ is the vector of Pauli matrices, $\bm{S}_i$ is a localized spin at site $i$, and $J_{\rm K}$ is the exchange coupling constant. 
As in the previous section, we treat $\bm{S}_i$ as the classical spin with $|\bm{S}_i|=1$, for which the sign of $J_{\rm K}$ is irrelevant. 
The Fourier transform of the model in (\ref{eq:Ham_KLM}) is expressed as 
\begin{eqnarray}
\label{eq:Ham_kspace}
\mathcal{H}^{\rm KLM}=\sum_{\bm{k},\sigma} \varepsilon_{\bm{k}} c^{\dagger}_{\bm{k}\sigma}c_{\bm{k}\sigma} +
\frac{J_{\rm K}}{\sqrt{N}} \sum_{\bm{k},\bm{q},\sigma, \sigma'} c^{\dagger}_{\bm{k}\sigma}\bm{\sigma}_{\sigma \sigma'} c_{\bm{k}+\bm{q}\sigma'} \cdot \bm{S}_{\bm{q}}, 
\end{eqnarray}
where $\varepsilon_{\bm{k}}$ is the energy dispersion of the electrons given by
\begin{equation}
\label{eq:dispersion}
\varepsilon_{\bm{k}} = -\sum_{ij} t_{ij} e^{-i \bm{q}\cdot (\bm{r}_i-\bm{r}_j)},
\end{equation}
and $c_{\bm{k} \sigma}^{\dagger}$ and $c_{\bm{k}\sigma}$ are the Fourier transform of $c_{i\sigma}^{\dagger}$ and $c_{i\sigma}$, respectively. 
In the second term in (\ref{eq:Ham_kspace}), $\bm{S}_{\bm{q}}$ is the Fourier transform of $\bm{S}_i$ and $N$ is the number of sites. 
This term represents the scattering of itinerant electrons by the localized spins with momentum transfer $\bm{q}$. 

The model in (\ref{eq:Ham_KLM}) and (\ref{eq:Ham_kspace}) is one of the fundamental models to describe the electronic and magnetic properties in rare-earth compounds~\cite{Stewart_RevModPhys.56.755,Hewson199704,Stewart_RevModPhys.73.797}. 
It is, however, also relevant to a wider range of itinerant magnetism, e.g., in transition metal compounds which are described by the Hubbard-type models~\cite{Cox,Imada_RevModPhys.70.1039,Fazekas1999,Khomskii2014}, when the mean-field approximation for the Coulomb interaction is justified~\cite{Martin_PhysRevLett.101.156402}.

Although the instability toward multiple-$Q$ topological spin crystals in (\ref{eq:Ham_KLM}) has been studied in both strong-coupling regime ($J_{\rm K}\gg t_{ij}$)~\cite{Agterberg_PhysRevB.62.13816,Kumar_PhysRevLett.105.216405,Sahinur_PhysRevB.91.140403,Shahzad_PhysRevB.96.224402,yambe2020double,reja2020skyrmion,Kathyat_PhysRevB.103.035111} and weak-coupling regime ($J_{\rm K}\ll t_{ij}$)~\cite{Martin_PhysRevLett.101.156402,Akagi_JPSJ.79.083711,Akagi_PhysRevLett.108.096401,Hayami_PhysRevB.90.060402,Ozawa_doi:10.7566/JPSJ.85.103703,Hayami_PhysRevB.95.224424}, we focus on the latter in the following. 
In the weak-coupling limit, the ground state can be elucidated by deriving effective magnetic interactions by the perturbation in terms of the second term in (\ref{eq:Ham_kspace}). 
As will be detailed in the next section~\ref{sec:Multi-spin interactions beyond the RKKY interaction}, the lowest-order contribution is written in the form of 
\begin{equation}
\label{eq:RKKY}
\mathcal{H}^{\rm RKKY}=   -J^2_{\rm K}\sum_{\bm{q}} \chi^{0}_{\bm{q}} \bm{S}_{\bm{q}} \cdot \bm{S}_{-\bm{q}}, 
\end{equation}
where $\chi^{0}_{\bm{q}}$ is the bare susceptibility of itinerant electrons [see (\ref{eq:chi0}) for the expression]. 
This is called the RKKY interaction~\cite{Ruderman,Kasuya,Yosida1957}. 
It is noteworthy that the lowest-order effective spin Hamiltonian in (\ref{eq:RKKY}) is formally equivalent to (\ref{eq:Ham_spinq}) by reading the coupling constant $ -J^2_{\rm K} \chi^{0}_{\bm{q}}$ as $J_{\bm{q}}$.
This correspondence harbors frustration similar to that discussed in section~\ref{sec:Frustration in insulating magnets}, as shown below.

\begin{figure}[h!]
\begin{center}
\includegraphics[width=0.8 \hsize]{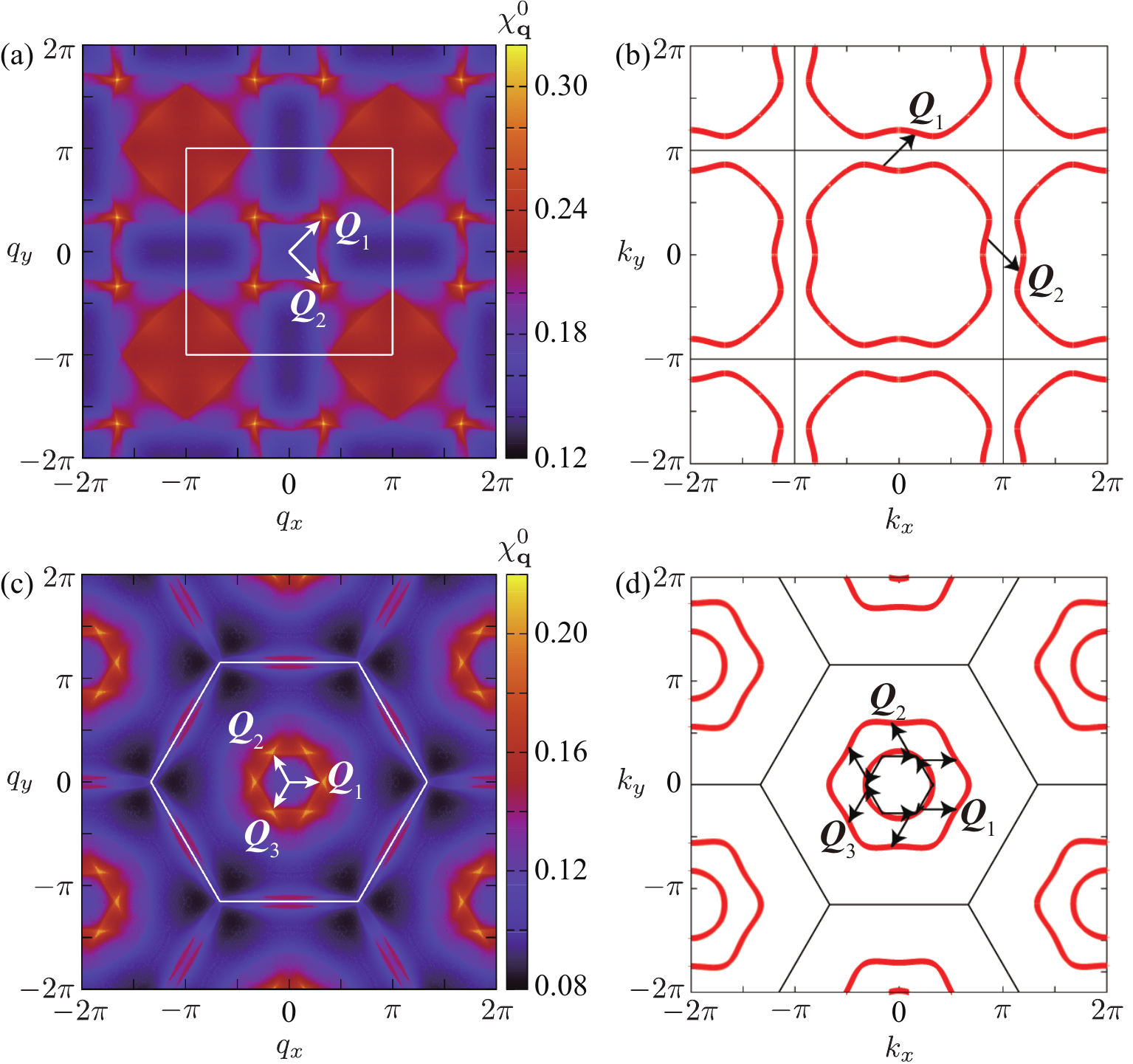} 
\caption{
\label{fig:chiq}
(a), (c) The contour plots of the bare susceptibility $\chi_{\bm{q}}^0$ as a function of $\bm{q}$ for (a) the square lattice model with $t_3=-0.5$ and $\mu=0.98$ and (c) the triangular lattice model with $t_3=-0.85$ and $\mu=-3.5$; we take $t_1=1$. 
The maxima of $\chi_{\bm{q}}^0$ are located at $\bm{Q}_1$ and $\bm{Q}_2$ in (a), while $\bm{Q}_1$, $\bm{Q}_2$, and $\bm{Q}_3$ in (c). 
In both cases, $\bm{Q}_\nu$ are connected with each other by the rotational symmetry of the lattice structure. 
The white square in (a) and hexagon in (c) represent the first Brillouin zone. 
(b) and (d) display the Fermi surfaces corresponding to (a) and (c), respectively. 
$\bm{Q}_\nu$ are the nesting vectors giving the maxima of $\chi^0_{\bm{q}}$ in (a) and (c). 
Figure is reprinted with permission from reference~\cite{Hayami_PhysRevB.95.224424}. Copyright 2017 by the American Physical Society.
}
\end{center}
\end{figure}

The magnetic ground state to optimize the RKKY interaction in (\ref{eq:RKKY}) is obtained by maximizing $\chi_{\bm{q}}^0$. 
Hence, the ordering vector is set by the peak position of $\chi_{\bm{q}}^0$, which depends on the dispersion $\varepsilon_{\bm{k}}$ in (\ref{eq:dispersion}) and the electron density. 
For example, when we consider the third-neighbor hopping $t_3$ in addition to the nearest-neighbor 
$t_1$, $\varepsilon_{\bm{k}}$ for the square lattice case is given by   
\begin{equation}
\varepsilon_{\bm{k}} = -2\sum_{l=1, 2}(t_1 \cos \bm{k}\cdot \bm{e}_{l} +t_3 \cos  2 \bm{k}\cdot \bm{e}_{l}), 
\end{equation}
where $\bm{e}_1=\hat{\bm{x}}=(1,0)$ and $\bm{e}_2=\hat{\bm{y}}=(0,1)$, and 
for the triangular lattice case,  
\begin{equation}
\varepsilon_{\bm{k}} = -2\sum_{l=1,2, 3}(t_1 \cos \bm{k}\cdot \bm{e}_{l} +t_3 \cos 2 \bm{k}\cdot \bm{e}_{l}), 
\label{eq:ek_triangular}
\end{equation}
where $\bm{e}_1=\hat{\bm{x}}$, $\bm{e}_2=-\hat{\bm{x}}/2+\sqrt{3}\hat{\bm{y}}/2$, and $\bm{e}_3=-\hat{\bm{x}}/2-\sqrt{3}\hat{\bm{y}}/2$.  
Here, we also set the lattice constant $a=1$ as the length unit for both cases. 
Figures~\ref{fig:chiq}(a) and~\ref{fig:chiq}(c) show $\chi_{\bm{q}}^0$ on the square lattice with $t_3=-0.5$ and $\mu=0.98$ and the triangular lattice with $t_3=-0.85$ and $\mu=-3.5$, respectively, where $\mu$ represents the chemical potential~\cite{Hayami_PhysRevB.95.224424}. 
The corresponding Fermi surfaces are shown in figures~\ref{fig:chiq}(b) and \ref{fig:chiq}(d). 
The bare susceptibility shows multiple peaks at the wave vectors for which the Fermi surfaces are nested, and the maxima are related by the rotational symmetry of the system: fourfold (sixfold) rotational symmetry of the square (triangular) lattice. 
In the square lattice case, the peaks are found at $\bm{Q}_1=(\pi/3, \pi/3)$ and $\bm{Q}_2=(\pi/3,-\pi/3)$, while those in the triangular lattice case are found at $\bm{Q}_1=(\pi/3,0)$, $\bm{Q}_2=(-\pi/6,\sqrt{3}\pi/6)$, and $\bm{Q}_3=(-\pi/6,-\sqrt{3}\pi/6)$.
Thus, the situation is similar to the case of the model for insulating magnets in figure~\ref{fig:Jq}; 
the RKKY interaction in (\ref{eq:RKKY}) leads to the degeneracy between different single-$Q$ spiral states with the wave vector $\bm{Q}_{\nu}$ (see also section~\ref{sec: Second-order RKKY interaction}). 
We call this the itinerant frustration, in analogy with the frustration in the insulating case. 

There is, however, a difference from the insulating case in the mechanism of lifting the degeneracy. 
In the insulating case, not only the original Heisenberg interactions but also the additional interactions which lift the degeneracy are usually short-ranged. 
In particular, further-neighbor interactions decay exponentially in distance, since they are derived by the perturbation in terms of the hopping of localized electrons. 
Also, higher-order multiple-spin interactions become small, as they are proportional to $1/U^{n-1}$, where $U$ is the onsite Coulomb repulsion and $n$ is the order of the interaction. 
On the other hand, in the itinerant case, additional higher-order contributions are also long-ranged, similar to the lowest-order RKKY interactions in (\ref{eq:RKKY}). 
Moreover, they are not necessarily small; the coefficients can be large depending on the electronic state since they are given by the products of Green's functions of the itinerant electrons; see section~\ref{sec:Multi-spin interactions beyond the RKKY interaction}. 
Indeed, several theoretical studies have shown that the ground state in the Kondo lattice model in (\ref{eq:Ham_KLM}) is not given by the single-$Q$ spiral state but by noncoplanar multiple-$Q$ topological spin crystals. 
The prominent example was obtained at a particular electron filling 
where the Fermi surface has perfect nesting~\cite{Martin_PhysRevLett.101.156402}. 
Similar attempts have been performed for the situations where the Fermi surface has multiple connections in the extended Brillouin zone~\cite{Akagi_JPSJ.79.083711,Akagi_PhysRevLett.108.096401,Hayami_PhysRevB.90.060402} and more generic situations where the Fermi surface has no special property except for the rotational symmetry~\cite{Ozawa_doi:10.7566/JPSJ.85.103703,Hayami_PhysRevB.95.224424}. 
These considerations have brought about theoretical findings of a plethora of multiple-$Q$ topological spin crystals in hexagonal systems~\cite{Martin_PhysRevLett.101.156402,Akagi_JPSJ.79.083711,Kato_PhysRevLett.105.266405,Barros_PhysRevB.88.235101,Ozawa_PhysRevLett.118.147205,Venderbos_PhysRevLett.108.126405,Venderbos_PhysRevB.93.115108,Barros_PhysRevB.90.245119,Ghosh_PhysRevB.93.024401,Ozawa_PhysRevB.96.094417,Chern_PhysRevB.97.035120,Hayami_PhysRevB.99.094420,Wang_PhysRevLett.124.207201}, tetragonal systems~\cite{Solenov_PhysRevLett.108.096403,hayami_PhysRevB.91.075104,Ozawa_doi:10.7566/JPSJ.85.103703,Okada_PhysRevB.98.224406,hayami2018multiple,Hayami_PhysRevLett.121.137202,Su_PhysRevResearch.2.013160,Hayami_doi:10.7566/JPSJ.89.103702}, and cubic systems~\cite{Chern_PhysRevLett.105.226403,hayami2014charge,Hayami_PhysRevB.89.085124,Okumura_PhysRevB.101.144416,okumura2020tracing}. 
We will review some of these studies in section~\ref{sec:Multiple-$Q$ topological spin crystals in itinerant magnets}. 
The fundamental mechanism common to this itinerant frustration is that the system tends to lift the degeneracy with respect to the rotational symmetry of the lattice structure through the higher-order multiple-spin interactions, as described in the following sections.

\subsection{Multiple-spin interactions in itinerant magnets}
\label{sec:Multi-spin interactions beyond the RKKY interaction}

In this section, we discuss the effective multiple-spin interactions in itinerant magnets. 
We briefly review a systematic derivation by the perturbative expansion with respect to the exchange coupling term in the Kondo lattice model in (\ref{eq:Ham_kspace}). 
After presenting the general framework of the perturbative expansion in section~\ref{sec:Perturbation expansion}, we present the second-order contribution in section~\ref{sec: Second-order RKKY interaction}, the fourth-order ones in section~\ref{sec: Fourth-order interaction}, and the higher-order ones in section~\ref{sec:Higher-order interactions}. 
In section~\ref{sec:Remarks}, we remark on some related studies of the effective multiple-spin interactions. 

\subsubsection{Perturbation expansion}
\label{sec:Perturbation expansion}

Suppose the exchange coupling $J_{\rm K}$ is small enough compared to the bandwidth of itinerant electrons in (\ref{eq:Ham_kspace}), one can expand the free energy of the system with respect to $J_{\rm K}$: 
\begin{eqnarray}
\label{eq:freeenergy_expand}
F-F^{(0)}&=-T \log 
\left\langle \mathcal{T} \exp  \left( 
-\int^\beta_0 \mathcal{H}' (\tau) d\tau  
\right)
 \right\rangle_{\rm con} \nonumber  \\
 &=-\frac{T}{2!} \int^{\beta}_0 d \tau_1 \int^{\beta}_0 d \tau_2 
 \langle \mathcal{T} \mathcal{H}' (\tau_1) \mathcal{H}' (\tau_2) \rangle_{\rm con} \nonumber \\
 &\ \ \ - \frac{T}{4!}\int^{\beta}_0 d \tau_1 \cdots  \int^{\beta}_0 d \tau_4 
 \langle \mathcal{T} \mathcal{H}' (\tau_1) \cdots \mathcal{H}' (\tau_4) \rangle_{\rm con} 
- \cdots \nonumber \\
\label{eq:freeenergy_expand3}
  &= F^{(2)}+F^{(4)}+\cdots, 
\end{eqnarray}
where $\mathcal{H}'$ represents the second term of (\ref{eq:Ham_kspace}), $\mathcal{T}$ is the time-ordering operator, $\tau$ is the imaginary time, $T$ is the temperature, and $\beta$ is the inverse temperature where the Boltzmann constant is set as unity. 
$\langle \cdots \rangle_{\rm con}$ stands for the averaged value over the connected Feynman diagrams. 
$F^{(0)}$ represents the free energy from the first term of (\ref{eq:Ham_kspace}). 
Note that there are no odd-order terms in the expansion due to the time-reversal symmetry in the system. 

The $2n$th-order contribution to the free energy can be expressed in the general form~\cite{Hayami_PhysRevB.95.224424,komarov2017effective}
\begin{eqnarray}
\label{eq:nth_freeenergy}
F^{(2n)}=& &\frac{T}{n}\left(\frac{J_{\rm K}}{\sqrt{N}}\right)^{2n} 
\sum_{\bm{k}, \omega_p}\sum_{\bm{q}_1,\cdots,\bm{q}_{2n},l} 
G_{\bm{k}}G_{\bm{k}+\bm{q}_1}\cdots G_{\bm{k}+\bm{q}_1+\cdots+\bm{q}_{2n-1}}   
\delta_{\bm{q}_1+\bm{q}_2+\cdots+\bm{q}_{2n},l\bm{G}}\nonumber  \\ 
&\times& \sum_{\{P\}}(-1)^{\lambda_P}
\prod_{\nu,\nu'} 
\bm{S}_{\bm{q}_\nu} \cdot \bm{S}_{\bm{q}_{\nu'}}, 
\end{eqnarray}
where $G_{\bm{k}} (i \omega_p) = \left[ i \omega_p -(\varepsilon_{\bm{k}}-\mu) \right]^{-1}$ is noninteracting  spin-independent Green's function, $\omega_p$ is the Matsubara frequency, $\mu$ is the chemical potential, $\delta$ is the Kronecker delta, and $\bm{G}$ is the reciprocal lattice vector ($l$ is an integer). 
Here and hereafter, the Matsubara frequency dependence of Green's function is not explicitly written for notational simplicity. 
The sum of $\{P\}$ is taken for all the combinations of $\nu$ and $\nu'$ [the number of the combinations is $_{2n}\mathrm{C}_{2} \cdot {}_{2n-2}\mathrm{C}_{2} \cdots {}_{2}\mathrm{C}_{2}/(n!)$], and $\lambda_P$ 
is $+1$ ($-1$) for an even (odd) permutation.
The product is taken for $1\leq\nu'<\nu \leq 2n$.

\begin{figure}[htb!]
\begin{center}
\includegraphics[width=1.0 \hsize]{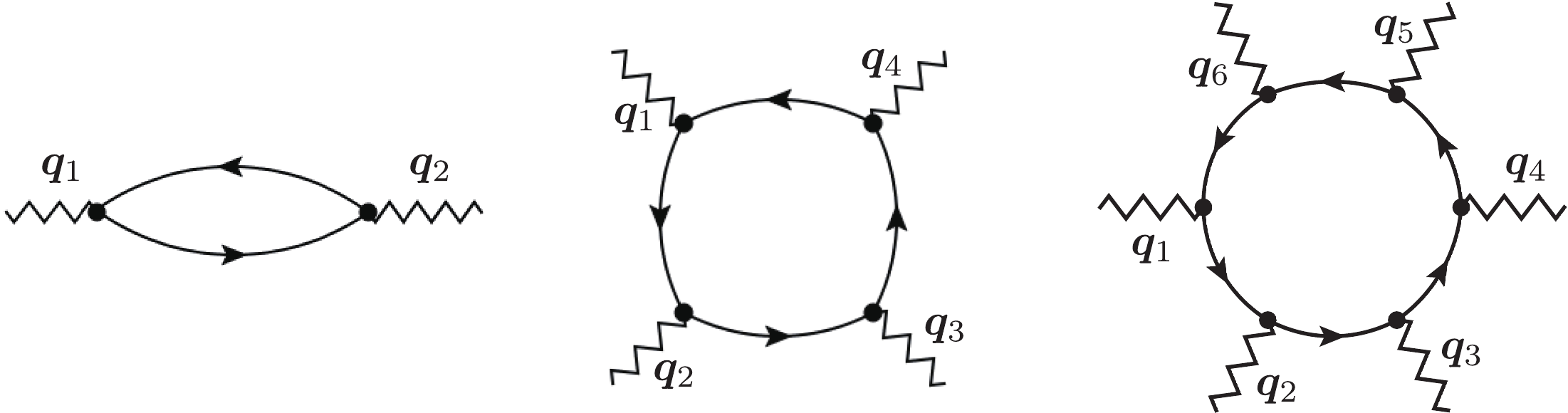} 
\caption{
\label{fig:diagram_general}
Feynman diagrams in the perturbative expansion of the free energy for $n=1$, $2$, and $3$ in (\ref{eq:nth_freeenergy}) from left to right. 
The vertices with wavy lines denote the scattering by localized spins and the solid curves represent Green's functions of itinerant electrons, $G_{\bm{k}}$. 
Figure is reprinted with permission from reference~\cite{Hayami_PhysRevB.95.224424}. Copyright 2017 by the American Physical Society.
}
\end{center}
\end{figure}

Figure~\ref{fig:diagram_general} represents the Feynman diagrams for $n=1$, $2$, and $3$ in (\ref{eq:nth_freeenergy})~\cite{Hayami_PhysRevB.95.224424}. 
$F^{(2n)}$ in (\ref{eq:nth_freeenergy}) gives the effective multiple-spin interaction at the $2n$th order of $J_{\rm K}$. 
In the following, we discuss the specific form of such interactions at the second order (section~\ref{sec: Second-order RKKY interaction}), fourth order (section~\ref{sec: Fourth-order interaction}), and higher orders (section~\ref{sec:Higher-order interactions}) of $J_{\rm K}$. 

\subsubsection{Second-order RKKY interaction}
\label{sec: Second-order RKKY interaction}

The lowest-order contribution in (\ref{eq:nth_freeenergy}) is given by the second-order one in terms of $J_{\rm K}$ ($n=1$), which is expressed as 
\begin{equation}
\label{eq:RKKYHam_G}
F^{(2)}=T
\frac{J_{\rm K}^2}{N}  \sum_{\bm{k}, \bm{q}, \omega_p} G_{\bm{k}+\bm{q}} G_{\bm{k}} \bm{S}_{\bm{q}}\cdot \bm{S}_{-\bm{q}}. 
\end{equation}
By taking the summation of $\omega_p$, (\ref{eq:RKKYHam_G}) turns into (\ref{eq:RKKY}) in section~\ref{sec:Frustration in itinerant magnets}, which is reexpressed as 
\begin{equation}
\label{eq:RKKYHam}
F^{(2)}=-J_{\rm K}^2 \sum_{\bm{q}}\chi_{\bm{q}}^0 \bm{S}_{\bm{q}}\cdot \bm{S}_{-\bm{q}}, 
\end{equation}
where $\chi_{\bm{q}}^0$ is the bare susceptibility of itinerant electrons, 
\begin{eqnarray}
\chi_{\bm{q}}^0 &=-\frac{T}{N}\sum_{\bm{k},\omega_p} G_{\bm{k}+\bm{q}}G_{\bm{k}}  = \frac{1}{N} \sum_{\bm{k}} \frac{f(\varepsilon_{\bm{k}})-f(\varepsilon_{\bm{k}+\bm{q}})}{\varepsilon_{\bm{k}+\bm{q}}-\varepsilon_{\bm{k}}}. 
\label{eq:chi0}
\end{eqnarray}
Here, $f(\varepsilon_{\bm{k}})$ is the Fermi distribution function.  
Thus, the second-order free energy gives a pairwise interaction between the localized spins, which is called the RKKY interaction~\cite{Ruderman,Kasuya,Yosida1957}. 
The coefficient of this bilinear interaction depends 
on the band structure and the electron density through (\ref{eq:chi0}), as mentioned in section~\ref{sec:Frustration in itinerant magnets}.

The magnetic state that optimizes the RKKY energy in (\ref{eq:RKKYHam}) is a single-$Q$ spiral state, whose spin structure is represented by 
\begin{equation}
\label{eq:spin_helical}
\bm{S}_i=(\cos \bm{Q}\cdot \bm{r}_i,\sin \bm{Q}\cdot \bm{r}_i,0).
\end{equation}
Here, $\bm{Q}$ is the ordering vector defining the pitch and direction of the spiral, which is dictated by the peak of $\chi_{\bm{q}}^0$ in (\ref{eq:chi0}). 
This is because the state with $|\bm{S}_{\bm{Q}}|^2=|\bm{S}_{-\bm{Q}}|^2=N/2$ and $\bm{S}_{\bm{q}}=0$ for $\bm{q}\neq \pm\bm{Q}$ gives the lowest energy of (\ref{eq:RKKYHam}) under the constraint $\sum_{\bm{q}}|\bm{S}_{\bm{q}}|^2=N$; 
any spiral with other $\bm{q}$ or any superpositions of spirals with different wave vectors, i.e., multiple-$Q$ states, lead to an energy cost. 
Therefore, at the lowest order, the system has the degeneracy between the different single-$Q$ spiral states when there are several $\bm{Q}_\nu$ which maximize $\chi_{\bm{q}}^0$, as discussed in section~\ref{sec:Frustration in itinerant magnets}. 
The free energy for the degenerate states is given by
\begin{equation}
\label{eq:F(2)_helical}
F^{(2)}=-2J_{\rm K}^2 
\chi_{\bm{Q}_{\nu}}^0 S_{\bm{Q_{\nu}}} \cdot S_{-\bm{Q_{\nu}}}.  
\end{equation} 
This is represented by the Feynman diagram in the left panel of figure~\ref{fig:diagram_general} with replacing $\bm{q}_1$ and $\bm{q}_2$ by $\bm{Q}_\nu$ and $-\bm{Q}_\nu$, respectively.

\subsubsection{Fourth-order interaction}
\label{sec: Fourth-order interaction}

The fourth-order contribution in (\ref{eq:nth_freeenergy}) is given by 
\begin{eqnarray}
\label{eq:4thfreeenergy_exact}
F^{(4)}&= &\frac{T}{2}\frac{J_{\rm K}^4}{N^2} \sum_{\bm{k}, \omega_p}\sum_{\bm{q}_1,\bm{q}_2,\bm{q}_3,\bm{q}_4,l} 
G_{\bm{k}}G_{\bm{k}+\bm{q}_1}G_{\bm{k}+\bm{q}_1+\bm{q}_2}G_{\bm{k}+\bm{q}_1+\bm{q}_2+\bm{q}_3} \delta_{\bm{q}_1+\bm{q}_2+\bm{q}_3+\bm{q}_4,l\bm{G}} 
\nonumber \\
&\times& \left[ 
(\bm{S}_{\bm{q}_1}\cdot \bm{S}_{\bm{q}_2})
(\bm{S}_{\bm{q}_3}\cdot \bm{S}_{\bm{q}_4})
+
(\bm{S}_{\bm{q}_1}\cdot \bm{S}_{\bm{q}_4})
(\bm{S}_{\bm{q}_2}\cdot \bm{S}_{\bm{q}_3})
-
(\bm{S}_{\bm{q}_1}\cdot \bm{S}_{\bm{q}_3})
(\bm{S}_{\bm{q}_2}\cdot \bm{S}_{\bm{q}_4})
\right].  
\end{eqnarray}
The corresponding Feynman diagram is shown in the middle of figure~\ref{fig:diagram_general}. 
This gives four-spin interactions, which may lift the degeneracy between the single-$Q$ spiral states mentioned above. 
Specifically, the relevant contributions arise from the wave vectors $\bm{Q}_\nu$ where the bare susceptibility shows the maxima. 
For the case satisfying $\bm{q}_1+\bm{q}_2+\bm{q}_3+\bm{q}_4 =\bm{0}$ ($l=0$), the fourth-order free energy is given by the sum of five types of the four-spin interactions: 
\begin{eqnarray}
\label{eq:F41}
F^{(4)}_1&=\frac{J^4}{N} \sum_{\nu}
(2A_1- A_2)
(\bm{S}_{\bm{Q}_\nu}\cdot \bm{S}_{\bm{Q}_\nu})
(\bm{S}_{-\bm{Q}_\nu}\cdot \bm{S}_{-\bm{Q}_\nu}),  \\
\label{eq:F42}
F^{(4)}_2&=\frac{J^4}{N} \sum_{\nu}
(2A_2)
(\bm{S}_{\bm{Q}_\nu}\cdot \bm{S}_{-\bm{Q}_\nu})^2,  
  \\
  \label{eq:F43}
F^{(4)}_3&=4\frac{J^4}{N} \sum_{\nu, \nu'}
(B_1+ B_2 -  B_3)
(\bm{S}_{\bm{Q}_\nu}\cdot \bm{S}_{-\bm{Q}_\nu})
(\bm{S}_{\bm{Q}_{\nu'}}\cdot \bm{S}_{-\bm{Q}_{\nu'}}), \\
\label{eq:F44}
F^{(4)}_4&=4\frac{J^4}{N} \sum_{\nu, \nu'}
(-B_1+ B_2 +   B_3)
(\bm{S}_{\bm{Q}_\nu}\cdot \bm{S}_{\bm{Q}_{\nu'}})
(\bm{S}_{-\bm{Q}_\nu}\cdot \bm{S}_{-\bm{Q}_{\nu'}}), \\
\label{eq:F45}
F^{(4)}_5&=4\frac{J^4}{N} \sum_{\nu, \nu'}
(B_1- B_2 + B_3)
(\bm{S}_{\bm{Q}_\nu}\cdot \bm{S}_{-\bm{Q}_{\nu'}})
(\bm{S}_{-\bm{Q}_{\nu'}}\cdot \bm{S}_{\bm{Q}_\nu}), 
\end{eqnarray}
where the sums in (\ref{eq:F43})-(\ref{eq:F45}) are taken for $\nu>\nu'$. The coefficients are given by 
\begin{eqnarray}
\label{eq:A1}
A_1 &= \frac{T}{N}\sum_{\bm{k}, \omega_p}(G_{\bm{k}})^2 G_{\bm{k}-\bm{Q}_\nu}G_{\bm{k}+\bm{Q}_\nu}, 
\quad A_2 
= \frac{T}{N}\sum_{\bm{k}, \omega_p}(G_{\bm{k}})^2 (G_{\bm{k}+\bm{Q}_\nu})^2, \\
B_1 &= \frac{T}{N}\sum_{\bm{k}, \omega_p}(G_{\bm{k}})^2 G_{\bm{k}+\bm{Q}_\nu}G_{\bm{k}+\bm{Q}_{\nu'}}, 
\quad B_2 
= \frac{T}{N}\sum_{\bm{k}, \omega_p}(G_{\bm{k}})^2 G_{\bm{k}+\bm{Q}_\nu}G_{\bm{k}-\bm{Q}_{\nu'}}, \nonumber \\
\label{eq:B3}
B_3 &= \frac{T}{N}\sum_{\bm{k}, \omega_p} G_{\bm{k}} G_{\bm{k}+\bm{Q}_\nu}G_{\bm{k}+\bm{Q}_{\nu'}}G_{\bm{k}+\bm{Q}_\nu+\bm{Q}_{\nu'}}.   
\end{eqnarray}
The sign and amplitude of the coefficients depend on the band structure and electron, but their dependences are different from that in the RKKY interaction. 
Similarly, the free energy can be derived for the cases with $\bm{q}_1+\bm{q}_2+\bm{q}_3+\bm{q}_4 = \bm{G}$ to satisfy $2\bm{Q}_{\nu} = \bm{G}$~\cite{Martin_PhysRevLett.101.156402,Akagi_JPSJ.79.083711,Akagi_PhysRevLett.108.096401,Hayami_PhysRevB.90.060402,hayami_PhysRevB.91.075104} (see section~\ref{sec:Partial nesting}) and $4\bm{Q}_{\nu} = \bm{G}$ ($\nu=1,2,3$)~\cite{Hayami_PhysRevB.94.024424}.

When the bare susceptibility has multiple peaks at symmetry-related $\bm{Q}_\nu$ as exemplified in figures~\ref{fig:chiq}(a) and \ref{fig:chiq}(c), the coefficient $A_2$ takes a positive value and becomes dominant among the contributions in (\ref{eq:A1}) and (\ref{eq:B3}) at low temperature~\cite{Akagi_PhysRevLett.108.096401,Ozawa_doi:10.7566/JPSJ.85.103703,Hayami_PhysRevB.95.224424}.  
This indicates that $F_2^{(4)}$ in (\ref{eq:F42}), which is the biquadratic interaction in momentum space with the positive coefficient, is the most important contribution among the fourth-order multiple-spin interactions. 
In the following sections, we will show that the positive biquadratic interaction plays a crucial role in stabilizing multiple-$Q$ topological spin crystals.

\subsubsection{Higher-order interactions}
\label{sec:Higher-order interactions}

The higher-order $2n$th contributions describe the scattering processes by $2n$ localized spins in (\ref{eq:nth_freeenergy}). 
Extending the fourth-order argument straightforwardly, one may expect the dominant contribution as 
\begin{equation}
\label{eq:F2nessential}
F^{(2n)}_{(\bm{Q},-\bm{Q})} = \frac{2^n T}{n} \left(\frac{J_{\rm K}}{\sqrt{N}}\right)^{2n} 
  \sum_{\bm{k}, \omega_p,\nu}
(G_{\bm{k}})^n (G_{\bm{k}+\bm{Q}_\nu})^n 
(\bm{S}_{\bm{Q}_\nu} \cdot \bm{S}_{-\bm{Q}_\nu})^n. 
\end{equation}
This indicates that the $(4m+2)$th-order terms with $G^{2m+1}_{\bm{k}}G^{2m+1}_{\bm{k}+\bm{Q}_{\nu}}<0$ tend to favor a single-$Q$ spiral state as the lowest-order RKKY interaction in (\ref{eq:RKKYHam_G}), while the $4m$th-order ones with $G^{2m}_{\bm{k}}G^{2m}_{\bm{k}+\bm{Q}_{\nu}}>0$ tend to favor a multiple-$Q$ state as the fourth-order biquadratic interaction in (\ref{eq:F42}) ($m$ is an integer). 
This suggests that the higher-order contributions in (\ref{eq:F2nessential}) can be renormalized into the lower-order bilinear and biquadratic interactions; see (\ref{eq:effHam_spin}) in section~\ref{sec:Bilinear-biquadratic model in momentum space}. 

Meanwhile, the higher-order contributions also include qualitatively different interactions from the bilinear and biquadratic interactions. 
For instance, the sixth-order contribution may include the term proportional to $[\bm{S}_{\bm{Q}_1} \cdot (\bm{S}_{\bm{Q}_2}\times \bm{S}_{\bm{Q}_3})]^2 $ when $\bm{Q}_1+\bm{Q}_2+\bm{Q}_3= \bm{0}$. 
Recently, the authors and the collaborator pointed out that this scalar-chirality-type interaction can drive a phase shift among the constituent waves of multiple-$Q$ topological spin crystals~\cite{hayami2020phase}.

\subsubsection{Remark}
\label{sec:Remarks}

Similar multiple-spin interactions in itinerant magnets have been discussed also in a different context. 
For instance, many studies based on the first-principles calculations were made to explain the origin of noncollinear and noncoplanar magnetic textures in bulk, surfaces, and heterostructures, which indicated the relevance of a variety of effective multiple-spin interactions, such as the four-spin interactions as $(\bm{S}_i\cdot \bm{S}_j)(\bm{S}_k\cdot \bm{S}_l)$~\cite{Kurz_PhysRevLett.86.1106,heinze2011spontaneous,Yoshida_PhysRevLett.108.087205,ueland2012controllable,Mankovsky_PhysRevB.101.174401,paul2020role,Brinker_PhysRevResearch.2.033240,lounis2020multiple,Spethmann_PhysRevLett.124.227203,Simon_PhysRevMaterials.4.084408,Mendive-Tapia_PhysRevB.103.024410}, the chiral biquadratic interaction as $(\bm{S}_i \times \bm{S}_j)(\bm{S}_i\cdot \bm{S}_j)$~\cite{brinker2019chiral,Laszloffy_PhysRevB.99.184430,Mankovsky_PhysRevB.101.174401,Brinker_PhysRevResearch.2.033240,lounis2020multiple}, and the chiral-chiral interaction as $[\bm{S}_i \cdot (\bm{S}_j \times \bm{S}_k)]^2$~\cite{grytsiuk2020topological,Bomerich_PhysRevB.102.100408}. 
Note that all these multiple-spin interactions in the literatures are basically short-ranged in real space, in contrast to the long-ranged ones in the previous sections derived by the perturbation theory in momentum space.

\section{Multiple-$Q$ topological spin crystals in itinerant magnets}
\label{sec:Multiple-$Q$ topological spin crystals in itinerant magnets}

In this section, we review the instabilities toward the multiple-$Q$ topological spin crystals discovered in the Kondo lattice model, which indicate the importance of the itinerant frustration and the inherent multiple-spin interactions originating from the spin-charge coupling discussed in the previous section. 
We here discuss three categories from the viewpoint of the nesting property of the Fermi surfaces. 
The first one is the perfect nesting case in section~\ref{sec:Perfect nesting}, and the second one is the case with multiple connections of the Fermi surfaces in the extended Brillouin zone in section~\ref{sec:Partial nesting}. 
These two cases occur for particular electronic band structures and at particular electron fillings. The last one in section~\ref{sec:General case} is a more generic situation where the bare susceptibility has multiple maxima according to the symmetry of the system.

\subsection{Perfect nesting}
\label{sec:Perfect nesting}

The nesting property of the Fermi surfaces is fundamental to understand the instabilities in itinerant electron systems~\cite{peierls1955quantum,Gruner_RevModPhys.60.1129,Gruner_RevModPhys.66.1}. 
For instance, instabilities toward electronic ordering, such as charge and spin density waves, occur predominantly at the nesting wave vector. 
In the case of the perfect nesting, in which all the points on the Fermi surface are connected with others by a single nesting vector, the magnetic susceptibility $\chi^0_{\bm{q}}$ in (\ref{eq:chi0}) has a delta-functional peak at the nesting vector. 
In this case, the system is unstable against an infinitesimal perturbation since the electronic state gains an energy by gap opening on the entire Fermi surface.
For example, in the tight-binding model with nearest-neighbor hopping on a square lattice at half filling where the square-shaped Fermi surface is perfectly nested, a N\'eel order with the ($\pi,\pi$) wave vector is induced and the system becomes an insulator immediately when the Coulomb interaction is turned on. 

\begin{figure}[htb!]
\begin{center}
\includegraphics[width=0.9 \hsize]{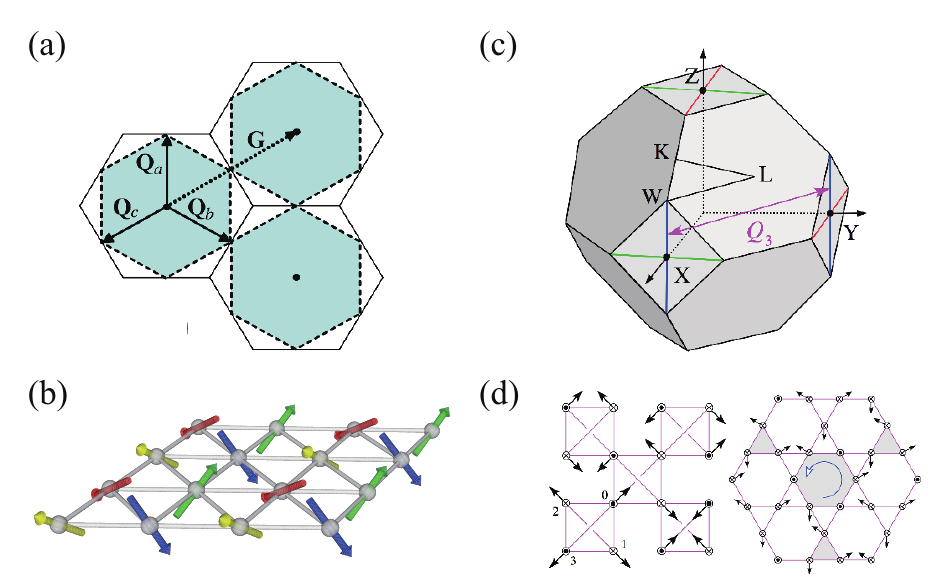} 
\caption{
\label{fig:perfect_nesting}
Perfect nesting of the Fermi surfaces and the magnetic instabilities toward multiple-$Q$ topological spin crystals in the (a)(b) triangular and (c)(d) pyrochlore lattice systems. 
(a) The Brillouin zone (black hexagons) and the Fermi surface (shaded hexagons) at 3/4 filling of the tight-binding model with nearest-neighbor hopping on the triangular lattice. 
$\bm{Q}_a$, $\bm{Q}_b$, and $\bm{Q}_c$ are the nesting wave vectors, while $\bm{G}$ is the reciprocal lattice vector. 
(b) Schematic picture of the four-sublattice noncoplanar order stabilized by the perfect nesting in (a). 
(c) The Brillouin zone in the pyrochlore case. 
The red, green, and blue lines show the Fermi surfaces at 1/4 filling, which are connected by three wave vectors (one of them is shown by $\bm{Q}_3$). 
(d) Schematic picture of the sixteen-sublattice noncoplanar order stabilized by the perfect nesting in (c). 
The left panel shows a projection from the [001] direction of the pyrochlore lattice, while the right one is a [111] slice on the kagome layer. 
Figure (a) is reprinted with permission from reference~\cite{Martin_PhysRevLett.101.156402}. Copyright 2008 by the American Physical Society.
Figures (c) and (d) are reprinted with permission from reference~\cite{Chern_PhysRevLett.105.226403}. Copyright 2010 by the American Physical Society. 
}
\end{center}
\end{figure}

The perfect nesting also leads to multiple-$Q$ topological spin crystals when the Fermi surface is nested by more than a single wave vector and $\chi^0_{\bm{q}}$ is divergent at the multiple nesting vectors. 
An example was found in a two-dimensional triangular lattice system at 3/4 filling~\cite{Martin_PhysRevLett.101.156402}. 
In this case, the Fermi surface is perfectly nested by three wave vectors, as shown in figure~\ref{fig:perfect_nesting}(a). 
This special nesting leads to an instability toward triple-$Q$ magnetic ordering by gap opening on the entire Fermi surface. 
Interestingly, this state composed of a superposition of three spin spirals has the noncoplanar magnetic texture in figure~\ref{fig:perfect_nesting}(b), which exhibits a nonzero net scalar chirality. 
Consequently, the system becomes a magnetic Chern insulator showing a quantized anomalous Hall effect. 
Another example was found in a three-dimensional pyrochlore lattice system at 1/4 filling~\cite{Chern_PhysRevLett.105.226403}. 
In this case, the Fermi surface consists of lines on the Brillouin zone boundaries, which are perfectly nested by three wave vectors, as shown in figure~\ref{fig:perfect_nesting}(c). 
This line-type perfect nesting also leads to an instability toward complicated triple-$Q$ noncoplanar spin ordering shown in figure~\ref{fig:perfect_nesting}(d).

\subsection{$(d-2)$-dimensional connections of Fermi surfaces}
\label{sec:Partial nesting}

\begin{figure}[htb!]
\begin{center}
\includegraphics[width=1.0 \hsize]{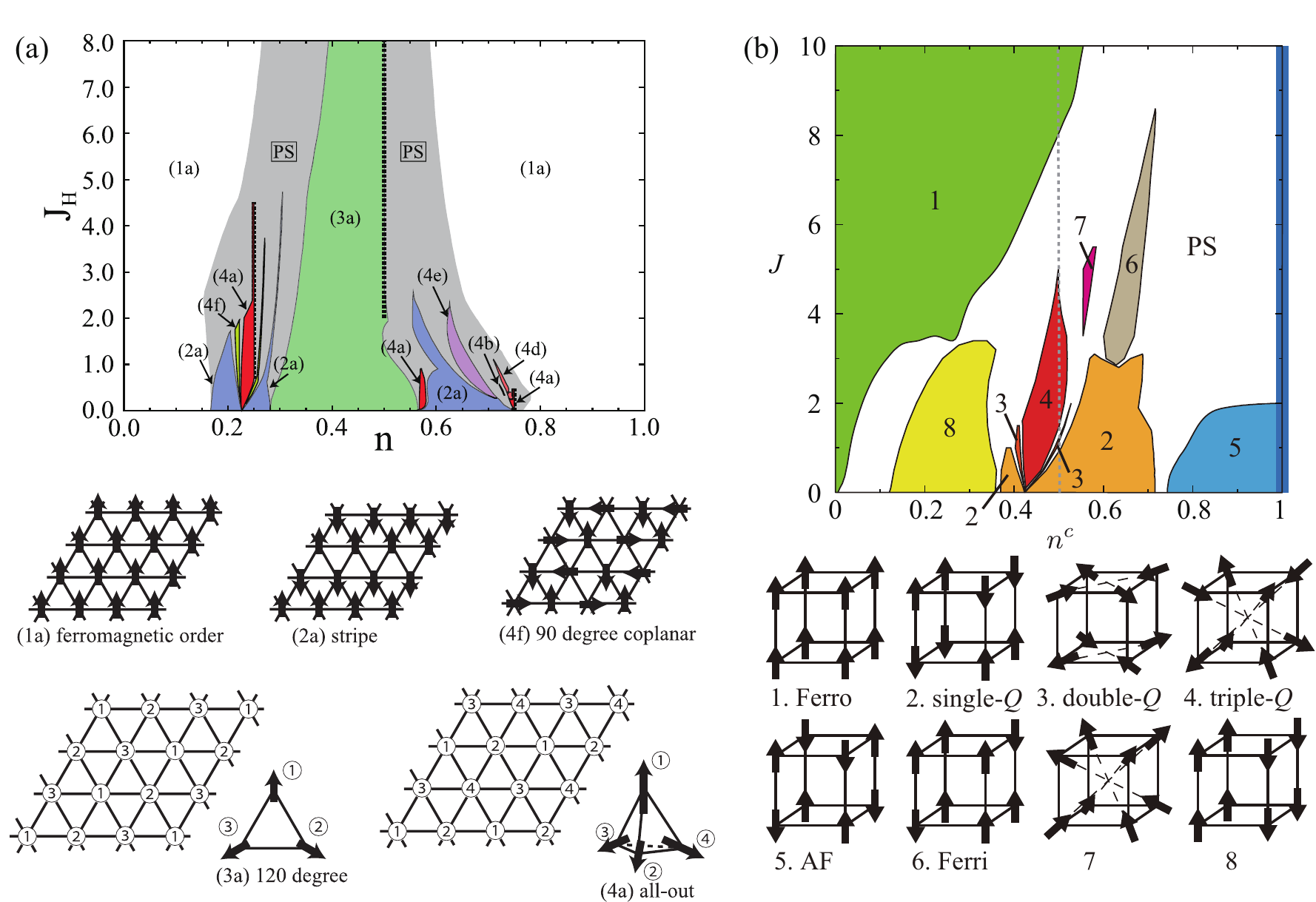} 
\caption{
\label{fig:multipleQ_prestudy}
Instabilities toward multiple-$Q$ topological spin crystals in the (a) triangular and (b) cubic lattice systems. 
(a) Ground-state phase diagram of the Kondo lattice model with nearest-neighbor hopping on the triangular lattice obtained by variational calculations. 
The horizontal and vertical axes are the electron filling $n$ and the spin-charge coupling $J_{{\rm H}} \equiv J_{\rm K}$ in (\ref{eq:Ham_KLM}), respectively. 
The lower panels display several representative magnetic orders. 
The phases represented by (4a) (red areas) show the triple-$Q$ noncoplanar magnetic order. 
The large area near 1/4 filling is the one induced by the $(d-2)$-dimensional connections of the Fermi surfaces, while the small one near 3/4 filling is by the perfect nesting in figures~\ref{fig:perfect_nesting}(a) and \ref{fig:perfect_nesting}(b). 
The area near $n=0.57$ is replaced by a single-$Q$ state when long-period spirals are included in the variational calculations~\cite{Azhar_PhysRevLett.118.027203}. 
(b) Ground-state phase diagram in the cubic lattice case ($n^{c} \equiv  n$ and $J \equiv J_{\rm K}$). 
The phase $4$ in red represents the noncoplanar triple-$Q$ state induced by the $(d-2)$-dimensional connections.
Figure (a) is reprinted with permission from reference~\cite{Akagi_JPSJ.79.083711}. Copyright 2010 by the Physical Society of Japan.
Figure (b) is reprinted with permission from reference~\cite{Hayami_PhysRevB.89.085124}. Copyright 2014 by the American Physical Society. 
}
\end{center}
\end{figure}

A different type of the multiple-$Q$ instability, which is more nontrivial than the perfect nesting case, was found at particular electronic states. 
This occurs when $(d-2)$-dimensional portions of the Fermi surfaces are connected by the multiple-$Q$ wave vectors in the $d$-dimensional extended Brillouin zone. 
This is a weaker nesting compared to the perfect nesting in section~\ref{sec:Perfect nesting} 
which is regarded as a $(d-1)$-dimensional connection of the Fermi surfaces in general
\footnote{
The pyrochlore case in figure~\ref{fig:perfect_nesting}(c) is special since the Fermi surfaces are the $(d-2)$-dimensional lines. 
Their connections are also $(d-2)$-dimensional ones, but we categorize it to the perfect nesting case since the entire portions of the Fermi surfaces are connected.}. 

A representative of such multiple-$Q$ instabilities was found for the Kondo lattice model on the triangular lattice~\cite{Akagi_JPSJ.79.083711}. 
As shown in the phase diagram in figure~\ref{fig:multipleQ_prestudy}(a), the same noncoplanar triple-$Q$ state as that by the perfect nesting at 3/4 filling in figure~\ref{fig:perfect_nesting}(b) was found to be stabilized near 1/4 filling. 
A similar triple-$Q$ state was also found for the periodic Anderson model~\cite{hayami_PhysRevB.91.075104}. 
Notably, this noncoplanar state is stable in a much wider region in the phase diagram compared to that by the perfect nesting. 
It also remains robust against thermal fluctuations (as a quasi-long-range order)~\cite{Kato_PhysRevLett.105.266405} 
and quantum fluctuations~\cite{Akagi_doi:10.7566/JPSJ.82.123709}. 
Later, a different noncoplanar triple-$Q$ state was found also for the cubic lattice case near 1/4 filling, as shown in figure~\ref{fig:multipleQ_prestudy}(b)~\cite{Hayami_PhysRevB.89.085124}.  
Besides, a variety of multiple-$Q$ states were obtained on various lattice structures, such as the checkerboard~\cite{Venderbos_PhysRevLett.109.166405}, honeycomb~\cite{Jiang_PhysRevLett.114.216402,Venderbos_PhysRevB.93.115108}, kagome~\cite{Barros_PhysRevB.90.245119,Ghosh_PhysRevB.93.024401}, square~\cite{Agterberg_PhysRevB.62.13816,hayami_PhysRevB.91.075104}, face-centered-cubic (fcc)~\cite{Shindou_PhysRevLett.87.116801}, and Shastry-Sutherland lattices~\cite{Shahzad_PhysRevB.96.224402}. 
These topological spin crystals have common features: They appear at a particular electron filling far from that for perfect nesting, and the periods of the magnetic structures are very short. 
These results suggest an underlying common mechanism despite the lack of perfect nesting.

\begin{figure}[hbt!]
\begin{center}
\includegraphics[width=1.0 \hsize]{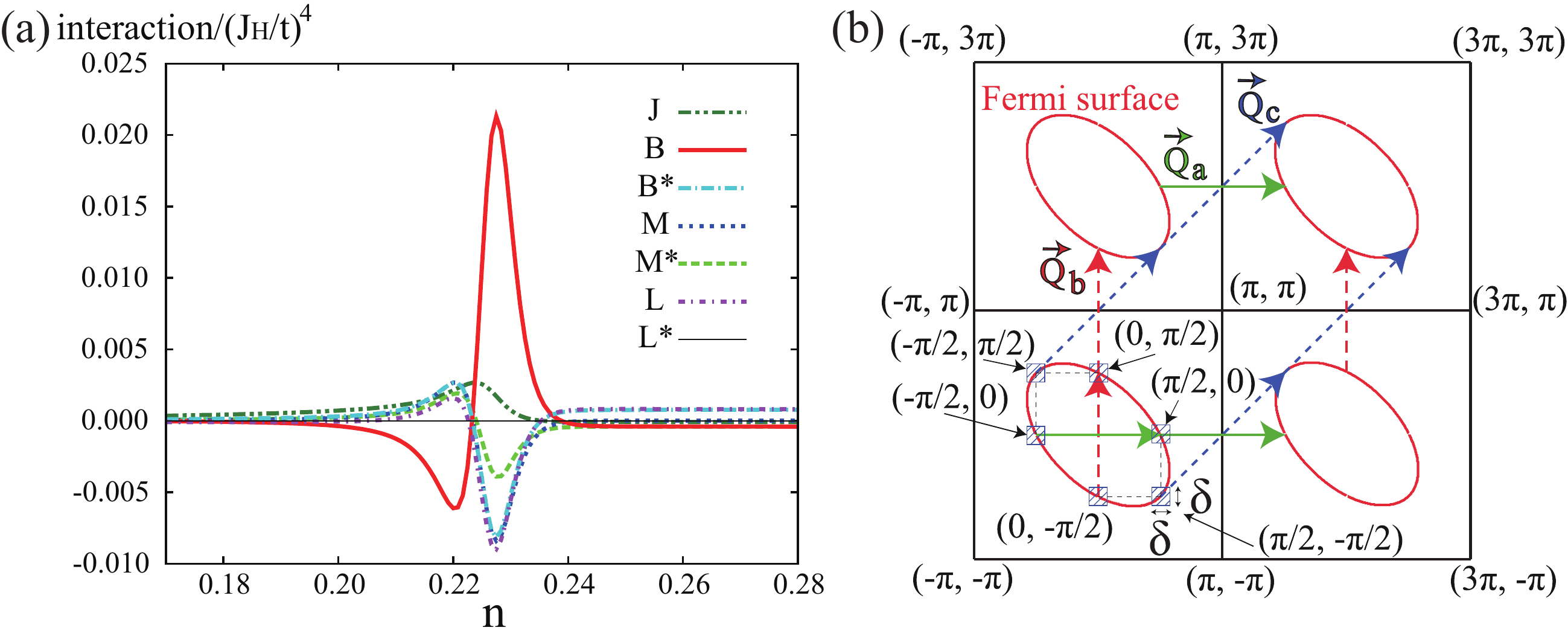} 
\caption{
\label{fig:multiple_connections_tri}
(a) Coefficients of different contributions to the fourth-order free energy as functions of the electron filling $n$.  
The most enhanced $B$ corresponds to the coefficient for the positive biquadratic interaction. 
(b) The Fermi surfaces at $n=0.225$ in the extended Brillouin zone scheme. 
The triangular lattice is defined as the square lattice with diagonal bonds. 
The six points on the Fermi surfaces (the hatched small squares) are multiply connected by the wave vectors ${\bm Q}_a=(\pi,0)$, ${\bm Q}_b=(0,\pi)$, and ${\bm Q}_c=(\pi,\pi)$. 
Figure is reprinted with permission from reference~\cite{Akagi_PhysRevLett.108.096401}. Copyright 2012 by the American Physical Society. 
}
\end{center}
\end{figure}

The mechanism was first discussed for the triple-$Q$ state on the triangular lattice~\cite{Akagi_PhysRevLett.108.096401}. 
By using the perturbation expansion in section~\ref{sec:Perturbation expansion}, it was pointed out that the positive biquadratic interaction in the fourth-order contribution is critically enhanced at the particular electron filling $n\simeq 0.225$, as shown in figure~\ref{fig:multiple_connections_tri}(a). 
At this filling, the Fermi surface is almost circular but has a special property; namely, six points on the Fermi surface are multiply connected by the three wave vectors ${\bm Q}_a$, ${\bm Q}_b$, and ${\bm Q}_c$ in the extended Brillouin zone, as shown in figure~\ref{fig:multiple_connections_tri}(b). 
This is the $(d-2)$-dimensional connections of the Fermi surfaces ($d-2=0$ in this two-dimensional case, namely, $0$-dimensional point connections). 
At this filling, $\chi_{\bm q}^0$ shows multiple peaks at the three wave vectors which are divergent in the limit of zero temperature. 
This leads to the critical enhancement of the biquadratic interaction and the instability toward the triple-$Q$ noncoplanar spin state with a local gap formation in the electronic state at the connected points on the Fermi surface
\footnote{
This is a lifting of the degeneracy at the second-order RKKY level. 
In this case, however, the degeneracy appears not only among the single-$Q$ states but also including the double- and triple-$Q$ states~\cite{Akagi_PhysRevLett.108.096401}. 
The situation is different from the general case in section~\ref{sec:General case} where the degeneracy by the RKKY interaction appears only among the single-$Q$ states and the multiple-$Q$ states have higher energies.}.

\begin{table}[htb!]
\begin{tabular}{ccc}
\hline \hline
lattice  & multiple-$Q$ wave vectors& symmetry \\
 \hline
square & $(\pi,0)$, $(0,\pi)$ & $C_{4}$ \\
triangular & $(\pi,0)$, $(0,\pi)$, $(\pi,\pi)$ & $C_{6}$ \\
cubic & $(0,\pi,\pi)$, $(\pi,0,\pi)$, $(\pi,\pi,0)$ & $C_{3} 
 $ \\
fcc & $(\pi,0,0)$, $(0,\pi,0)$, $(0,0,\pi)$ & $C_{3} 
 $ \\
 \hline
 \hline
 \end{tabular}
 \caption{
Some representative examples of the lattice structures, multiple-$Q$ wave vectors, and the symmetries relevant to the multiple-$Q$ topological spin crystals induced by the $(d-2)$-dimensional connections of the Fermi surfaces. 
The lattice constant is set to be unity in all the cases. 
Corresponding multiple-$Q$ spin patterns are schematically shown in figure~\ref{fig:perfect_nesting}(b) for the triangular lattice case and in the insets of the right panels in figure~\ref{fig:multiple_connections} for the other cases. 
Table is reprinted with permission from reference~\cite{Hayami_PhysRevB.90.060402}. Copyright 2014 by the American Physical Society.
 }
 \label{tab:multipleQ}
 \end{table}

\begin{figure}[hbt!]
\begin{center}
\includegraphics[width=0.73 \hsize]{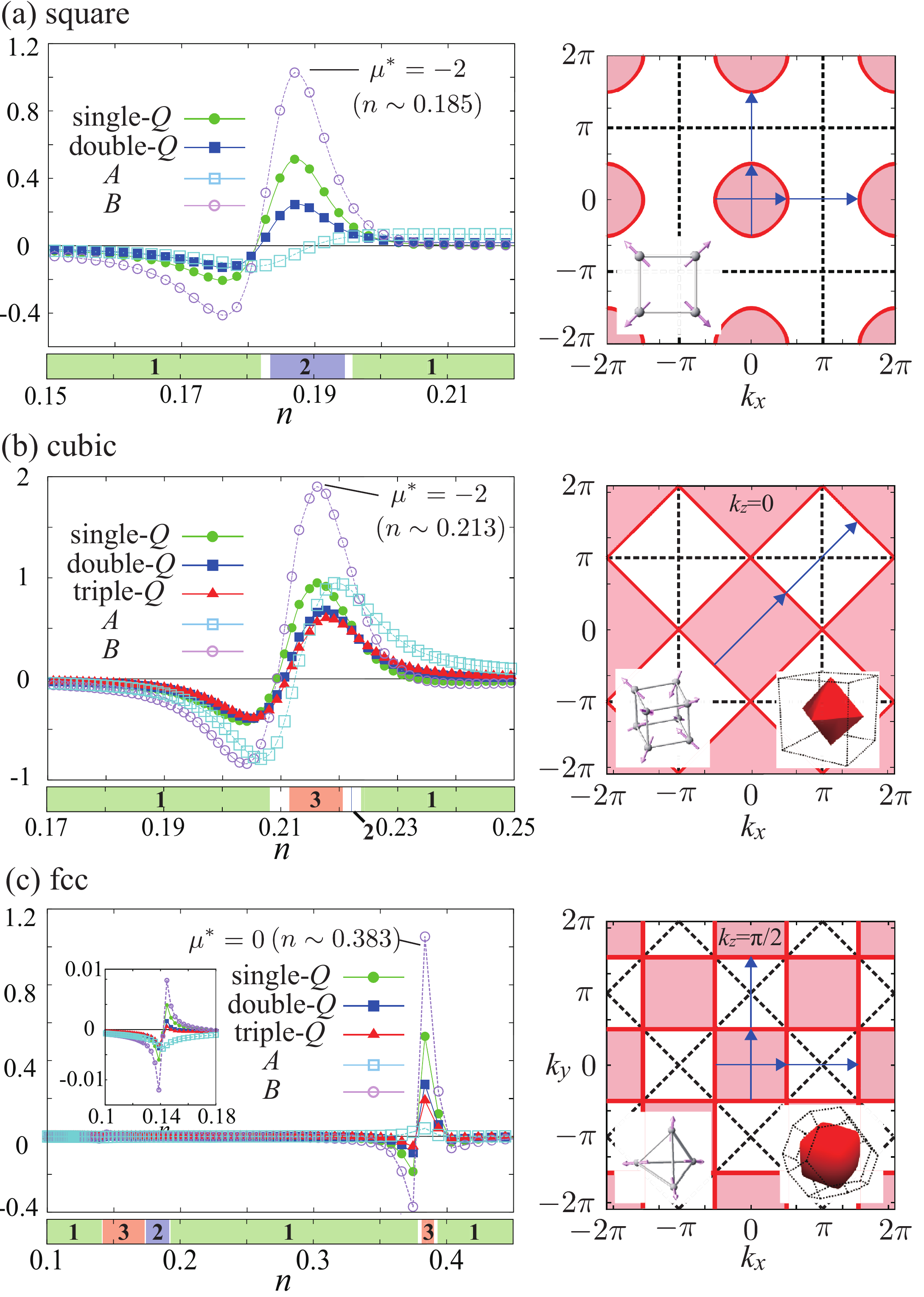} 
\caption{
\label{fig:multiple_connections}
Enhancement of the coefficient of the effective biquadratic interaction (left panels) and the Fermi surfaces multiply connected by the wave vectors in table~\ref{tab:multipleQ} (right panels) for the (a) square, (b) cubic, and (c) fcc lattice cases. 
$B$ stands for the biquadratic interaction corresponding to (\ref{eq:F42}). 
The inset of the left panel of (c) shows the enlarged plot near $n=0.14$. 
The bottom strip in each left panel represents the variational ground state at $J_{\rm K}=0.1$; 
1, 2, and 3 denote the single-, double-, and triple-$Q$ states, respectively. 
The schematic figures of the multiple-$Q$ states are presented in the insets of the right panels. 
The red lines and the blue arrows in the right panels represent the Fermi surfaces and the connecting vectors, respectively, at the electron fillings where $B$ is critically enhanced; the dashed lines are the Brillouin zone boundaries. 
(b) and (c) show the slices at $k_z=0$, and $k_z=\pi/2$, respectively; the three-dimensional Fermi surfaces in the first Brillouin zone are presented in each inset. 
Figure is reprinted with permission from reference~\cite{Hayami_PhysRevB.90.060402}. Copyright 2014 by the American Physical Society. 
}
\end{center}
\end{figure}

The idea was generalized to other lattices and the $(d-2)$-dimensional connections are shown to be a universal mechanism for stabilizing the multiple-$Q$ states~\cite{Akagi2011,Hayami_PhysRevB.90.060402}. 
To establish the multiple connections of the Fermi surfaces, we need commensurate and rather large wave vectors, like $(\pi,0)$, $(0,\pi)$, and $(\pi,\pi)$. 
Table~\ref{tab:multipleQ} summarizes such wave vectors for several lattice structures. 
Note that these wave vectors satisfy the condition $2\bm{Q}_\nu = \bm{G}$ (i.e., $\bm{q}_1+\bm{q}_2+\bm{q}_3+\bm{q}_4 = \bm{G}$) in the fourth-order free energy discussed in section~\ref{sec: Fourth-order interaction}. 
Indeed, at the particular electron fillings where the Fermi surfaces are multiply connected by these wave vectors as shown in the right panels of figure~\ref{fig:multiple_connections}, the fourth-order multiple-spin interactions derived by the perturbation in section~\ref{sec:Perturbation expansion} are critically enhanced ubiquitously in the different lattice systems as shown in the left panels of figure~\ref{fig:multiple_connections}. 
Such multiple-$Q$ instabilities are indeed found in the variational ground state of the Kondo lattice model, as shown in the bottom strips in the left panels of figure~\ref{fig:multiple_connections} and as mentioned above.

\subsection{General case}
\label{sec:General case}

In this section, we discuss general cases with neither perfect nesting nor $(d-2)$-dimensional connections of the Fermi surfaces. 
Surprisingly, even in such seemingly featureless cases, the system has an instability toward multiple-$Q$ noncoplanar spin states. 
In section~\ref{sec:Chiral stripe}, we illustrate the mechanism by taking an example of the double-$Q$ noncoplanar state discovered in the Kondo lattice model on a square lattice~\cite{Ozawa_doi:10.7566/JPSJ.85.103703}.
In section~\ref{sec:nsk2 skyrmion crystal}, we introduce other examples, two types of the SkXs with the skyrmion number of one and two, which are stabilized on the triangular lattice~\cite{Ozawa_PhysRevLett.118.147205}.  

\subsubsection{Double-$Q$ chiral stripe}
\label{sec:Chiral stripe}

\begin{figure}
\includegraphics[width=1.0 \hsize]{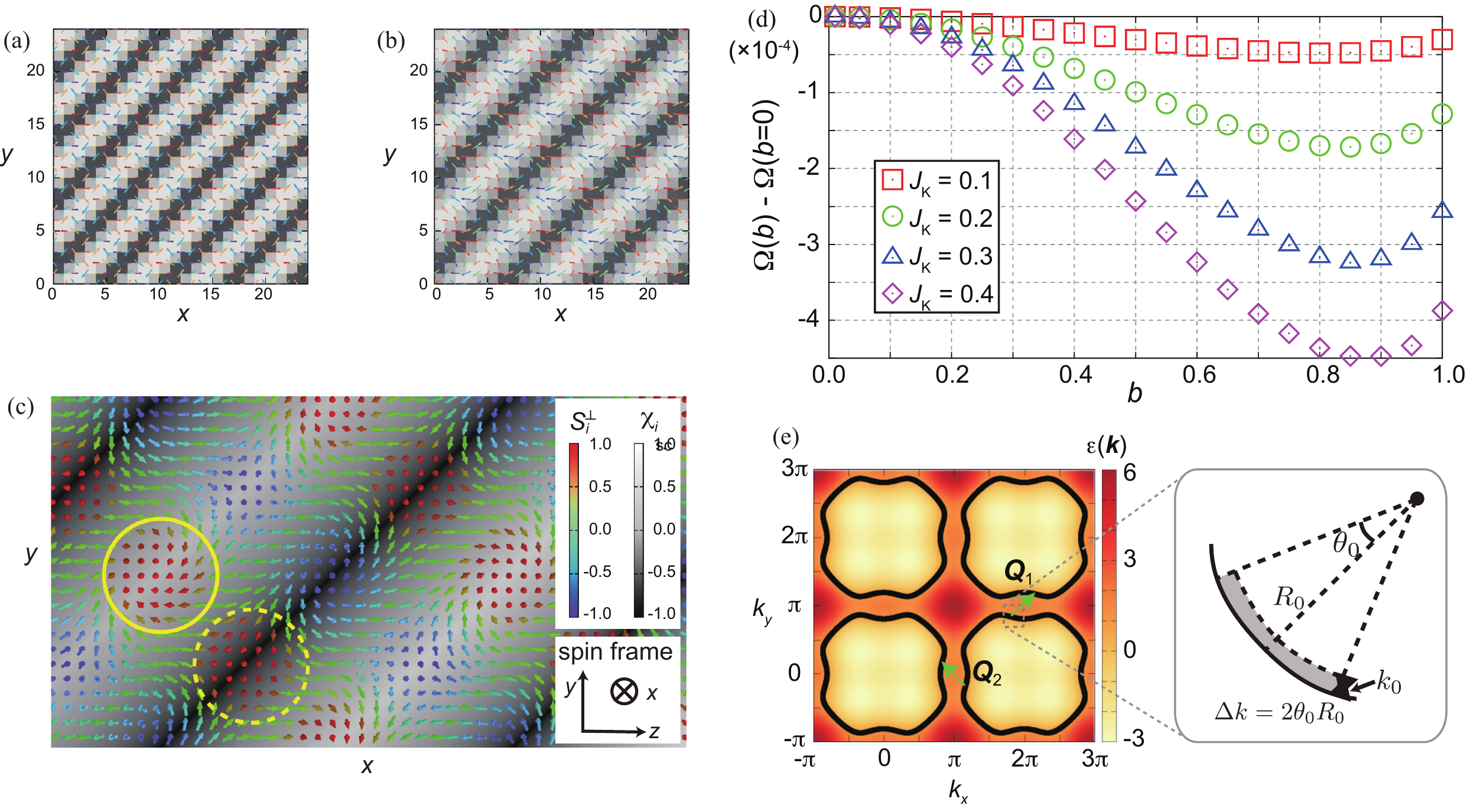}
\caption{
(a) and (b) Real-space spin and chirality configurations in the double-$Q$ CS state obtained from numerical simulations for the Kondo lattice model at different electron fillings with $t_1=1$, $t_3=-0.5$, and $J_{\rm K}=0.1$; the chemical potential is taken as (a) $\mu= 0.98$ and (b) $1.26$.
The arrows represent the in-plane spin component and their colors indicate the out-of-plane spin component.
The gray-scale background shows the striped modulation of the spin scalar chirality. 
(c) The double-$Q$ CS state in (\ref{eq:Spin_CS}) with $\bm{Q}_1=(\pi/12, \pi/12)$ and $\bm{Q}_2=(\pi/12, -\pi/12)$ for $b=1$. 
The spin frame is rotated to better visibility. 
The solid and dashed circles show vortex (meron) and antivortex (antimeron). 
(d) Grand potential of the double-$Q$ CS state in (\ref{eq:Spin_CS}) measured from that of the single-$Q$ helical state for $J_{\rm K}=0.1$, $0.2$, $0.3$, and $0.4$; we take $t_1=1$, $t_3=-0.5$, and $\mu=0.98$.  
(e) The Fermi surfaces connected by the ordering vectors $\bm{Q}_1$ and $\bm{Q}_2$ [the same as figure~\ref{fig:chiq}(b)].  
The color contour plots the energy in (\ref{eq:ek_triangular}). 
The right panel is the schematic of the Fermi surface near the hot spot in the cylindrical coordinate ($R_0$, $\theta_0$).  
Figure is reprinted with permission from reference~\cite{Ozawa_doi:10.7566/JPSJ.85.103703}. Copyright 2016 by the Physical Society of Japan.
}
\label{fig:CS}
\end{figure}

For a generic form of the Fermi surfaces, the RKKY interaction in (\ref{eq:RKKY}) favors a single-$Q$ spiral state with the wave vector maximizing $\chi_{\bm{q}}^0$, rather than multiple-$Q$ states, as the multiple-$Q$ superpositions inevitably have the higher harmonics, e.g., at $\bm{Q}_1+\bm{Q}_2$ and $2\bm{Q}_1$, and lead to the energy loss compared to the single-$Q$ state.  
Nevertheless, it was discovered by numerical simulations of the Kondo lattice model on a square lattice that a noncoplanar spin state is stabilized in the weak $J_{\rm K}$ regime~\cite{Ozawa_doi:10.7566/JPSJ.85.103703}. 
The typical spin and scalar chirality configurations are presented in figures~\ref{fig:CS}(a) and \ref{fig:CS}(b); 
the spins form a two-dimensional periodic array of noncoplanar vortices, and the scalar chirality shows a one-dimensional stripe in a diagonal direction. 
The magnetic period changes with the electron filling, namely, the size of the Fermi surface. 

The spin configuration found in the simulation is well approximated by a double-$Q$ state given by~\cite{Ozawa_doi:10.7566/JPSJ.85.103703} 
\begin{eqnarray}
\bm{S}_i=
	\left(\begin{array}{c}
	\sqrt{(1-b^2) + b^2\cos^2(\bm{Q}_2\cdot \bm{r}_i) } \cos( \bm{Q}_1\cdot \bm{r}_i)\\
	\sqrt{(1-b^2) + b^2\cos^2( \bm{Q}_2\cdot \bm{r}_i) } \sin( \bm{Q}_1\cdot \bm{r}_i)\\
	b\sin( \bm{Q}_2\cdot \bm{r}_i)
	\end{array}
	\right)^{\rm T}, 
\label{eq:Spin_CS}
\end{eqnarray}
where $\bm{Q}_1$ and $\bm{Q}_2$ are symmetry-related wave vectors at which $\chi_{\bm{q}}^0$ shows peaks [$C_4$ rotational symmetry in the square lattice case; see figures~\ref{fig:chiq}(a) and \ref{fig:chiq}(b)], and $b$ represents the amplitude of the $\bm{Q}_2$ component; the superscript T denotes the transpose of the vector. 
Note that the spin configuration is continuously connected to the single-$Q$ spiral state by taking $b \to 0$. 
The real-space spin configuration in (\ref{eq:Spin_CS}) is shown in figure~\ref{fig:CS}(c) with $\bm{Q}_1=(\pi/12, \pi/12)$ and $\bm{Q}_2=(\pi/12, -\pi/12)$ for $b=1$.
This indicates that the spin configuration is given by a periodic array of vortices and antivortices. 
Interestingly, the spins in each vortex and antivortex are noncoplanar and wrap half of a sphere. 
These half-skyrmion and half-antiskyrmion are called meron and antimeron, respectively~\cite{Lin_PhysRevB.91.224407}, and hence, the spin configuration in (\ref{eq:Spin_CS}) can be viewed as a meron-antimeron crystal.  
This state is termed as the double-$Q$ chiral stripe (CS) state owing to the stripe pattern in the scalar chirality~\cite{Ozawa_doi:10.7566/JPSJ.85.103703}
\footnote{
A similar but different double-$Q$ state with chiral stripe has also been discussed in reference~\cite{Solenov_PhysRevLett.108.096403}, although it has not been confirmed in numerical simulations.}. 

Figure~\ref{fig:CS}(d) presents the variational energy (grand potential) of the spin state in (\ref{eq:Spin_CS}) measured from that for the single-$Q$ state with $b=0$ for several values of $J_{\rm K}$. 
In all cases, the energy is lowered by introducing the second $\bm{Q}_2$ component, and optimized at relatively large value of $b$. 
The optimized energy agrees well with those obtained by the numerical simulations, indicating that (\ref{eq:Spin_CS}) describes well the spin states obtained numerically~\cite{Ozawa_doi:10.7566/JPSJ.85.103703}.

The mechanism of the instability toward this double-$Q$ CS state was again discussed by using the perturbation in terms of $J_{\rm K}$ presented in section~\ref{sec:Perturbation expansion}. 
Assuming (\ref{eq:Spin_CS}) with $b \ll 1$, the free energy up to the fourth order in (\ref{eq:RKKYHam}) and (\ref{eq:4thfreeenergy_exact}) gives the energy gain by forming the double-$Q$ CS state as~\cite{Ozawa_doi:10.7566/JPSJ.85.103703} 
\begin{equation}
\Delta F_{\rm CS}(b) =
\alpha_1 J_{\rm K}^2b^4 - \alpha_2 J_{\rm K}^4b^2,
\label{eq:del_energy_Jb_direct_simple}
\end{equation}
where
\begin{eqnarray}
& \alpha_1 
= \frac{1}{32}\left( 
\chi^0_{\bm{Q}_1} - \chi^0_{\bm{Q}_1 + 2\bm{Q}_2} \right), 
\ \ 
\alpha_2 
=\frac{1}{2}\left( A_{\bm{Q}_1} - 2B_{\bm{Q}_1, \bm{Q}_2} + W_{\bm{Q}_1, \bm{Q}_2}\right),  
\end{eqnarray}
and
\begin{eqnarray}
& A_{\bm{Q}_1} 
=T\sum_{\bm{k}, \omega_p}G_{\bm{k}}^2G_{\bm{k}+\bm{Q}_1}^2, 
\ \ B_{\bm{Q}_1, \bm{Q}_2} 
= T\sum_{\bm{k}, \omega_p} G_{\bm{k}} G_{\bm{k}+\bm{Q}_1}^2 G_{\bm{k}+\bm{Q}_1+\bm{Q}_2}, \nonumber \\
& W_{\bm{Q}_1, \bm{Q}_2} 
= T\sum_{\bm{k}, \omega_p}G_{\bm{k}}G_{\bm{k}+\bm{Q}_1}G_{\bm{k}+\bm{Q}_2}G_{\bm{k}+\bm{Q}_1+\bm{Q}_2}. 
\end{eqnarray}
(\ref{eq:del_energy_Jb_direct_simple}) indicates that there is energy competition between the second-order RKKY contribution $\sim J_{\rm K}^2 b^4$ and the fourth-order one $\sim J_{\rm K}^4 b^2$ in the small $b$ limit. 
The coefficient $\alpha_1$ for the former is always positive because $\chi^0_{\bm{q}}$ 
is maximized at $\bm{q}=\bm{Q}_1$ and $\bm{Q}_2$.
Hence, when $\alpha_2$ for the latter is positive, the optimal value of $b$ to minimize $\Delta F_{\rm CS}(b)$ is given by $b_{\rm opt} = \sqrt{\alpha_2/(2 \alpha_1)} J_{\rm K}$.  
Indeed, after explicit evaluation of the coefficients  $A_{\bm{Q}_1}$,  $B_{\bm{Q}_1, \bm{Q}_2}$, and 
$W_{\bm{Q}_2, \bm{Q}_2}$, one can find that $\alpha_2 \rightarrow +\infty$ in the low-temperature limit. 
While this indicates the breakdown of the perturbative expansion, it suggests that the system has an instability toward the double-$Q$ CS state and the amplitude $b$ of the second component can be in the order of one even for very small $J_{\rm K}$, seemingly supporting the results in figure~\ref{fig:CS}(d). 

The breakdown of the perturbative expansion can be avoided by taking the local reference frame for the itinerant electron spins along the localized spins in one of the single-$Q$ spiral states~\cite{Ozawa_doi:10.7566/JPSJ.85.103703}. 
Skipping the details of the derivation, the most dominant contributions arising from the regions around the Fermi surface points connected by $\bm{Q}_1$ and $\bm{Q}_2$ (hot spots) can be summarized into the energy cost $\Delta E_1$ and the energy gain $\Delta E_2$ given by~\cite{Ozawa_doi:10.7566/JPSJ.85.103703} 
\begin{eqnarray}
\label{eq:DeltaE_1}
\Delta E_1 &= - \frac{\Delta k}{4 \pi^2} \frac{J_{\rm K}^2b^2}{16 v_{\rm F}} \ln{ \left [ \frac{x + \sqrt{\left(\frac{J_{\rm K}}{2 v_{\rm F}}\right)^2 +x^2}}
{ k_0+ x + \sqrt{\left(\frac{J_{\rm K}}{2 v_{\rm F}}\right)^2 + (k_0+x)^2}} \right ]},
\\
\label{eq:DeltaE_2}
\Delta E_2 &=  \frac{\Delta k}{4 \pi^2} \frac{J_{\rm K}^2b^2}{16 v_{\rm F}} \ln{ \left[ \frac{x + \sqrt{\left(\frac{J_{\rm K}b}{4 v_{\rm F}}\right)^2  +x^2}}
{ k_0+ x + \sqrt{\left(\frac{J_{\rm K}b}{4 v_{\rm F}}\right)^2 + (k_0+x)^2}} \right],}
\end{eqnarray} 
respectively, where $v_{\rm F}$ is the Fermi velocity at the hot spot, $\Delta k$ and $k_0$ define the circular rectangle of integration around the hot spots, 
and $x={\Delta k}^2/8 R_0$; see figure~\ref{fig:CS}(e). 
Note that, in this rotated local frame, the divergence at the fourth order in the original frame is renormalized and incorporated in the second-order contributions. 
The results in (\ref{eq:DeltaE_1}) and (\ref{eq:DeltaE_2}) indicate that $\Delta E_1 + \Delta E_2 <0$ for $b \ll 1$, which explains the instability toward the double-$Q$ CS state in the weak coupling limit of $J_{\rm K}\to 0$.  
Indeed, the perturbative arguments were confirmed quantitatively by careful comparison with the variational calculations with (\ref{eq:Spin_CS})~\cite{Ozawa_doi:10.7566/JPSJ.85.103703}.

\subsubsection{Triple-$Q$ skyrmion crystal}
\label{sec:nsk2 skyrmion crystal}

\begin{figure*}[h!]
\begin{center}
	 \includegraphics[width=0.95 \hsize]{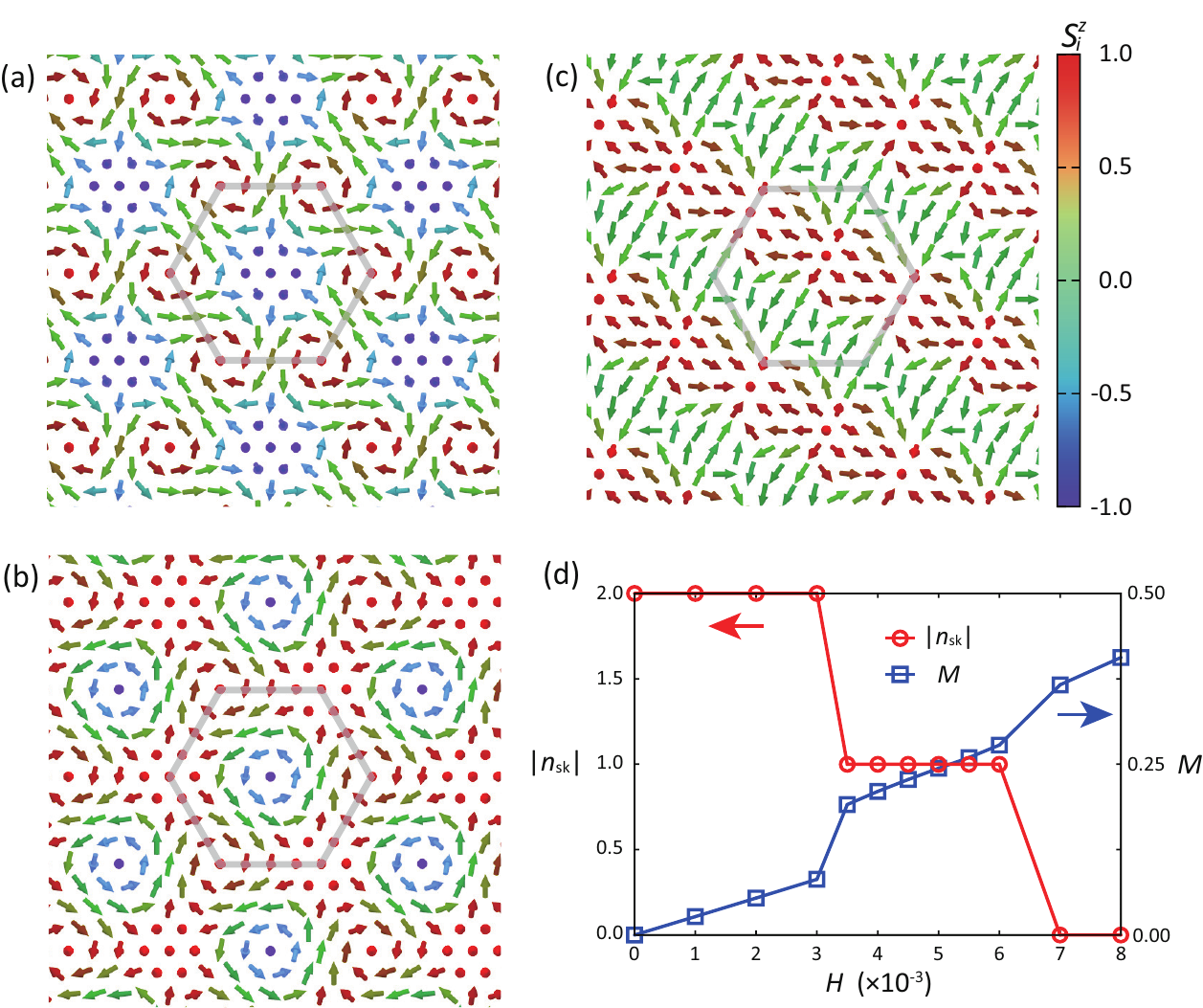}
	 \caption{
	 \label{fig:nsk2_SkX} 
Spin textures in (a) the $n_{\rm sk}=2$ SkX at $H=0$, (b) the $n_{\rm sk}=1$ SkX at $H=0.005$, and (c) the $n_{\rm sk}=0$ state at $H=0.008$ obtained by numerical simulations for the Kondo lattice model on the triangular lattice with $t_1=1$, $t_3 = -0.85$, $J_{\rm K}=0.5$, and $\mu=-3.5$.
The gray hexagons represent the magnetic unit cell.
(d) $H$ dependences of the skyrmion number $|n_{\rm sk}|$ and the magnetization of the localized spin per site, $M$. 
Figure is reprinted with permission from reference~\cite{Ozawa_PhysRevLett.118.147205}. Copyright 2017 by the American Physical Society.
	}
\end{center}
\end{figure*}

In the square lattice case above, there are two wave vectors related with the $C_4$ rotational symmetry, and the system becomes unstable toward the double-$Q$ CS state by making a superposition of the two components. 
Under the hexagonal symmetry, however, there are three wave vectors related by $C_3$. 
In this case also, similar instability toward a double-$Q$ state composed of two wave vectors out of three occurs in the Kondo lattice model in the weak $J_{\rm K}$ region for $0<J_{\rm K}\lesssim 0.11$, but in addition a different instability was found in the larger $J_{\rm K}$ region~\cite{Ozawa_PhysRevLett.118.147205,Hayami_PhysRevB.95.224424}. 
Figure~\ref{fig:nsk2_SkX}(a) shows the spin configuration obtained by numerical simulation of the Kondo lattice model on the triangular lattice with $t_1=1$, $t_3=-0.85$, $J_{\rm K}=0.5$, and $\mu=-3.5$ for which $\chi^0_{\bm{q}}$ has the peaks at $\bm{Q}_1=(\pi/3,0)$, $\bm{Q}_2=(-\pi/6,\sqrt{3}\pi/6)$, and $\bm{Q}_3=(-\pi/6,-\sqrt{3}\pi/6)$, as shown in figure~\ref{fig:chiq}(c).
The spin structure, which preserves the $C_3$ rotational symmetry, turns out to be a SkX with high skyrmion number of two in the magnetic unit cell~\cite{eriksson1990measure}, and hence, termed as the $n_{\rm sk}=2$ SkX
\footnote{
The sign of $n_{\rm sk}$ is irrelevant owing to the continuous rotational symmetry in spin space in the Kondo lattice model; namely, the SkXs with $\pm 2$ are energetically degenerate. 
The degeneracy is lifted, e.g., by the DM interaction and the bond-dependent anisotropic interaction in the magnetic field (see section~\ref{sec:Triangular lattice} for the latter case). 
This holds also for the $n_{\rm sk}=1$ case in the magnetic field.}~\cite{Ozawa_PhysRevLett.118.147205}. 
The spin configuration is well approximated by a superposition of three sinusoidal waves as
\begin{eqnarray}
\bm{S}_i\propto 
  (\cos \bm{Q}_1\cdot \bm{r}_i, \cos \bm{Q}_2\cdot \bm{r}_i, \cos \bm{Q}_3\cdot \bm{r}_i). 
  \label{eq:Qsk=2}
\end{eqnarray}
This is a triple-$Q$ state, where the spin structure factor has the peaks at $\bm{Q}_1$, $\bm{Q}_2$, and  $\bm{Q}_3$ with equal intensity.
It has a periodic array of vortices with vorticity $v=-2$ centered at downward spins and merons with $v=+1$
\footnote{
We note that this state is a relative of the triple-$Q$ state in figure~\ref{fig:perfect_nesting}(b) with a longer magnetic period.}. 

In contrast to the double-$Q$ CS state in section~\ref{sec:Chiral stripe}, the emergence of the $n_{\rm sk}=2$ SkX is not simply understood from the perturbation expansion with respect to $J_{\rm K}$ in section~\ref{sec:Multi-spin interactions beyond the RKKY interaction}. 
Indeed, as mentioned above, the double-$Q$ CS state is stabilized in the weak $J_{\rm K}$ limit, and it is taken over by the $n_{\rm sk}=2$ SkX when $J_{\rm K}$ is increased. 
Nevertheless, the stabilization mechanism of the $n_{\rm sk}=2$ SkX is well explained by the effective four-spin interactions derived by the perturbation expansion in section~\ref{sec: Fourth-order interaction}, as we will detail in the following section~\ref{sec:Effective spin model in itinerant magnets}.

Interestingly, by applying an external magnetic field to the present system, the $n_{\rm sk}=2$ SkX turns into the SkX with $n_{\rm sk}=1$~\cite{Ozawa_PhysRevLett.118.147205}. 
The effect of the magnetic field is introduced by adding the Zeeman coupling, 
\begin{equation}
\mathcal{H}^{\rm Z}=- H \sum_i S_i^z, 
\label{eq:H^Z}
\end{equation}
to (\ref{eq:Ham_KLM}), where the magnetic field $H$ is applied only to the localized spins for simplicity. 
The spin configuration obtained by the numerical simulation for $H=0.005$ is shown in figure~\ref{fig:nsk2_SkX}(b). 
This is again viewed as a periodic array of vortices with $v=-2$ and merons with $v=1$, but in a different manner from the $n_{\rm sk}=2$ SkX. 
Indeed, the skyrmion number is reduced to one. 
This is also a triple-$Q$ state with the same intensity at $\bm{q}=\bm{Q}_1$, $\bm{Q}_2$, and  $\bm{Q}_3$ in the spin structure factor. 
Hence, this state is called the $n_{\rm sk}=1$ SkX. 
For a larger magnetic field, the spin state becomes topologically trivial, i.e., $n_{\rm sk}=0$, whose spin structure is shown in figure~\ref{fig:nsk2_SkX}(c). 
Figure~\ref{fig:nsk2_SkX}(d) summarizes the changes of $|n_{\rm sk}|$ and the magnetization in localized spins per site $M =|\sum_i \bm{S}_i|/N$ in an applied magnetic field $H$~\cite{Ozawa_PhysRevLett.118.147205}.
The result indicates that the system exhibits two successive transitions with the changes in the skyrmion number $n_{\rm sk}$ from $2$ to $1$, and to $0$, while increasing the magnetic field. 

\section{Effective spin model for itinerant frustration}
\label{sec:Effective spin model in itinerant magnets}

In the previous section, we have reviewed that a variety of topological spin crystals appear in the weak spin-charge coupling regime of the Kondo lattice model. 
Some of them suggest that the perturbation in terms of the spin-charge coupling can account for the instabilities toward the multiple-$Q$ states. 
In this section, we present that an effective spin model with the bilinear and biquadratic interactions in momentum space, which is constructed on the basis of the perturbation expansion, reproduces well not only the multiple-$Q$ phases in the weak coupling limit like the double-$Q$ CS state in section~\ref{sec:Chiral stripe} but also those in the intermediate coupling regime like the SkXs in section~\ref{sec:nsk2 skyrmion crystal}. 
This indicates that the effective spin model provides a powerful framework to study the itinerant frustration in a wide range of parameters. 
Indeed, as demonstrated in section~\ref{sec:Extensions of the effective spin model}, the effective model and its extensions have been shown to be useful for the comprehensive study of the phase diagram in a wide parameter range and the exploration of further exotic topological spin crystals, since the computational cost is much cheaper compared to that for the models including itinerant electrons explicitly. 
After introducing the Hamiltonian in section~\ref{sec:Bilinear-biquadratic model in momentum space}, we demonstrate that the model can reproduce the multiple-$Q$ topological spin crystals discovered in the original Kondo lattice model in section~\ref{sec:Multiple-$Q$ magnetic instability}. 

\subsection{Bilinear-biquadratic model in momentum space}
\label{sec:Bilinear-biquadratic model in momentum space}

The perturbation expansion in section~\ref{sec:Multi-spin interactions beyond the RKKY interaction} indicates that many different types of effective spin interactions can contribute to the magnetic ordering in itinerant magnets. 
The comparison between different terms, however, gives an insight that the positive biquadratic interaction may play an important role, in addition to the primary RKKY interaction, as discussed in section~\ref{sec: Fourth-order interaction}. 
Based on this observation, an effective spin model for the itinerant frustration was proposed in the form~\cite{Hayami_PhysRevB.95.224424} 
\begin{equation}
\label{eq:effHam_spin}
\mathcal{H}^{\rm BBQ}=  2\sum_\nu
\left[ -J\bm{S}_{\bm{Q_{\nu}}} \cdot \bm{S}_{-\bm{Q_{\nu}}}
+\frac{K}{N} \left(\bm{S}_{\bm{Q_{\nu}}} \cdot \bm{S}_{-\bm{Q_{\nu}}}\right)^2 \right], 
\end{equation}
where the sum is taken for a set of $\bm{Q}_\nu$ giving the multiple peaks in the bare susceptibility $\chi_{\bm{q}}^{0}$; 
$J$ is set to be an energy unit and $K$ is taken to be positive. 
The first term originates from the second-order RKKY interaction in (\ref{eq:F(2)_helical}) and the second one is from one of the fourth-order contributions in (\ref{eq:F42}). 
Therefore, in the sense of the perturbation, $J$ is proportional to $J_{\rm K}^2$ and dominant over $K$ proportional to $J_{\rm K}^4$. 
However, the coefficients include Green's functions of itinerant electrons, which depend on the band structure and the electron filling. 
In addition, the model can be regarded to include a series of all the higher-order contributions, as discussed in section~\ref{sec:Higher-order interactions}. 
In the following, we therefore do not limit ourselves to $J\gg K$ and discuss the instabilities toward multiple-$Q$ topological spin crystals in a wide range of parameters of $J$ and $K$.  

The model in (\ref{eq:effHam_spin}) has the bilinear and biquadratic interactions defined in momentum space, which is in contrast to the bilinear-biquadratic model with short-range interactions in real space used for magnetic insulators
~\cite{takahashi1977half,yoshimori1978fourth,Hoffmann_PhysRevB.101.024418}. 
The difference arises from the different origin of the effective interactions; 
the effective spin interactions in the magnetic insulators are derived by the perturbation in terms of the electron hopping of localized electrons, and hence, they decay exponentially in real space (see section~\ref{sec:Frustration in insulating magnets}), but those for the itinerant magnets with weak spin-charge coupling are caused by the Fermi surface instability in momentum space, and hence, they can be long-ranged in nature. 
Reflecting such a difference, the coupling constant $K$ for the biquadratic interaction becomes positive in the present case, while it is negative in most cases for the magnetic insulators. 

\subsection{Multiple-$Q$ magnetic instability}
\label{sec:Multiple-$Q$ magnetic instability}

In this section, we present that the effective spin model in (\ref{eq:effHam_spin}) well reproduces the instabilities toward the multiple-$Q$ states found in the Kondo lattice model, following reference~\cite{Hayami_PhysRevB.95.224424}. 
In section~\ref{sec:Phase diagram on a square lattice}, we discuss the double-$Q$ CS state on the square lattice, and in section~\ref{sec:Phase diagram on a triangular lattice}, we discuss the double-$Q$ CS and the SkXs with $n_{\rm sk}=1$ and $2$ on the triangular lattice.

\subsubsection{On the square lattice}
\label{sec:Phase diagram on a square lattice}

\begin{figure}[htb!]
\begin{center}
\includegraphics[width=1.0 \hsize]{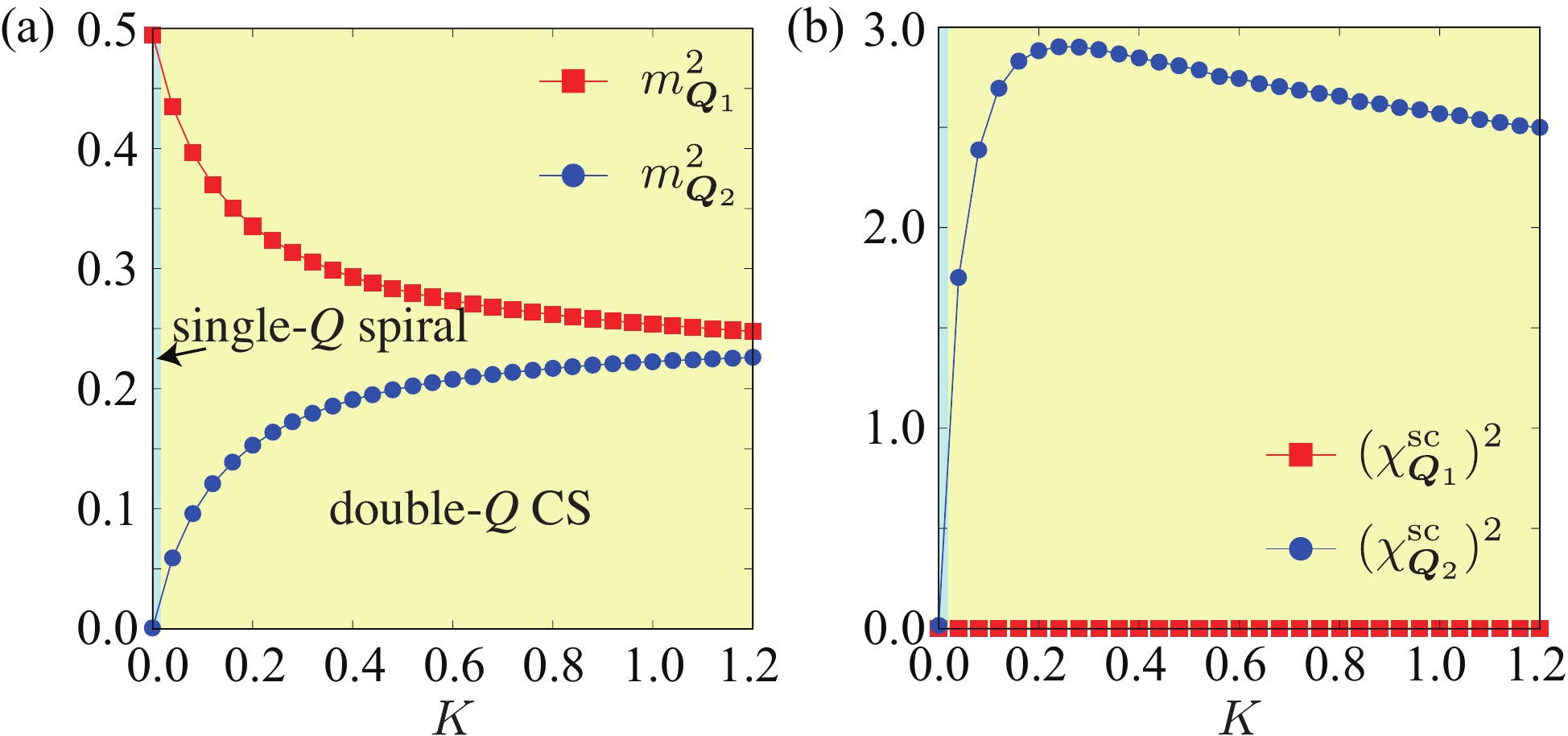} 
\caption{
\label{fig:SL_JKmodel}
Instability toward the double-$Q$ CS state in the effective spin model in (\ref{eq:effHam_spin}) on the square lattice. 
$K$ dependences of the $\bm{Q}_1=(\pi/3,\pi/3)$ and $\bm{Q}_2=(\pi/3,-\pi/3)$ components of (a) the squared magnetization and (b) the squared scalar chirality obtained by the simulated annealing for the model in (\ref{eq:effHam_spin}). 
Figure is reprinted with permission from reference~\cite{Hayami_PhysRevB.95.224424}. Copyright 2017 by the American Physical Society.
}
\end{center}
\end{figure}

First, we discuss the result for the effective bilinear-biquadratic model in (\ref{eq:effHam_spin}) on the square lattice by assuming the maxima in the bare susceptibility at $\bm{Q}_1=(\pi/3, \pi/3)$ and $\bm{Q}_2=(\pi/3, -\pi/3)$~\cite{Hayami_PhysRevB.95.224424}.  
Figure~\ref{fig:SL_JKmodel} shows the spin and scalar chirality as functions of $K$ obtained by simulated annealing:  
the $\bm{Q}_\nu$ components of (a) the magnetization, 
\begin{equation}
m_{\bm{Q}_\nu}=
\frac1N \sqrt{\sum_{i,j} \bm{S}_i \cdot \bm{S}_j  e^{i \bm{Q}_\nu \cdot (\bm{r}_i-\bm{r}_j)}}, 
\label{eq:mQ}
\end{equation}
and (b) the scalar chirality, 
\begin{equation}
\chi^{\rm sc}_{\bm{Q}_\nu}=\frac1N \sqrt{\sum_{i,j} \chi^{\rm sc}_{i} \chi^{\rm sc}_{j} e^{i \bm{q}\cdot (\bm{r}_i-\bm{r}_j)}}, 
\label{eq:chiQ}
\end{equation}
where $\nu=1$ and $2$. 
$\chi^{\rm sc}_{i}$ in (\ref{eq:chiQ}) is the local scalar chirality at site $i$ calculated by 
$\chi^{\rm sc}_i = \bm{S}_{i} \cdot (\bm{S}_{i+\hat{x}}\times \bm{S}_{i+\hat{y}})
+\bm{S}_{i} \cdot (\bm{S}_{i-\hat{x}}\times \bm{S}_{i-\hat{y}})
-\bm{S}_{i} \cdot (\bm{S}_{i-\hat{x}}\times \bm{S}_{i+\hat{y}})
-\bm{S}_{i} \cdot (\bm{S}_{i+\hat{x}}\times \bm{S}_{i-\hat{y}})$, 
where $\hat{x}$ and $\hat{y}$ denote the shifts by the lattice constant in the $x$ and $y$ directions, respectively. 
At $K=0$, the RKKY interaction stabilizes the single-$Q$ spiral state with $\bm{Q}_1$, but the introduction of $K$ induces the double-$Q$ state by mixing the $\bm{Q}_2$ component, as shown in figure~\ref{fig:SL_JKmodel}(a)
\footnote{The state with interchanging $\bm{Q}_1$ and $\bm{Q}_2$ is energetically degenerate as expected from the symmetry. Here and hereafter, the wave vectors are ordered appropriately for better visibility.}. 
In this state for $K>0$, the scalar chirality becomes nonzero only for the $\bm{Q}_2$ component, as shown in figure~\ref{fig:SL_JKmodel}(b). 
This noncoplanar double-$Q$ state is basically the same as the double-$Q$ CS state found for the Kondo lattice model in section~\ref{sec:Chiral stripe}.

In a similar manner to (\ref{eq:del_energy_Jb_direct_simple}), the energy difference between the single-$Q$ spiral and the double-$Q$ CS states is evaluated for the model in (\ref{eq:effHam_spin}) in the limit of $b \ll 1$ as~\cite{Hayami_PhysRevB.95.224424}
\begin{eqnarray}
E^{2Q} - E^{1Q}
 \label{eq:compare1Q2Q}
\sim \frac{Jb^4}{32}-\frac{Kb^2}{2},  
\end{eqnarray}
where $E^{1Q}=- J + K/2$ is the energy per site for the single-$Q$ spiral state.
Thus, the condition to stabilize the double-$Q$ CS state, i.e., $E^{2Q} < E^{1Q}$, reads 
\begin{equation}
0< b^2 < \frac{16K}{J}. 
\end{equation}
This means that an infinitesimal $K$ makes the single-$Q$ spiral state unstable by introducing the second component with the amplitude $b$. 

The above argument is generic and applicable to any lattices, such as the triangular and cubic lattices~\cite{Hayami_PhysRevB.95.224424}.
Thus, the effective bilinear-biquadratic model in (\ref{eq:effHam_spin}) well reproduces the instability toward the double-$Q$ CS state in the weak coupling limit of the Kondo lattice model.

\subsubsection{On the triangular lattice}
\label{sec:Phase diagram on a triangular lattice}

\begin{figure}[t!]
\begin{center}
\includegraphics[width=1.0 \hsize]{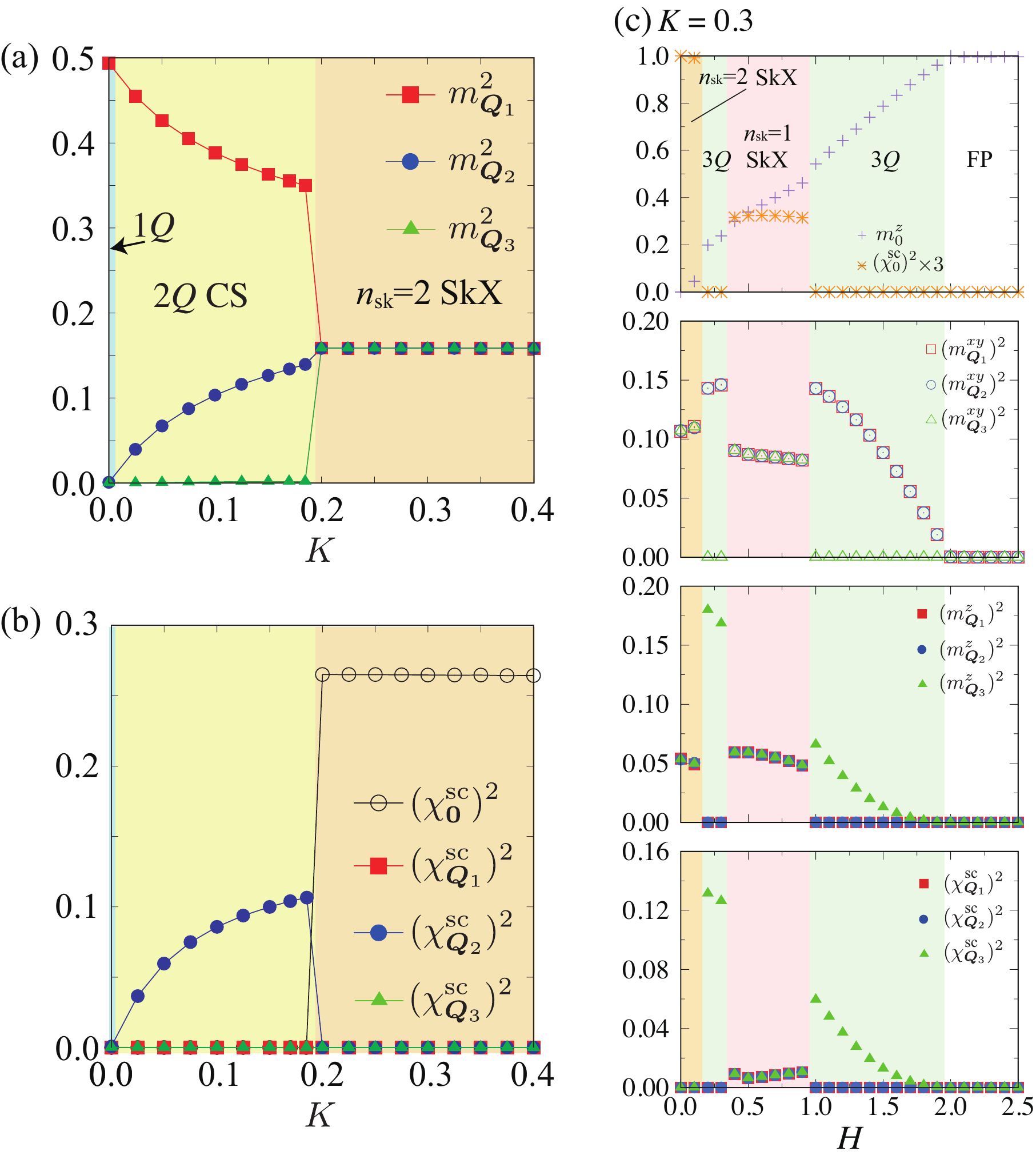} 
\caption{
\label{fig:TL_JKmodel}
(a, b) $K$ dependences of the $\bm{Q}_1=(\pi/3,0)$, $\bm{Q}_2=(-\pi/6,\sqrt{3}\pi/6)$, and $\bm{Q}_3=(-\pi/6,-\sqrt{3}\pi/6)$ components of (a) the squared magnetization and (b) the squared scalar chirality obtained by the simulated annealing for the model in (\ref{eq:effHam_spin}).
(c) $H$ dependences of $m^z_0$ and $(\chi^{\rm sc}_0)^2$, $(m^{xy}_{\bm{Q}_\nu})^2$, $(m^{z}_{\bm{Q}_\nu})^2$, and $ (\chi^{\rm sc}_{\bm{Q}_\nu})^2$ for $K=0.3$ from top to bottom. 
$3Q$ and FP represent the $n_{\rm sk}=0$ triple-$Q$ and fully polarized states, respectively.
Figures (a) and (b) are reprinted with permission from reference~\cite{Hayami_PhysRevB.95.224424}. Copyright 2017 by the American Physical Society.
Figure (c) is reprinted with permission from reference~\cite{Hayami_PhysRevB.103.054422}. Copyright 2021 by the American Physical Society.
}
\end{center}
\end{figure}

Next, we introduce the results on the triangular lattice, by choosing $\bm{Q}_1=(\pi/3,0)$, $\bm{Q}_2=(-\pi/6,\sqrt{3}\pi/6)$, and $\bm{Q}_3=(-\pi/6,-\sqrt{3}\pi/6)$.
The spin and scalar chirality are shown in figures~\ref{fig:TL_JKmodel}(a) and \ref{fig:TL_JKmodel}(b), respectively~\cite{Hayami_PhysRevB.95.224424}. 
The scalar chirality at $\bm{Q}_\nu$ components on the triangular lattice is defined in a similar manner to (\ref{eq:chiQ}) with the summation over the local scalar chirality $\chi^{\rm sc}_{\bm{R}}= \bm{S}_j \cdot (\bm{S}_k \times \bm{S}_l)$, where 
$\bm{R}$ is the position vector at the center of triangle and $j,k,l$ are three sites on the triangle in the counterclockwise order. 
Similar to the square lattice case in section~\ref{sec:Phase diagram on a square lattice}, the single-$Q$ state at $K=0$ turns into the double-$Q$ CS state by introducing $K$. 
While increasing $K$, however, a phase transition from the double-$Q$ CS state occurs at $K \simeq 0.19$. 
At the transition, the magnetic moment changes discontinuously, as shown in figure~\ref{fig:TL_JKmodel}(a); all $m_{\bm{Q}_\nu}$ become nonzero with equal intensity for $K \gtrsim 0.19$. 
In the scalar chirality sector, the uniform ($\bm{q}=\bm{0}$) component is induced, while the $\bm{Q}_\nu$ components all vanish, as shown in figure~\ref{fig:TL_JKmodel}(b). 
This noncoplanar triple-$Q$ state is the $n_{\rm sk}=2$ SkX obtained in the Kondo lattice model in section~\ref{sec:nsk2 skyrmion crystal}. 
Thus, the effective model in (\ref{eq:effHam_spin}) reproduces the phase sequence from the double-$Q$ CS state to the triple-$Q$ $n_{\rm sk}=2$ SkX found in the Kondo lattice model, which indicates that the increase of $K$ mimics the increase of $J_{\rm K}$.

Next, we present the effect of the magnetic field applied along the $z$ direction for nonzero $K$ by considering the model Hamiltonian $\mathcal{H}^{\rm BBQ}+ \mathcal{H}^{\rm Z}$~\cite{Hayami_PhysRevB.95.224424,Hayami_PhysRevB.103.054422}. 
Figure~\ref{fig:TL_JKmodel}(c) shows the magnetic field dependence of the spin and chirality components at $K=0.3$. 
In this case, to distinguish the magnetization perpendicular and parallel to the magnetic field, the squared magnetizations are plotted by decomposing into the $xy$ and $z$ components, 
\begin{eqnarray}
\label{eq:m^xy}
m^{xy}_{\bm{Q}_\nu}&=&
\frac1N \sqrt{\sum_{i,j}  \left(S^x_i S^x_j+S^y_i  S^y_j\right)  e^{i \bm{Q}_\nu \cdot (\bm{r}_i-\bm{r}_j)}}, \\
\label{eq:m^z}
m^{z}_{\bm{Q}_\nu}&=&
\frac1N \sqrt{\sum_{i,j}  S^z_i S^z_j  e^{i \bm{Q}_\nu \cdot (\bm{r}_i-\bm{r}_j)}}, 
\end{eqnarray} 
respectively, in addition to the uniform component $m^z_0=|\sum_i S^z_i|/N$. 
While increasing the magnetic field from zero, the $n_{\rm sk}=2$ SkX changes into a triple-$Q$ state at $H \simeq 0.2$, where the uniform component of the scalar chirality $(\chi^{\rm sc}_0)^2$ vanishes. 
This triple-$Q$ state has double-$Q$ peaks in the $xy$ component and a single-$Q$ peak in the $z$ component of the magnetic moments, while it accompanies the single-$Q$ chirality density wave with $\bm{Q}_3$.
This state resembles the high-field triple-$Q$ state with $n_{\rm sk}=0$ in the Kondo lattice model in figure~\ref{fig:nsk2_SkX}(c) in section~\ref{sec:nsk2 skyrmion crystal}. 
While further increasing $H$, this state turns into the $n_{\rm sk}=1$ SkX at $H \simeq 0.4$ with a finite jump of $(\chi^{\rm sc}_0)^2$. 
For larger $H$, the system undergoes a phase transition to a triple-$Q$ state with $n_{\rm sk}=0$ at $H \simeq 1$, which is similar to the state for $0.2 \lesssim H \lesssim 0.4$. 
This triple-$Q$ state turns into the fully polarized state at $H= 2$.

Thus, the phase sequence from the $n_{\rm sk}=2$ SkX to the $n_{\rm sk}=1$ SkX, and to the triple-$Q$ state found in the Kondo lattice model in figure~\ref{fig:nsk2_SkX} is reproduced in the results for the effective spin model in (\ref{eq:effHam_spin}), except for the narrow window of the triple-$Q$ state appearing between the $n_{\rm sk}=2$ and $n_{\rm sk}=1$ SkXs. 
The difference might be attributed to the factors omitted in the effective model, such as the interactions at wave vectors other than $\bm{Q}_\nu$ and other types of magnetic interactions dropped off in the perturbation expansion. 
The good agreement again indicates that the instabilities toward the multiple-$Q$ topological spin crystals in the Kondo lattice model are well captured by the effective spin model.

\section{Extensions of the effective spin model}
\label{sec:Extensions of the effective spin model}

In the previous section, we have reviewed the previous studies showing that the effective spin model with the bilinear and biquadratic interactions in momentum space well reproduces the instabilities toward multiple-$Q$ topological spin crystals found in the Kondo lattice model. 
It opens the way for further exploration of exotic spin states by smaller computational costs than those for the original itinerant electron problems. 
Indeed, a plethora of topological spin crystals have been found in extensions of the effective model by additionally including, e.g., anisotropic interactions, single-ion anisotropy, and the Dzyaloshinskii-Moriya interaction, and some of them are shown to be relevant to experiments. 
In this section, we introduce a collection of such recent theoretical studies for centrosymmetric lattice systems in section~\ref{sec:Centrosymmetric systems} and noncentrosymmetric lattice systems in section~\ref{sec:Noncentrosymmetric systems}.

\subsection{Centrosymmetric systems}
\label{sec:Centrosymmetric systems}

In this section, we review the topological spin crystals stabilized in the presence of the magnetic anisotropy in  two centrosymmetric lattice systems. 
One is the square SkX in a square lattice system, which is stabilized by synergy between the positive biquadratic, bond-dependent anisotropic, and easy-axis anisotropic interactions~\cite{Hayami_PhysRevB.103.024439} (section~\ref{sec:Square lattice}). 
The other is the triangular SkXs and the meron crystals in a triangular lattice system, which appear in the presence of the bond-dependent anisotropic interaction and the single-ion anisotropy~\cite{Hayami_PhysRevB.103.054422} (section~\ref{sec:Triangular lattice}). 

\subsubsection{Square lattice}
\label{sec:Square lattice}

An extension of the effective model in (\ref{eq:effHam_spin}) was studied by including the effect of spin-orbit coupling on a centrosymmetric square lattice. 
The spin-orbit coupling brings the anisotropy in the magnetic interactions that satisfies the fourfold rotational symmetry of the system~\cite{Hayami_PhysRevB.95.224424,Hayami_PhysRevLett.121.137202,Hayami_PhysRevB.103.054422,Su_PhysRevResearch.2.013160,Yasui2020imaging,Hayami_PhysRevB.103.024439}. 
In the bilinear-biquadratic model, such bond-dependent anisotropy is incorporated by adding the Hamiltonian given by 
\begin{eqnarray}
\label{eq:Model}
\mathcal{H}^{\rm BA
}=  &2\sum_{\nu} 
\left[ -J  \sum_{\alpha \beta}\Gamma^{\alpha\beta}_{\bm{Q}_\nu} S^\alpha_{\bm{Q}_\nu} S^\beta_{-\bm{Q}_\nu} 
+\frac{K}{N} \left(\sum_{\alpha \beta}\Gamma^{\alpha\beta}_{\bm{Q}_\nu} S^\alpha_{\bm{Q}_\nu} S^\beta_{-\bm{Q}_\nu}\right)^2 \right] ,
\end{eqnarray}
where $\Gamma^{\alpha\beta}_{\bm{q}}$ is a $\bm{q}$-dependent dimensionless form factor.  
The precise form of $\Gamma^{\alpha\beta}_{\bm{q}}$ is set by the relativistic spin-orbit coupling under the crystalline electric field and the details of the electronic band structure~\cite{shibuya2016magnetic,Hayami_PhysRevLett.121.137202,takagi2018multiple}. 
This type of anisotropy is present even in the system with inversion symmetry, in contrast to the DM interaction which is active only when the inversion symmetry is broken. 
Similar bond-dependent anisotropy has been discussed for short-range interactions in magnetic insulators, such as the compass and Kitaev interactions~\cite{Shekhtman_PhysRevB.47.174,khomskii2003orbital,jackeli2009mott,Li_PhysRevB.94.035107,Maksimov_PhysRevX.9.021017,Matsumoto_PhysRevB.101.224419,Motome_2020}. 
In the following, we discuss the case with the relevant wave vectors along the $x$ and $y$ directions: $\bm{Q}_1=(Q,0)$ and $\bm{Q}_2=(0,Q)$. 
In this case, $\Gamma^{\alpha\beta}_{\bm{Q}_\nu}$ has the form of 
\begin{eqnarray}
\label{eq:Gamma1}
\Gamma_{\bm{Q}_1}&=\left(
\begin{array}{ccc}
-I^{\rm BA} & 0 & 0\\
0 & I^{\rm BA} &0 \\
0 & 0 & I^{z}
\end{array}
\right), 
\quad
\Gamma_{\bm{Q}_2}&=\left(
\begin{array}{ccc}
I^{\rm BA} & 0 & 0\\
0 & -I^{\rm BA} &0 \\
0 & 0 & I^{z}
\end{array}
\right). 
\end{eqnarray}
This anisotropy prefers a specific spiral plane according to the sign of $I^{\rm BA}$: 
A positive (negative) $I^{\rm BA}$ favors the spiral plane perpendicular (parallel) to $\bm{Q}_\nu$. 
In the following, we introduce the result for $Q=\pi/3$, $I^{\rm BA}>0$, and $I^z=0.2$; 
similar results are obtained for $I^{\rm BA}<0$ by exchanging the $x$ and $y$ spin components.

\begin{figure}[t!]
\begin{center}
\includegraphics[width=1.0 \hsize]{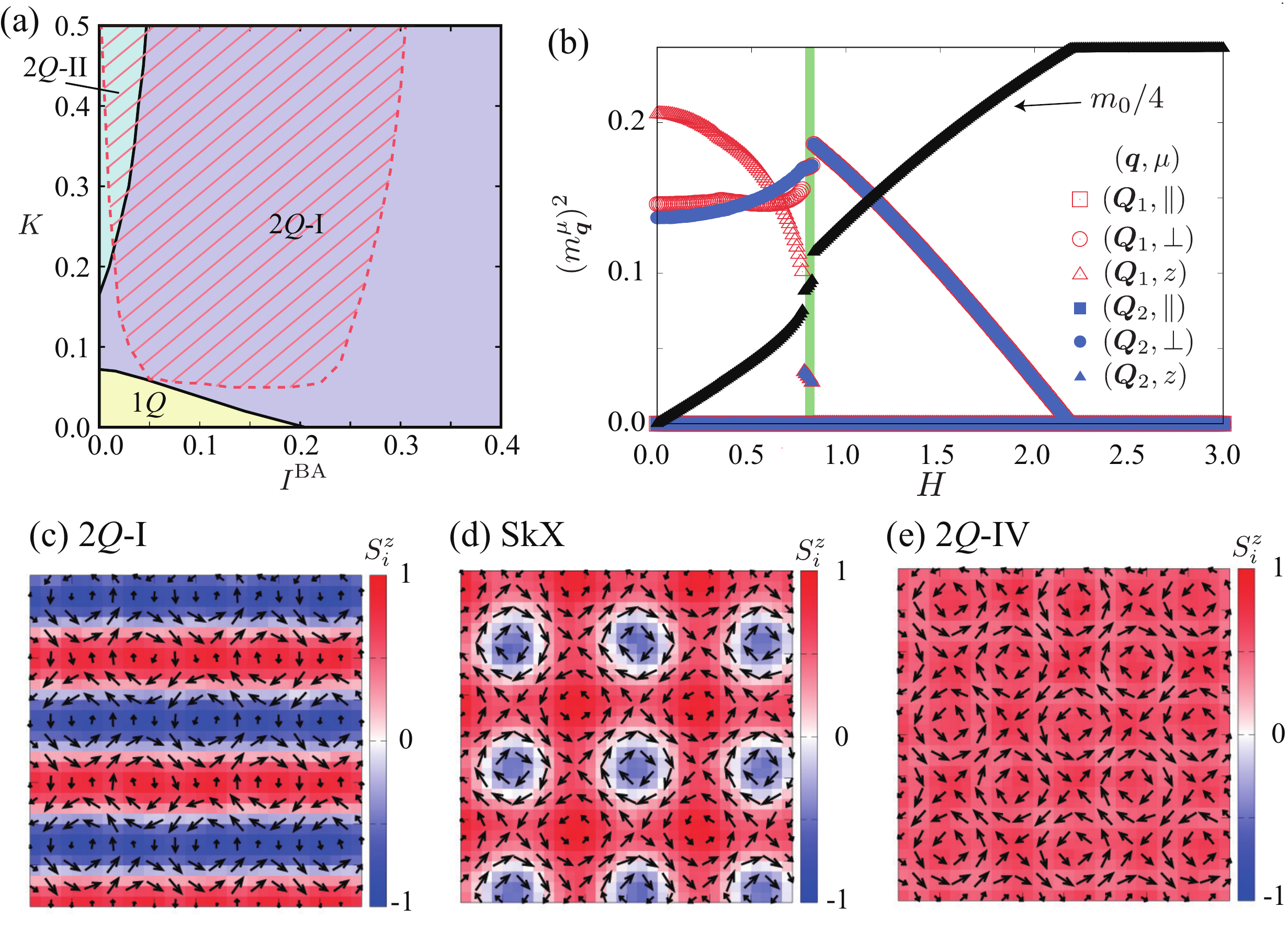} 
\caption{
\label{fig:SL}
(a) Magnetic phase diagram at zero magnetic field for the model given by $\mathcal{H}^{\rm BBQ} + \mathcal{H}^{\rm BA} + \mathcal{H}^{\rm Z}$ obtained by the simulated annealing. 
The parameters are set as $\bm{Q}_1=(Q,0)$ and $\bm{Q}_2=(0,Q)$ with $Q=\pi/3$ and 
$I^{z}=0.2$. 
2$Q$-I and 2$Q$-II stand for two different double-$Q$ states, while $1Q$ is for the single-$Q$ state. 
The hatched area shows the parameter region where the system undergoes a phase transition to a double-$Q$ state with nonzero scalar chirality including the square SkX in an applied magnetic field. 
(b) Magnetic field dependences of the magnetizations $m_0$ and $(m^{\lambda}_{\bm{Q}_\nu})^2$ ($\lambda=\parallel, \perp, z$) at $I^{\rm BA}=0.1$ and $K=0.2$; the other parameters are common to (a). 
The green region indicates the state with nonzero scalar chirality, which is identified as the square SkX. 
(c)-(e) Snapshots of the spin configurations in (c) the 2$Q$-I state at $H=0$, (d) the square SkX at $H=0.78$, and (e) the 2$Q$-IV state at $H=1$. 
The arrows and the contour show the $xy$ and $z$ components of the spin moment, respectively. 
Figures (a), (b), (d) and (e) are reprinted with permission from reference~\cite{Hayami_PhysRevB.103.024439}. Copyright 2021 by the American Physical Society.
}
\end{center}
\end{figure}

Figure~\ref{fig:SL}(a) shows the zero-field magnetic phase diagram of the model Hamiltonian given by $\mathcal{H}^{\rm BBQ}+\mathcal{H}^{\rm BA}+\mathcal{H}^{\rm Z}$ while varying $I^{\rm BA}$ and $K$, obtained by the simulated annealing~\cite{Hayami_PhysRevB.103.024439}. 
There are three magnetic phases: the single-$Q$ state for small $I^{\rm BA}$ and $K$ denoted as 1$Q$, the double-$Q$ state for large $I^{\rm BA}$ and $K$ denoted as 2$Q$-I, and the the other double-$Q$ state for the large $I^{\rm BA}$ and small $K$ denoted as 2$Q$-II. 
The 1$Q$ state is a simple proper-screw spiral whose spiral plane is perpendicular to $\bm{Q}_\nu$. 
On the other hand, the 2$Q$-I state is given by a superposition of the proper-screw spiral and the sinusoidal wave, whose real-space spin configuration is represented in figure~\ref{fig:SL}(c); the $xy$ spin component has a double-$Q$ structure with different intensities, leading 
to a periodic array of vortices. 
This state exhibits a chirality density wave along the $\bm{Q}_1$ direction, similar to the double-$Q$ CS state discussed in sections~\ref{sec:Chiral stripe} and \ref{sec:Phase diagram on a square lattice}. 
Meanwhile, in the 2$Q$-II state, both $xy$ and $z$ spin components have the single-$Q$ sinusoidal structures. 
This state also exhibits a chirality density wave along the $\bm{Q}_1$ direction. 

Although these three states at zero field show no net scalar chirality $\chi^{\rm sc}_0$, the system undergoes a phase transition to a double-$Q$ state with $\chi^{\rm sc}_0\neq 0$ under the magnetic field in the hatched area in figure~\ref{fig:SL}(a), lying across the $2Q$-I and $2Q$-II states~\cite{Hayami_PhysRevB.103.024439}. 
Figure~\ref{fig:SL}(b) exemplifies such behavior by plotting the magnetic field dependences of the squared magnetizations at $I^{\rm BA}=0.1$, $K=0.2$, and $I^z=0.2$~\cite{Hayami_PhysRevB.103.024439}. 
Here, $m^{\parallel}_{\bm{Q}_\nu}$ and $m^{\perp}_{\bm{Q}_\nu}$ are the in-plane parallel and perpendicular components of the magnetization with $\bm{Q}_\nu$, respectively [cf. (\ref{eq:m^xy}) and (\ref{eq:m^z})]. 
Three double-$Q$ states are obtained while increasing $H$, in addition to the fully polarized state above $H\simeq 2.2$. 
The low-field state below $H\simeq 0.775$ corresponds to the 2$Q$-I state connected to that at $H=0$, while the high-field state before entering the fully polarized state corresponds to a different double-$Q$ state dubbed 2$Q$-IV, whose spin structure is characterized by a superposition of two sinusoidal waves along the $\bm{Q}_1$ and $\bm{Q}_2$ directions as shown in figure~\ref{fig:SL}(e). 
The intermediate-field state, which appears in the narrow region between the 2$Q$-I and 2$Q$-IV states, shows nonzero $\chi^{\rm sc}_0$, as shown in figure~\ref{fig:SL}(b).  
The spin structure of the intermediate state is shown in figure~\ref{fig:SL}(d), which represents 
the square SkX with $n_{\rm sk}=1$ satisfying fourfold rotational symmetry with the equal weights for $\bm{Q}_1$ and $\bm{Q}_2$ in both $xy$ and $z$ spin components~\cite{Hayami_PhysRevB.103.024439}. 
Note that this SkX is energetically degenerate with the antiskyrmion counterpart in the present model; the degeneracy can be lifted by taking into account the contributions from higher harmonics, as discussed in the reference~\cite{Hayami_doi:10.7566/JPSJ.89.103702}.

The parameter region of $I^{\rm BA}$ and $K$ where the square SkX is stabilized by the magnetic field is drastically extended down to small $K$ by taking into account $I^{\rm BA}$, as shown by the hatched region in figure~\ref{fig:SL}(a)
\footnote{
In the narrow window for large $I^{\rm BA}$, a meron crystal or a topologically trivial double-$Q$ state with nonzero $\chi^{\rm sc}_0$ but vanishing $n_{\rm sk}$ appears to be stabilized, instead of the square SkX~\cite{Hayami_PhysRevB.103.024439}.}; 
the region is limited to $K \gtrsim 0.58$ at $I^{\rm BA}=0$, whereas the boundary comes down to $K \simeq 0.07$ for $I^{\rm BA} \simeq 0.05$. 
This indicates that the bond-dependent anisotropic interaction $I^{\rm BA}$ plays an important role in the stabilization of the square SkX. 
Recently, the effect of similar bond-dependent interactions including the dipolar interaction on the square SkX has been discussed in the literature~\cite{Wang_PhysRevB.103.104408,Utesov_PhysRevB.103.064414}.

It is noteworthy that the phase sequence against $H$ in figure~\ref{fig:SL}(b) well reproduces the experimental results for a centrosymmetric material GdRu$_2$Si$_2$, which was recently discovered to host the square SkX~\cite{khanh2020nanometric,Yasui2020imaging}. 
GdRu$_2$Si$_2$ is a layered material with square lattices of the localized Gd moments which couple with the itinerant electrons. 
In this compound, three distinct phases were observed besides the fully polarized state at high fields through the resonant x-ray scattering and Lorentz transmission electron microscopy measurements~\cite{khanh2020nanometric}.
From the detailed comparison between experiment and theory, it was concluded that the three phases are well explained by the 2$Q$-I, the square SkX, and 2$Q$-IV obtained for the effective spin model. 
Interestingly, the magnetic period of the square SkX in this compound is extremely short compared to those in noncentrosymmetric materials, which also supports the importance of the itinerant frustration. 
The importance of the spin-charge coupling was also confirmed by the observation of the charge density wave in the scanning tunneling microscopy experiment~\cite{Yasui2020imaging}.

\subsubsection{Triangular lattice}
\label{sec:Triangular lattice}
 
Next, we discuss multiple-$Q$ topological spin crystals found in the effective spin model on the triangular lattice~\cite{Hayami_PhysRevB.103.054422}. 
In this case, respecting the sixfold rotational symmetry and the mirror symmetry of the triangular lattice, $\bm{Q}_{\nu}$ are set as $\bm{Q}_1 = (\pi/3,0,0)$, $\bm{Q}_2 = (\pi/6,\sqrt{3}\pi/6,0)$, and $\bm{Q}_3 = (-\pi/6,-\sqrt{3}\pi/6,0)$, and the form factor in the bond-dependent anisotropy described by (\ref{eq:Model}) is taken as 
\begin{eqnarray}
\label{eq:Gamma1_TL}
\Gamma_{\bm{Q}_1}&=\left(
\begin{array}{ccc}
-I^{\rm A} & 0 & 0\\
0 & I^{\rm A} &0 \\
0 & 0 & 0
\end{array}
\right), 
\quad 
\Gamma_{\bm{Q}_2}
=\left(
\begin{array}{ccc}
\displaystyle \frac{I^{\rm A}}{2} & \displaystyle \frac{\sqrt{3}I^{\rm A}}{2} & 0\\
\displaystyle \frac{\sqrt{3}I^{\rm A}}{2} & -\displaystyle \frac{I^{\rm A}}{2} &0 \\
0 & 0 & 0
\end{array}
\right), \nonumber \\
\Gamma_{\bm{Q}_3}&=\left(
\begin{array}{ccc}
\displaystyle \frac{I^{\rm A}}{2} & -\displaystyle \frac{\sqrt{3}I^{\rm A}}{2} & 0\\
-\displaystyle \frac{\sqrt{3}I^{\rm A}}{2} & -\displaystyle \frac{I^{\rm A}}{2} &0 \\
0 & 0 & 0
\end{array}
\right). 
\end{eqnarray}
This interaction prefers a specific spiral plane according to the sign of $I^{\rm A}$, similar to that in the square lattice case in (\ref{eq:Model}) and (\ref{eq:Gamma1}). 
In addition, we discuss the effect of the local single-ion anisotropy, which also arises from the spin-orbit coupling, given by 
\begin{equation}
\label{eq:Ham_SIA}
\mathcal{H}^{\rm SIA}= -A \sum_{i} \left(S^z_i\right)^2, 
\end{equation} 
where the positive (negative) $A$ represents the easy-axis (easy-plane) anisotropy. 

Similar to the square lattice case in section~\ref{sec:Square lattice}, the effects of magnetic anisotropy have been investigated theoretically for the triangular lattice systems. 
For the bond-dependent anisotropy, the instabilities toward multiple-$Q$ states by short-range anisotropic interactions of Kitaev type were discussed in the Mott insulators with strong spin-orbit coupling~\cite{Michael_PhysRevB.91.155135,Rousochatzakis2016} and in the Ni-halide monolayer~\cite{amoroso2020spontaneous}. 
On the other hand, the effect of the single-ion anisotropy has been discussed for frustrated~\cite{leonov2015multiply,Lin_PhysRevB.93.064430,Hayami_PhysRevB.93.184413} and itinerant magnets~\cite{Hayami_PhysRevB.99.094420,Wang_PhysRevLett.124.207201}. 
In the following, we introduce the results for these two types of magnetic anisotropy in the context of the itinerant frustration.

\begin{figure*}[htb!]
\begin{center}
\includegraphics[width=0.97\hsize]{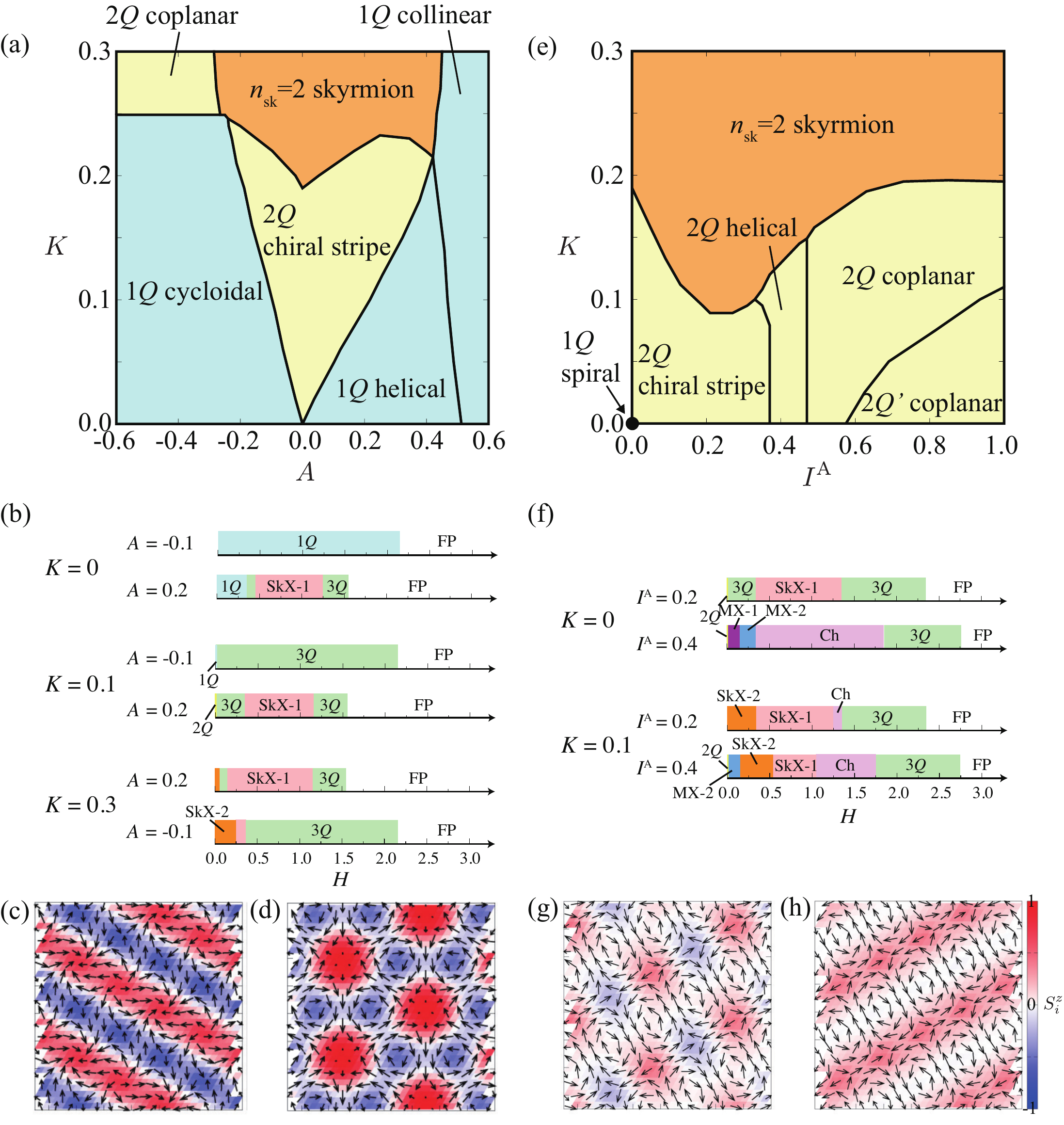} 
\caption{
\label{fig:TL}
(a, e) Magnetic phase diagrams at zero field for (a) $\mathcal{H}^{\rm BBQ}+\mathcal{H}^{\rm SIA}$ and (e) $\mathcal{H}^{\rm BBQ}+\mathcal{H}^{\rm BA}$ obtained by the simulated annealing. 
(b, f) Phase diagrams for several values of (b) $K$ and $A$, and (f) $K$ and $I^{\rm A}$ in the [001] magnetic field $H$. 
$1Q$, $2Q$, $3Q$, SkX-1, SkX-2, MX-1, MX-2, Ch, and FP stand for the single-$Q$ state, double-$Q$ state, triple-$Q$ state, $n_{\rm sk}=1$ SkX, $n_{\rm sk}=2$ SkX, $n_{\rm sk}=1$ meron crystal, $n_{\rm sk}=2$ meron crystal, multiple-$Q$ states with nonzero uniform scalar chirality, and the fully polarized state, respectively. 
(c, d) Real-space spin configurations of the $n_{\rm sk}=2$ skyrmion crystals at (c) $H=0.1$ and (d) $H=0.2$ for $K=0.3$ and $A=-0.1$.
(g, h) Real-space spin configurations of (g) the $n_{\rm sk}=1$ meron crystal at $H=0.1$ and (h) the $n_{\rm sk}=2$ meron crystal at $H=0.3$ for $K=0$ and $I^{\rm A}=0.4$. 
In (c), (d), (g), and (h), the contour shows the $z$ component of the spin moment, and the arrows represent the $xy$ components. 
Figure is reprinted with permission from reference~\cite{Hayami_PhysRevB.103.054422}. Copyright 2021 by the American Physical Society.
}
\end{center}
\end{figure*}

\paragraph{Single-ion anisotropy.}
We first discuss the effect of the single-ion anisotropy, by taking the model Hamiltonian $\mathcal{H}^{\rm BBQ}+\mathcal{H}^{\rm SIA}+\mathcal{H}^{\rm Z}$. 
Figure~\ref{fig:TL}(a) shows the zero-field magnetic phase diagram while varying $A$ and $K$, obtained by the simulated annealing~\cite{Hayami_PhysRevB.103.054422}. 
There are six phases in the phase diagram. 
Three of them are single-$Q$ states: 
The 1$Q$ cycloidal (helical) state for $A<0$ ($A>0$) has a spin spiral in the $xy$ ($xz$ or $yz$) plane, while in the 1$Q$ collinear state for large positive $A$, all the spins are aligned along the $\pm z$ direction. 
The rest three are multiple-$Q$ states appearing in the presence of the biquadratic interaction $K$. 
One is the double-$Q$ CS state in the small $|A|$ region, which is a relative of that found for $A=0$ in section~\ref{sec:Phase diagram on a triangular lattice}, with the spiral plane is fixed depending on the sign of $A$ similar to the single-$Q$ states. 
The second one is the triple-$Q$ state appearing in the larger $K$ region of the double-$Q$ CS state, which is the $n_{\rm sk}=2$ SkX with an anisotropic spin structure depending on $A$: 
The $xy$ spin component always shows the double-$Q$ structure, and becomes larger (smaller) than the single-$Q$ $z$ spin component for $A<0$ ($A>0$). 
By increasing $A$, the $xy$ spin component vanishes, and then the $n_{\rm sk}=2$ SkX turns into the $1Q$ collinear state continued from the smaller $K$ region. 
Meanwhile, by decreasing $A$, the $z$ spin component vanishes and the $n_{\rm sk}=2$ SkX changes into the third multiple-$Q$ state, the double-$Q$ coplanar state. 

When introducing the magnetic field, a further variety of the multiple-$Q$ states are  obtained~\cite{Hayami_PhysRevB.103.054422}. 
As an example, the result for the magnetic field along the [001] direction is presented in figure~\ref{fig:TL}(b) for several parameter sets of $A$ and $K$. 
In the small $K$ region where the single-$Q$ or double-$Q$ CS state is stabilized at $H=0$, the $n_{\rm sk}=1$ SkX is stabilized in the intermediate field region in the presence of the easy-axis anisotropy $A>0$
\footnote{
The $n_{\rm sk}=1$ SkX appears also for the easy-plane anisotropy $A<0$, but it is much more fragile against the anisotropy compared to the easy-axis case.}, in addition to the topologically trivial triple-$Q$ states, as shown in the results for $K=0$ and $0.1$ in figure~\ref{fig:TL}(b). 
On the other hand, in the large $K$ region, the $n_{\rm sk}=2$ SkX at zero field remains stable against both easy-axis and easy-plane anisotropy as shown in figure~\ref{fig:TL}(b). 
These behaviors of the $n_{\rm sk}=1$ and $n_{\rm sk}=2$ SkXs are qualitatively consistent with those obtained for the Kondo lattice model~\cite{Hayami_PhysRevB.99.094420}. 
Interestingly, there are two variants of the $n_{\rm sk}=2$ SkX depending on $A$ and $H$: One is a superposition of the magnetic vortices in the $xy$ spin component and the sinusoidal wave in the $z$ spin component, which breaks threefold rotational symmetry as shown in figure~\ref{fig:TL}(c), and the other retains the sytmmetry in both $xy$ and $z$ spin components as shown in figure~\ref{fig:TL}(d).

\paragraph{Bond-dependent anisotropy.}
Next, we discuss the effect of the bond-dependent anisotropy $I^{\rm A}$ for the model Hamiltonian $\mathcal{H}^{\rm BBQ}+\mathcal{H}^{\rm BA}+\mathcal{H}^{\rm Z}$. 
Figure~\ref{fig:TL}(e) shows the zero-field magnetic phase diagram obtained by the simulated annealing~\cite{Hayami_PhysRevB.103.054422}. 
The result is drastically different from that for the single-ion anisotropy $A$ in figure~\ref{fig:TL}(a). 
The difference appears already at $K=0$; the single-$Q$ states for $A$ are all replaced by the double-$Q$ states for $I^{\rm A}$. 
The double-$Q$ CS state for $0 < I^{\rm A} \lesssim 0.37$ is given by a superposition of the proper-screw spiral and the sinusoidal wave similar to that in section~\ref{sec:Phase diagram on a triangular lattice}. 
Meanwhile, the double-$Q$ helical state for $0.37 \lesssim I^{\rm A} \lesssim 0.47$ and the double-$Q$ coplanar state for $I^{\rm A} \gtrsim 0.47$ are given by a superposition of the two proper-screw spirals and two in-plane sinusoidal waves, respectively
\footnote{
The double-$Q$ coplanar state for $I^{\rm A} \gtrsim 0.47$ is further classified into two types: the isotropic one with $(m_{\bm{Q}_1})^2 = (m_{\bm{Q}_2})^2$ for $0.47 \lesssim I^{\rm A} \lesssim 0.58$ and the anisotropic one with $(m_{\bm{Q}_1})^2 > (m_{\bm{Q}_2})^2$ for $I^{\rm A} \gtrsim 0.58$, denoted as $2Q$ coplanar and $2Q'$ coplanar in figure~\ref{fig:TL}(e), respectively. 
}. 
All the double-$Q$ states change into the $n_{\rm sk}=2$ SkX while increasing $K$, as shown in figure~\ref{fig:TL}(e). 
It is worthy to note that, in contrast to the case with the single-ion anisotropy, the $n_{\rm sk}=2$ SkX for $I^{\rm A}>0$ exhibits a uniform magnetization along the $z$ direction even at zero field. 
This means that the degeneracy between the skyrmion with $n_{\rm sk}=+2$ and the antiskyrmion with $n_{\rm sk}=-2$ is lifted under the magnetic field. 

In the presence of the magnetic field, further intriguing topological spin crystals are stabilized, as shown in figure~\ref{fig:TL}(f). 
When the bond-dependent anisotropy is relatively weak, as exemplified in the results for $I^{\rm A}=0.2$, the $n_{\rm sk}=1$ SkX is stabilized irrespective of $K$, while the $n_{\rm sk}=2$ SkX and a noncoplanar state with net scalar chirality but zero skyrmion number [denoted as Ch in figure~\ref{fig:TL}(f)] are stabilized by introducing $K$. 
The interesting feature is found in the $n_{\rm sk}=1$ SkXs: 
The bond-dependent anisotropy lifts the degeneracy between the skyrmion and the antiskyrmion similar to the $n_{\rm sk}=2$ case above, and furthermore, the positive (negative) $I^{\rm A}$ stabilizes the Bloch(N\'eel)-type SkX (see figure~\ref{fig:Intro_crystal}). 
This is because the bond-dependent anisotropy under the magnetic field breaks the chiral symmetry of the system, and selects a particular vorticity and helicity
\footnote{
There remains the degeneracy between the Bloch(N\'eel)-type SkX with the helicity $\pm \pi/2$ ($0$ and $\pi$).}. 
In addition to these SkXs, while increasing the bond-dependent anisotropy, the other topological spin crystals, meron crystals, are stabilized in the weak field region, as exemplified in the results for $I^{\rm A}=0.4$ in figure~\ref{fig:TL}(f). 
There are two types of meron crystals with different skyrmion number per magnetic unit cell, $n_{\rm sk}$: One is the $n_{\rm sk}=1$ meron crystal composed of the periodic array of 
one meron-like and three antimerion-like spin textures as shown in figure~\ref{fig:TL}(g), 
the other is the $n_{\rm sk}=2$ one with four moron-like textures as shown in 
figure~\ref{fig:TL}(h).

The above results indicate that the interplay between the biquadratic interaction and the magnetic anisotropy gives rise to a plethora of topological spin crystals.
Owing to the small computational cost, the extended effective spin model is useful for a comprehensive study of the multiple-$Q$ instabilities in a wide parameter region. 
Indeed, it was found that the above model with a fine balance between the easy-plane anisotropy and the bond-dependent anisotropy accounts for the SkX with nanometer size in Gd$_3$Ru$_4$Al$_{12}$~\cite{Hirschberger_10.1088/1367-2630/abdef9}. 
Furthermore, it predicts new topological spin crystals which have never been observed in experiments, such as the $n_{\rm sk}=1$ and $n_{\rm sk}=2$ meron crystals, as shown above. 
The findings would encourage further exploration of exotic topological states.

\subsection{Noncentrosymmetric systems}
\label{sec:Noncentrosymmetric systems}

In the centrosymmetric systems discussed above, the spin-orbit coupling gives rise to the anisotropic interactions which are symmetric with respect to the spin components. 
In the noncentrosymmetric systems where the spatial inversion symmetry is broken, antisymmetric interactions can also arise from the spin-orbit coupling. 
Such antisymmetric interactions are derived by the perturbation expansion for the Kondo lattice Hamiltonian in (\ref{eq:Ham_kspace}) with the antisymmetric spin-orbit coupling described by 
\begin{eqnarray}
\label{eq:Ham_krep}
\mathcal{H}^{\rm ASOC}=  \sum_{\bm{k}} \bm{g}_{\bm{k}} \cdot c^{\dagger}_{\bm{k}\sigma}\bm{\sigma}_{\sigma \sigma'}c_{\bm{k}\sigma'},    
\end{eqnarray}
where $\bm{g}_{\bm{k}}$ is the antisymmetric vector with respect to $\bm{k}$.
By similar procedure to section~\ref{sec:Perturbation expansion}, the DM-type antisymmetric interactions, which are described by the outer products of two spins, are obtained in the first order of $\mathcal{H}^{\rm ASOC}$
\footnote{The second-order contribution leads to symmetric interactions, which include the bond-dependent interactions discussed in section~\ref{sec:Centrosymmetric systems}.}. 
In the following, we review the topological spin crystals stabilized in the presence of the DM-type interactions by focusing on the square lattice system with asymmetry (polarity) perpendicular to the plane in section~\ref{sec:Square lattice_NC}~\cite{Hayami_PhysRevLett.121.137202} and the chiral cubic lattice system in section~\ref{sec:Cubic lattice_NC}~\cite{Okumura_PhysRevB.101.144416}.

\subsubsection{Square lattice}
\label{sec:Square lattice_NC}

We first review the multiple-$Q$ topological spin crystals on a square lattice with polarity perpendicular to the plane
\footnote{
The situation is realized by the asymmetric environment between the upper and lower sides of the square plane, such as on surfaces or in heterostructures.}, where the antisymmetric spin-orbit coupling in (\ref{eq:Ham_krep}) has the form of the Rashba-type spin-orbit coupling as $\bm{g}_{\bm{k}} =(g_{\bm{k}}^x, g_{\bm{k}}^y) \propto (\sin k_y,-\sin k_x)$~\cite{Hayami_PhysRevLett.121.137202}. 
In this case, the effective spin model is given by  
\begin{eqnarray}
\label{eq:Hameff}
\mathcal{H}^{\rm polar-square} = -2\sum_{\nu} 
\left[\sum_{\alpha \beta}
J_{\nu}^{\alpha \beta} S^{\alpha}_{\bm{Q}_{\nu}} S^{\beta}_{-\bm{Q}_{\nu}}+i \bm{D}_{\nu} \cdot \left(\bm{S}_{\bm{Q}_{\nu}} \times \bm{S}_{-\bm{Q}_\nu}\right)
\right] - H \sum_i S_i^z, 
\end{eqnarray}
where $J^{\alpha\beta}_{\nu}$ and $\bm{D}_{\nu}$ are the coupling constants for the symmetric and antisymmetric exchange interactions in momentum space ($\alpha, \beta=x, y, z$); the biquadratic interaction $K$ is ignored for simplicity. 
Below, we discuss the results for $\bm{Q}_1=(0,\pi/4)$ and $\bm{Q}_2=(\pi/4,0)$, for which $J^{\alpha\beta}_{\nu}$ and $\bm{D}_{\nu}$ can be taken as $J_1^{xx}=J_2^{yy}\equiv J^{xx}$, $J_1^{yy}=J_2^{xx}\equiv J^{yy}$, $J_1^{zz}=J_2^{zz} \equiv J^{zz}$, and $D_{1}^x=-D_2^y \equiv D$ without loss of generality (all other components are zero).

\begin{figure}[t]
\begin{center}
\includegraphics[width=1.0 \hsize]{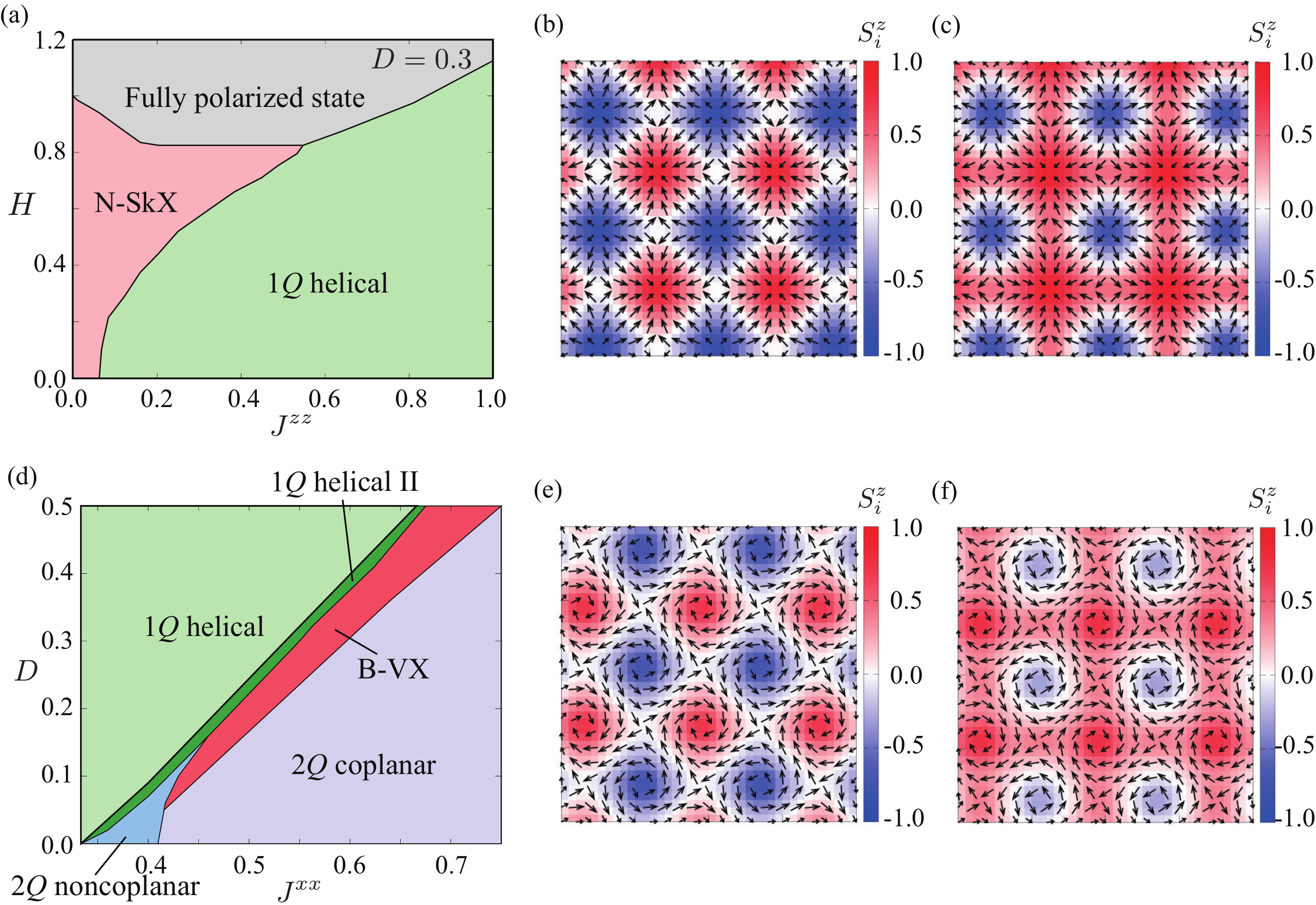} 
\caption{
\label{fig:SL_SOC}
(a) Magnetic phase diagram of the model in (\ref{eq:Hameff}) while changing $J^{zz}$ and $H$ at $D=0.3$ with $J^{xx}=J^{yy}$ ($J^{xx}+J^{yy}+J^{zz}=1$). 
$1Q$ helical and N-SkX represent the single-$Q$ spiral the N\'eel-type skyrmion crystal, respectively. 
(b, c) Real-space spin configurations of (b) the N\'eel VX at $H=0$ (see the main text) and (c) the N\'eel SkX at $H=0.3$ for $J^{zz}=0$ and $D=0.3$. 
The arrows and contour denote the $xy$ and $z$ components of the spin moments, respectively. 
(d) Magnetic phase diagram while changing $J^{xx}$ and $D$ at $H=0$ with $J^{yy}=J^{zz}$. 
B-VX represents the Bloch-type VX state. 
(e, f) Real-space spin configurations of (e) the Bloch VX at $H=0$ and (f) the Bloch SkX at $H=0.235$ for $J^{xx}=0.5625$ and $D=0.3$.  
Figure is reprinted with permission from reference~\cite{Hayami_PhysRevLett.121.137202}. Copyright 2018 by the American Physical Society.
}
\end{center}
\end{figure}

Figure~\ref{fig:SL_SOC}(a) shows the magnetic phase diagram of the effective spin model in (\ref{eq:Hameff}) by performing the simulated annealing for $J^{xx}+J^{yy}+J^{zz}=1$, $J^{xx}=J^{yy}$, and $D=0.3$~\cite{Hayami_PhysRevLett.121.137202}. 
Besides the single-$Q$ (1$Q$) helical state for large $J^{zz}$ and the fully polarized state for large $H$, the square N\'eel-type SkX is stabilized in the small $J^{zz}$ and $H$ region. 
This state is given by a superposition of the two proper-screw spirals, forming a periodic array of magnetic vortices. 
At $H=0$, the spatial regions of the vortices with $S^z_i>0$ and the antivortices with $S^z_i<0$ have the same size and shape as shown in figure~\ref{fig:SL_SOC}(b), resulting in the cancellation of the scalar chirality. 
Thus, this spin state is regarded as a N\'eel-type VX (N\'eel VX) or meron-antimeron crystal. 
While introducing $H$, the vortex regions are extended and the antivortex regions are shrunk, which turns the state into the N\'eel-type SkX with $n_{\rm sk}=1$, as shown in figure~\ref{fig:SL_SOC}(c). 
This is reasonable since the Rashba-type DM interaction is known to favor a similar N\'eel-type SkX also in the Heisenberg model with the short-range DM interaction for polar insulating magnets~\cite{nagaosa2013topological}. 

Interestingly, however, a Bloch-type SkX can also be stabilized in the effective spin model (\ref{eq:Hameff}) by tuning the symmetric anisotropic interaction. 
Figure~\ref{fig:SL_SOC}(d) shows the zero-field phase diagram in the plane of $J^{xx}$ and $D$ with $J^{yy}=J^{zz}$. 
In the intermediate region between the single-$Q$ helical state for small $J^{xx}$ and large $D$ (denoted as $1Q$ helical and $1Q$ helical II) and the double-$Q$ state for large $J^{xx}$ and small $D$ (denoted as $2Q$ coplanar), double-$Q$ noncoplanar states are stabilized by the competition between the DM interaction and the symmetric anisotropic interaction. 
One of them in the red region in figure~\ref{fig:SL_SOC}(d) has the spin configuration with a periodic array of vortices as exemplified in figure~\ref{fig:SL_SOC}(e), where the spins near the 
vortex core rotate in the tangential planes when moving from the core to periphery; this corresponds to the Bloch-type VX or meron-antimeron crystal. 
Similar to the N\'eel-type VX and SkX in figures~\ref{fig:SL_SOC}(b) and \ref{fig:SL_SOC}(c), this Bloch-type VX evolves into a Bloch-type SkX by introducing $H$, as shown in figure~\ref{fig:SL_SOC}(f). 
The result indicates that the itinerant frustration can stabilize the Bloch-type SkX even in the presence of the Rashba-type DM interaction, contrary to the conventional wisdom that such Bloch-type SkXs are stabilized by the chiral-type DM interaction~\cite{rossler2006spontaneous,nagaosa2013topological}. 

Thus, the above results show that the types of the SkXs can be controlled by not only the spin-orbit coupling but also the electronic band structure.
The former would be designed by making surfaces and heterostructures, and controlled by an external electric field, while the latter would be changed by chemical doping and an external pressure. 
Such systematic studies will give an insight into the origin of topological spin crystals in bulk, thin films, and heterostructures of magnetic metallic systems.

\subsubsection{Cubic lattice}
\label{sec:Cubic lattice_NC}

\begin{figure}[htp]
\centering
\includegraphics[width=1.0 \hsize]{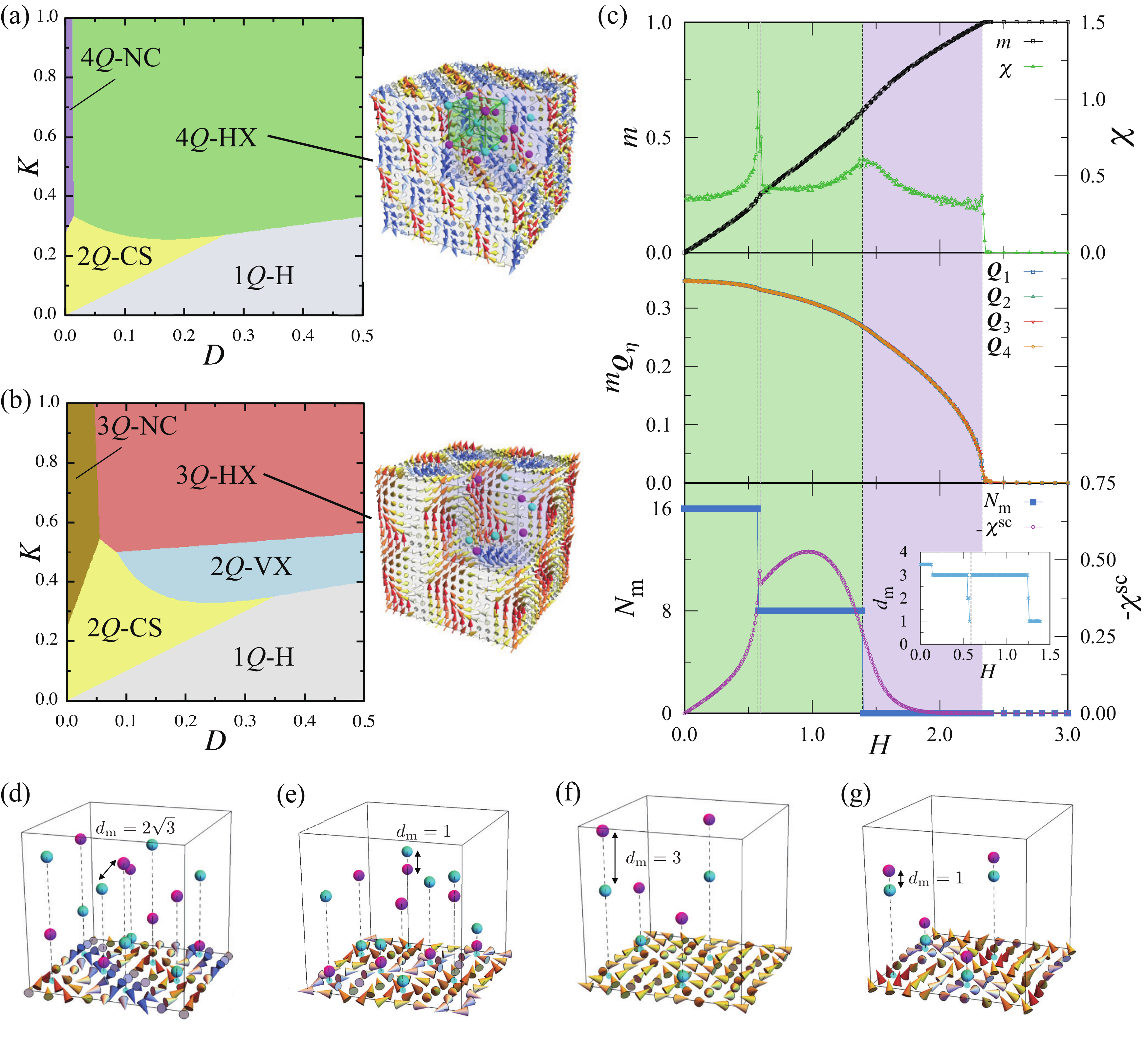}
\caption{
(a, b) Phase diagrams of the model in (\ref{eq:Ham_CL}) at zero field for the (a) 4$Q$ and (b) 3$Q$ cases obtained by the variational calculations.
HX, NC, VX, CS, and H represent the hedgehog crystal, the nonchiral, the vortex crystal, the chiral stripe, and the helical states, respectively.
Spin textures of the 4$Q$ and 3$Q$ HXs are shown in the insets of (a) and (b), respectively, where the color of the arrows represents the $z$ component of spins and the magenta (cyan) balls represent the (anti)hedgehogs corresponding to the (anti)monopoles in terms of the emergent magnetic field. 
(c) $H$ dependences of the magnetization $m$, the magnetic susceptibility $\chi$, the magnetization with wave vector $\bm{Q}_\nu$, $m_{\bm{Q}_\nu}$, the number of monopoles and antimonopoles, $N_\mathrm{m}$, and the net scalar chirality $\chi^{\mathrm{sc}}$ in the $4Q$ case with $D=0.3$ and $K=0.6$. 
The magnetic field is applied in the [001] direction. 
The black-dashed vertical lines represent the topological transitions by pair annihilation of monopoles and antimonopoles, while the gray ones represent other nontopological phase transitions.
The inset shows the minimum distance between the monopoles and antimonopoles, $d_\mathrm{m}$.
(d-g) Positions of monopoles and antimonopoles in the magnetic unit cell at (d) $H=0.00$, (e) $0.57$, (f) $0.60$, and (g) $1.39$.
The arrows at the bottom show the slice of the spin texture on the plane just below some of the monopoles and antimonopoles.
Figure is reprinted with permission from reference~\cite{Okumura_PhysRevB.101.144416}. Copyright 2020 by the American Physical Society.
}
\label{fig:CL}
\end{figure}

Finally, let us discuss three-dimensional topological spin crystals, HXs (see figure~\ref{fig:Intro_crystal}), in a noncentrosymmetric chiral system on the cubic lattice. 
The HXs are found to be stabilized by the itinerant frustration with the interplay between the biquadratic interaction and the DM-type interaction~\cite{Okumura_PhysRevB.101.144416}. 
The effective spin Hamiltonian is given by  
\begin{eqnarray}
\mathcal{H}^{\rm chiral-cubic} = \sum_{\nu}
&&\left[-J\bm{S}_{\bm{Q}_\nu}\cdot\bm{S}_{-\bm{Q}_\nu}+\frac{K}{N}\left({\bm S}_{\bm{Q}_\nu}\cdot{\bm S}_{-\bm{Q}_\nu}\right)^2 \right.
\nonumber \\
&& \qquad\qquad\quad \left.-i{\bm D}_\nu\cdot\left({\bm S}_{\bm{Q}_\nu}\times{\bm S}_{-\bm{Q}_\nu}\right)
\right]-H\sum_{i}S^z_i.\, 
\label{eq:Ham_CL}
\end{eqnarray}
where $J$ and $K$ are taken to be symmetric for simplicity; $\bm{D}_\nu$ is parallel to $\bm{Q}_\nu$ by assuming $\bm{g}_{\bm{k}}  \propto (\sin k_x,\sin k_y, \sin k_z)$ in (\ref{eq:Ham_krep}). 
This model realizes the HXs composed of multiple-$Q$ helices, as shown below: 
The HX composed of a superposition of four spin helices (4$Q$ HX) is stabilized by taking four tetrahedral $\bm{Q}_\nu$ as $\bm{Q}_1=(Q,-Q,-Q)$, $\bm{Q}_2=(-Q,Q,-Q)$, $\bm{Q}_3=(-Q,-Q,Q)$, and $\bm{Q}_4=(Q,Q,Q)$ in (\ref{eq:Ham_CL}), and the HX composed of three helices (3$Q$ HX) is stabilized by taking three cubic $\bm{Q}_\nu$ as $\bm{Q}_1=(Q,0,0)$, $\bm{Q}_2=(0,Q,0)$, and $\bm{Q}_3=(0,0,Q)$. 
The spin textures of the $4Q$ and $3Q$ HXs are shown in the insets of figures~\ref{fig:CL}(a) and \ref{fig:CL}(b), respectively. 
Experimentally, similar $4Q$ and $3Q$ HXs were discovered in the B20 compound MnSi$_{1-x}$Ge$_{x}$~\cite{tanigaki2015real,kanazawa2017noncentrosymmetric,fujishiro2019topological,Kanazawa_PhysRevLett.125.137202}. 
In the following, we discuss the results with $Q=\pi/4$.

Figures~\ref{fig:CL}(a) and \ref{fig:CL}(b) display the magnetic phase diagrams at zero field for the 4$Q$ and 3$Q$ cases, respectively, obtained by the variational calculations while changing $D=|\bm{D}_\nu|$ and $K$~\cite{Okumura_PhysRevB.101.144416}.
In both $4Q$ and $3Q$ cases, the HXs in figures~\ref{fig:CL}(a) and \ref{fig:CL}(b) are stabilized in the wide parameter range of nonzero $D$ and $K$, which indicates the importance of the interplay between the biquadratic interaction from the spin-charge coupling and the DM-type interaction from the spin-orbit coupling for the stabilization of the HXs.

Both the 4$Q$- and $3Q$-HXs have a periodic array of the topological defects where the spin length vanishes
\footnote{
On the discrete lattice, the defects were found to prefer the interstitial positions to avoid the singularities~\cite{Okumura_PhysRevB.101.144416}.}. 
The spins around the defects form hyperbolic hedgehogs and antihedgehogs, whose noncoplanar spin textures are the sources and sinks of the fictitious magnetic field emergent from the spin Berry phase mechanism, respectively. 
Hence, the hedgehogs and antihedgehogs are regarded as magnetic monopoles and antimonopoles, respectively~\cite{kanazawa2016critical}. 
The monopoles and antimonopoles are characterized by the topological numbers called the monopole charges calculated by the solid angles of the spins around the defects~\cite{Yang2016,okumura2020tracing}. 
The 4$Q$-HX has eight pairs of monopoles and antimonopoles in the magnetic unit cell so as to form two interpenetrating body-centered-cubic lattices, while the 3$Q$-HX has four pairs of monopoles and antimonopoles, which comprise spirals running in the [100], [010], and [001] directions.
The positions of monopole and antimonopoles in each HX are schematically shown as the magenta and cyan balls in the insets of figures~\ref{fig:CL}(a) and \ref{fig:CL}(b).

When the magnetic field is applied, the spin textures of the HXs are modulated, and accordingly, the monopole and antimonopoles move and may cause pair annihilation, which results in the topological phase transition. 
Figure~\ref{fig:CL}(c) shows such behavior in the case of the magnetic field applied along the [001] direction for the 4$Q$ case obtained by the simulated annealing~\cite{Okumura_PhysRevB.101.144416}. 
The field dependences of the magnetization $m$ and the magnetic susceptibility $\chi$ in the top panel of figure~\ref{fig:CL}(c) indicate that the system exhibits four phase transitions at $H\simeq0.575$, $0.595$, $1.395$, and $2.335$. 
The magnetizations with wave vector $\bm{Q}_\nu$, $m_{\bm{Q}_\nu}$, plotted in the second panel of figure~\ref{fig:CL}(c) have the equal amplitudes for the four components $\nu=1$-$4$ except for the fully polarized state for $H\gtrsim2.335$, indicating that the three phases below $H\simeq 2.335$ are $4Q$ states. 
The phase transitions at $H\simeq0.575$ and $1.395$ are the topological transitions characterized by the pair annihilation of monopoles and antimonopoles, where the number of monopoles and antimonopoles, $N_\mathrm{m}$, decreases from $16$, to $8$ and to $0$ successively, as shown in the bottom panel of figure~\ref{fig:CL}(c)
\footnote{The other phase transition at $H\simeq0.595$ is non-topological, where the higher-harmonic spin components show small changes~\cite{Okumura_PhysRevB.101.144416}.}.
The motions of the monopoles and antimonopoles while increasing $H$ are shown in figures~\ref{fig:CL}(d)-\ref{fig:CL}(g); 
the minimum distance between the monopoles and antimonopoles, $d_\mathrm{m}$, is plotted in the inset of the bottom panel of figure~\ref{fig:CL}(c). 
The change of $d_\mathrm{m}$ is related to the behavior of the net scalar chirality $\chi^\mathrm{sc}$ plotted in the bottom panel of figure~\ref{fig:CL}(c): 
The rapid changes of $|\chi^\mathrm{sc}|$ when approaching the topological transition at $H\simeq 0.575$ and $1.395$ are owing to the decrease of $d_\mathrm{m}$
\footnote{
Whether $\chi^\mathrm{sc}$ increases or decreases depends on the directions of the flows of local scalar chirality connecting the monopole-antimonopole pairs~\cite{Okumura_PhysRevB.101.144416}.
}.

The above results indicate that the long-range biquadratic and DM-type interactions are key ingredients for the stabilization of the HXs. 
This implies the importance of itinerant frustration for understanding the origin of the HXs recently discovered in MnSi$_{1-x}$Ge$_{x}$~\cite{tanigaki2015real,kanazawa2017noncentrosymmetric,fujishiro2019topological}.
The short period of the magnetic textures in experiments also supports the relevance of itinerant frustration. 
In experiments, the $3Q$ HX in MnGe is turned into the $4Q$ HX by Si doping~\cite{fujishiro2019topological}, suggesting a change in the electronic structure and the Fermi surfaces that lead to the switching of the relevant wave vectors. 
We note that a different mechanism has been suggested based on a short-range chiral-chiral interaction~\cite{grytsiuk2020topological,Mendive-Tapia_PhysRevB.103.024410}.
For deeper understanding, it is desired to clarify the electronic structure in each material, especially the doping dependence, by, e.g., the angle-resolved photoemission spectroscopy, the de Haas-van Alphen effect, and the first-principles calculations. 
It was recently shown that the sample thickness modulates the spin textures of the HX in MnGe~\cite{Kanazawa_PhysRevB.96.220414}. 
In addition, a similar but different type of modulation was found in an external magnetic field~\cite{Kanazawa_PhysRevLett.125.137202}. 
Such modulations have also been studied theoretically~\cite{Shimizu_PhysRevB.103.054427,shimizu2020spin}. 
The effective spin model based on the itinerant frustration and its extensions would be useful for understanding the experimental results. 

\section{Summary and perspective}
\label{sec:summary}

To summarize, we have reviewed recent progress in theoretical understanding of the topological spin crystals in itinerant magnets.  
The central concept is the itinerant frustration, that is the competition between the effective long-range magnetic interactions mediated by itinerant electrons. 
It has an analogy with the conventional frustration for the short-range exchange interactions in insulating magnets: 
The degeneracy at the level of bilinear interactions is lifted by long-range multiple-spin interactions inherent to the itinerant nature of electrons, in a similar manner to the conventional frustration where the degeneracy for the two-spin exchange interactions can be lifted by multiple-spin ones. 
The difference lies in the range of interactions; the itinerant frustration arises in the long-range interactions which are better described in momentum space, while the conventional one is in the exponentially short-range interactions in real space. 
This leads to a further variety of the topological spin crystals than ever. 
In addition, it is noteworthy that the itinerant frustration may give rise to topological spin crystals with very short periods down to a few lattice sites, as it is set by the inverse of the relevant Fermi wave numbers; the other conventional mechanisms like the DM interaction hardly realize such short periods within the realistic model parameters. 

As reviewed in this article, the importance of the itinerant frustration has been suggested from the careful analysis of the origin of multiple-$Q$ topological spin crystals discovered in the numerical calculations for a fundamental model for itinerant magnets, the Kondo lattice model. 
It was shown that the effective spin model with long-range bilinear and biquadratic interactions, which is constructed based on the perturbation in terms of the spin-charge coupling, well reproduces the multiple-$Q$ instabilities found in the Kondo lattice model. 
Furthermore, several extensions of the effective spin model, e.g., by including the anisotropic interactions, single-ion anisotropy, and the Dzyaloshinskii-Moriya interaction, uncovered more exotic multiple-$Q$ topological spin crystals. 
These findings are relevant to understanding of a new generation of the multiple-$Q$ topological spin crystals with unusually short magnetic periods, experimentally discovered e.g., in GdRu$_2$Si$_2$, Gd$_3$Ru$_4$Al$_{12}$, and MnSi$_{1-x}$Ge$_{x}$. 
Thus, the recent progress shows that the effective bilinear-biquadratic model in momentum space is a canonical model to discuss the itinerant frustration. 
It paves the way for further exploration of exotic topological spin crystals and associated quantum phenomena, since the computational cost is much cheaper than that for the models including itinerant electrons explicitly. 

There remain a number of interesting issues to be clarified from the concept of itinerant frustration. 
First of all, it is desired to construct the framework to evaluate the multiple-spin interactions in momentum space in a systematic way beyond the perturbative regime. 
Although the effective spin model with the bilinear and biquadratic interactions captures the instability toward the multiple-$Q$ states even for a relatively large spin-charge coupling, it is still unclear whether the other higher-order interactions can be dropped off or not in such a regime. 
More importantly, it should be clarified how the multiple-spin interactions as well as the anisotropic interactions are related with the electronic band structure. 
In particular, it would be intriguing to establish a guiding principle to enhance such interactions from the viewpoint of the band structure. 
This will give an insight to not only the exploration for further exotic topological spin crystals in experiments but also computational bottom-up engineering based on the electronic structure calculations.  

It is also interesting to investigate the possibility of multiple-$Q$ topological spin crystals arising from the competition and cooperation between the itinerant frustration in the long-range interactions and the conventional frustration in the short-range interactions. 
Some magnets including both itinerant and localized electrons may have short-range exchange interactions between the localized moments, in addition to the effective long-range interactions mediated by the itinerant electrons. 
As the conventional frustration can give rise to exotic magnetic ordered states, such as the partial disorder~\cite{Mekata_JPSJ.42.76,Motome2011,Hayami2011,Hayami2012,Ishizuka_PhysRevLett.108.257205}, the competition and cooperation with the itinerant frustration would lead to more exotic states with multiple-$Q$ modulations.

Another important issue is to construct the effective spin models for the materials hosting the multiple-$Q$ topological spin crystals. 
Although it was shown that the effective bilinear and biquadratic model and its extensions well explain the SkXs observed in GdRu$_2$Si$_2$~\cite{khanh2020nanometric,Yasui2020imaging}, and Gd$_3$Ru$_4$Al$_{12}$~\cite{hirschberger2019skyrmion,Hirschberger_10.1088/1367-2630/abdef9}, and the HXs in MnSi$_{1-x}$Ge$_{x}$~\cite{tanigaki2015real,kanazawa2017noncentrosymmetric,fujishiro2019topological,Kanazawa_PhysRevLett.125.137202}, there remain various topological spin crystals whose mechanisms are still missing, e.g., the SkXs in EuPtSi~\cite{kakihana2017giant,kaneko2018unique,tabata2019magnetic}, and Gd$_2$PdSi$_3$~\cite{kurumaji2019skyrmion,Hirschberger_PhysRevLett.125.076602,Hirschberger_PhysRevB.101.220401,Nomoto_PhysRevLett.125.117204,moody2020charge}, and the HX in SrFeO$_3$~\cite{Ishiwata_PhysRevB.84.054427,Ishiwata_PhysRevB.101.134406,Rogge_PhysRevMaterials.3.084404,Onose_PhysRevMaterials.4.114420}. 
The 4$f$-electron compound EuPtSi with the chiral lattice structure, which belongs to the same space group as MnSi, exhibits the SkX with extremely short magnetic period in the wide range of the temperature and the magnetic field~\cite{kakihana2017giant,kaneko2018unique}. 
The characteristic feature is fragility of the SkX depending on the field direction~\cite{tabata2019magnetic}, which suggests the importance of the magnetic anisotropy. 
Thus, it is expected that an extension of the effective spin model for the chiral cubic system discussed in section~\ref{sec:Cubic lattice_NC} by including other anisotropic interactions might be relevant to reproduce the experimental behaviors. 
Meanwhile, in the case of the 3$d$ perovskite SrFeO$_3$, because of the centrosymmetric lattice structure, it might be sufficient to take into account the cubic anisotropic interaction and omit the DM-type interaction for understanding the complicated phase diagram including the HX. 
In this case, however, the orbital degree of freedom, i.e., the hybridization between the Fe $3d$ and O $2p$ orbitals, might play an important role in stabilizing the multiple-$Q$ states~\cite{yambe2020double}. 
In such a situation, the extension of the effective spin model to the multi-orbital system is required. 
Furthermore, the effect of thermal fluctuations would also be important, as the HX is stabilized only at finite temperature in this system, in contrast to MnSi$_{1-x}$Ge$_{x}$~\cite{Ishiwata_PhysRevB.84.054427,Ishiwata_PhysRevB.101.134406}. 
The SkX in the 4$f$-electron compound Gd$_2$PdSi$_3$ is also worth studying based on the effective spin model with itinerant frustration, since the nesting of the Fermi surfaces has been suggested by the angle-resolved photoemission spectroscopy~\cite{Chaika_PhysRevB.64.125121,Inosov_PhysRevLett.102.046401}. 
As this compound has the centrosymmetric lattice structure, the effective spin model in section~\ref{sec:Triangular lattice} will be a good starting point to understand the origin of the SkX in this compound.  
After all, it is desired to carefully design the effective spin model compatible with the symmetry and the electronic structure of each material. 

Last but not least, it is worth examining the effect of the coupling with the other degrees of freedom in solids, such as charge, orbital, and lattice. 
As the spin textures may couple with the charge and orbital degrees of freedom in itinerant magnets, the charge and orbital in electrons can also exhibit the multiple-$Q$ density waves. 
Indeed, as mentioned in section~\ref{sec:Square lattice}, the charge density waves in the double-$Q$ square SkX as well as the other multiple-$Q$ states around it have been observed in GdRu$_2$Si$_2$ by the scanning tunneling microscopy experiment~\cite{Yasui2020imaging}. 
Furthermore, the lattice discreteness can affect the stability of the topological spin crystals. 
For example, the cores of the HXs prefer the interstitial positions of the lattice structure so as to avoid the singularity in the spin length, as discussed in section~\ref{sec:Cubic lattice_NC}. 
This in turn means that the topological spin crystals can couple with the lattice degree of freedom; namely, their formation can lead to lattice distortions, and vice versa. 
It also suggests the possibility to control the topological spin crystals not only by the electric and magnetic fields but also the lattice distortions and vibrations, the shape of samples, and the impurities and dislocations. 
The understanding of these couplings among the multiple degrees of freedom at the microscopic level will open the way to the Berry phase engineering of further intriguing quantum transports and multiferroic responses.

\ack
The authors thank Y. Akagi, K. Barros, C. D. Batista, T. Hanaguri, M. Hirschberger, Y. Kato, N. D. Khanh, S.-Z. Lin, T. Matsumoto, T. Misawa, K. Okada, T. Okubo, S. Okumura, R. Ozawa, S. Seki, K. Shimizu, Y. Su, R. Takagi, Y. Tokura, M. Udagawa, Y. Yamaji, R. Yambe, and Y. Yasui for fruitful collaborations and constructive discussions. 
The authors thank Y. Fujishiro, S. Ishiwata, H. Ishizuka, Y. Kamiya, N. Kanazawa, 
T. Kurumaji, and R. Takashima for their helpful discussions. 
This research was supported by JSPS KAKENHI Grants Numbers JP18K13488, JP19K03752, JP19H01834, JP19H05825, and by JST PREST (JPMJPR20L8) and JST CREST (JP-MJCR18T2). 
Parts of the numerical calculations were performed in the supercomputing systems in ISSP, the University of Tokyo.

\vspace{8mm}

\bibliographystyle{iopart-num}
\bibliography{ref}

\end{document}